\documentclass{article}

\usepackage{arxiv}

\usepackage[utf8]{inputenc} % allow utf-8 input
\usepackage[T1]{fontenc}    % use 8-bit T1 fonts
\usepackage{hyperref}       % hyperlinks
\usepackage{url}            % simple URL typesetting
\usepackage{booktabs}       % professional-quality tables
\usepackage{amsfonts}       % blackboard math symbols
\usepackage{nicefrac}       % compact symbols for 1/2, etc.
\usepackage{microtype} 
\usepackage{amsmath}
\usepackage{array}
\usepackage{geometry}
\usepackage{pgfplots}

\usepackage{multirow}
\usepackage{tikz}
\usetikzlibrary{shapes, arrows.meta, positioning}
\usetikzlibrary{decorations.pathreplacing}
\usetikzlibrary{shadows}
 \usetikzlibrary{calc}
 \usepackage{forest}

\usepackage{tikz}
\usepackage{tikz} \usetikzlibrary{calc, arrows.meta, intersections, patterns, positioning, shapes.misc, fadings, through,decorations.pathreplacing}

\usepackage{lipsum}
\usepackage{graphicx}
\graphicspath{ {./images/} }
\hypersetup{
  colorlinks,
  citecolor=blue,
  linkcolor=blue,
  urlcolor=blue}

\title{Shift-Left Techniques in Electronic Design Automation: A Survey}

\author{
 Xinyue Wu$^{\footnotemark[1]}$\\
  Global College\\
  Shanghai Jiao Tong University\\
  800 Dong Chuan Rd, Shanghai, China \\
  \texttt{wuxinyue@sjtu.edu.cn} \\
\And
 Zixuan Li$^{\footnotemark[1]}$\\
  Global College\\
  Shanghai Jiao Tong University\\
  800 Dong Chuan Rd, Shanghai, China \\
  \texttt{zixuanli@sjtu.edu.cn} \\
\And
 Fan Hu$^{\footnotemark[1]}$\\
  Global College\\
  Shanghai Jiao Tong University\\
  800 Dong Chuan Rd, Shanghai, China \\
  \texttt{hu-fan@sjtu.edu.cn} \\
\And
 Ting Lin\\
  Global College\\
  Shanghai Jiao Tong University\\
  800 Dong Chuan Rd, Shanghai, China \\
  \texttt{ting\_lin@sjtu.edu.cn} \\
\And
 Xiaotian Zhao\\
  Global College\\
  Shanghai Jiao Tong University\\
  800 Dong Chuan Rd, Shanghai, China \\
  \texttt{xiaotian.zhao@sjtu.edu.cn} \\
\And
 Runxi Wang\\
  Global College\\
  Shanghai Jiao Tong University\\
  800 Dong Chuan Rd, Shanghai, China \\
  \texttt{wangrunxi@sjtu.edu.cn} \\
\And
 Xinfei Guo$^{\footnotemark[2]}$\\
  Global College\\
  Shanghai Jiao Tong University\\
  800 Dong Chuan Rd, Shanghai, China \\
  \texttt{xinfei.guo@sjtu.edu.cn} \\
}

\begin{document}
%\maketitle
\fancyhf{}
\fancyhead[C]{\copyright {X. Wu, Z. Li, F. Hu et al.} {2025}. This is the author's version of the work only for personal use. The definitive version of Record is under review by ACM, copyright may be transferred without notice.}
\fancyfoot[C]{\thepage}

\footnotetext[1]{These authors contributed equally to this work.}
\footnotetext[2]{Corresponding author.}

\maketitle
\begin{abstract}
The chip design process involves numerous steps, beginning with defining product requirements and progressing through architectural planning, system-level design, and the physical layout of individual circuit blocks. As the enablers of large-scale chip development, Electronic Design Automation (EDA) tools play a vital role in helping designers achieve high-quality results. The Shift-Left methodology introduces a pathway toward creating digital twins and fusing multiple design steps, thereby transitioning traditionally sequential, physically-aware processes into virtual design environments. This shift allows designers to establish stronger correlations earlier and optimize designs more effectively. However, challenges remain, especially in accurately replicating downstream behaviors and determining the right scope and timing for adoption. These challenges, in turn, have revealed new opportunities for EDA vendors, physical designers, and logic designers alike. As the industry advances toward intelligent EDA tools and techniques, it is timely to reflect on Shift-Left progress made and the challenges that remain. The rise of AI techniques and the momentum of open-source design flows have significantly strengthened prediction and modeling capabilities, making data-driven methods increasingly relevant to the EDA community. This, in turn, enhances the ``Shift-Left'' features embedded in current tools. In this paper, we present a comprehensive survey of existing and emerging paradigms in Shift-Left research within EDA and the broader design ecosystem. Our goal is to provide a unique perspective on the state of the field and its future directions. Relevant papers mentioned are organized in~\href{https://github.com/iCAS-SJTU/Shift-Left-EDA-Papers}{https://github.com/iCAS-SJTU/Shift-Left-EDA-Papers}.
\end{abstract}

\keywords{Shift-Left techniques, Design efficiency, Design closure, Co-design}

%\received{XXX 2025}
%\received[revised]{XXX 20XX}
%\received[accepted]{XXX 20XX}

%\maketitle

%Added for saving space
\setlength\textfloatsep{\baselineskip}
\setlength\belowcaptionskip{-10pt}
%%%%%%%%%%%
\section{Introduction and Background}
The term ``Shift-Left'' \footnote{Some industries also termed it as ``Shift Up''.} emerged from the software testing industry in response to the escalating expenses associated with redesigns, aiming at discovering the bugs sooner than later to avoid surprises ~\cite{smith2001shift,hutchison2013shift,phan2023challenges}. The core principle behind the Shift-Left testing approach is to proactively identify and mitigate risks by conducting tests early and iteratively. This concept has transcended its origins and found application in diverse industries, such as manufacturing, where preemptive data-driven strategies are employed well before production begins. In general, the Shift-Left methodology encompasses a series of interconnected steps that expose the strong interdependency among them. Its adoption has facilitated expedited product and process design while enhancing cost-effectiveness for businesses.

%The key idea of shifting left testing method is to identify problems and minimize risks as early as possible by testing early and repeatedly. The concept has expanded into other industries including manufacturing where companies are using data to address concerns long before the manufacturing stage commences. In general, the shift left method applies to a sequence of steps that show strong dependency of each other, and has helped companies design the products and processes faster and with greater cost efficiency.

%The concept known as "Shift Left" (also referred to by some companies as "Shift Up") emerged from the software testing sector in response to the escalating expenses associated with redesigns. Its objective is to detect defects earlier in the development cycle, thereby averting unexpected issues. The core principle behind the shift left testing approach is to proactively identify and mitigate risks by conducting tests early and iteratively. This concept has transcended its origins and found application in diverse industries, such as manufacturing, where preemptive data-driven strategies are employed well before production begins. In essence, the shift left methodology encompasses a series of interconnected steps, emphasizing the interdependency among them. Its adoption has facilitated expedited product and process design while enhancing cost-effectiveness for businesses.
Recently, the Shift-Left methodology has been increasingly introduced to the Electronic Design Automation (EDA) industry to indicate tasks that were once performed sequentially must now be done concurrently. EDA softwares are the enablers of large-scale chips, and play a critical role in automating the major design tasks to ensure required performance, power and area (PPA) budgets. This is usually due to a tightening of dependencies between tasks in the chip design cycle and is driven by the escalating complexity of chips and the growing demands for superior design quality   ~\cite{shiftleft,smith2001shift,hutchison2013shift,bhardwaj2021shift}. Major EDA industries all started to push various types of Shift-Left updates in their tools to boost the engineering efficiency   ~\cite{synospsys,cadence,keysight,Siemens}. To understand the progress that has been made in this emerging field and recognize future opportunities to shift left, in this paper, we present a comprehensive study that summarizes the Shift-Left movement that has happened in the EDA field. The study picked up the isolated pieces of work that have hinted the Shift-Left update and chain them together in a systemic way following the order of design process. We sincerely hope that this study can serve as a starting point to enable more Shift-Left research in a more systematic way and foresee development opportunities that can help reduce the chip design cycle significantly. This survey is also the first one to formalize the problem of ``Shift-Left'' in the EDA field, with key metrics, challenges identified. \par

\section{Related Survey}
\label{relatedsurvey}
``Shift left'' in EDA field has been informally discussed across various industry forums, but it was not formally recognized as a research problem until recently.  ~\cite{bhardwaj2021shift} outlined several reasons why Shift-Left strategies were introduced by commercial EDA tool vendors over the years to address the design convergence problem.  ~\cite{10531906} discussed two key aspects of Shift-Left updates in EDA, specifically logic synthesis and macro placement. In  ~\cite{kahng2022machine}, the concept of ``shift left and up'' was mentioned to highlight its significant potential value, particularly in the detailed routing stage of physical design.  ~\cite{shiftleft} provided recent perspectives on the Shift-Left paradigm in EDA; however, it was written in Chinese, limiting its accessibility to a wider audience. Additionally, it focused more on the challenges rather than the progress made so far.

In summary, there is still a gap in the literature that formally defines the Shift-Left problem in EDA, tracks its progression, and outlines future directions. While EDA research is somewhat scattered, with individual tools or optimizations dispersed across various sources, Shift-Left has progressed rapidly. However, there is a need to consolidate past and recent advancements into a single, integrated view. Moreover, there are currently no methods to quantify Shift-Left; it remains more of a concept, which hinders its further development. This motivates this survey, which aims to provide a comprehensive overview of the Shift-Left movement in EDA, including perspectives on quantifying or evaluating its effectiveness and integrating existing Shift-Left advancements based on a typical design flow.

In this survey, we present a taxonomy, methodology, and comprehensive updates on Shift-Left strategies. With the rapid adoption of Machine Learning (ML) and Artificial Intelligence (AI) in EDA, these technologies have become key enablers behind many recent Shift-Left developments. Unlike prior AI-for-EDA (AI4EDA) surveys, which focus primarily on AI methods themselves, such as ML ~\cite{kahng2022machine,huang2021machine,fallon2020machine,ren2022machine}, graph neural networks (GNNs) ~\cite{lopera2021survey,sanchez2023comprehensive}, and large language models (LLMs) ~\cite{xu2025large,zhong2023llm4eda,chen2024dawn,he2024large,alsaqer2024potential,fang2025survey,pan2025survey}, this work provides a systematic view of how Shift-Left strategies have been applied in EDA. We cover both classical and recent works, offering a horizontal classification by prediction tasks and architectural studies, and discuss open challenges and future directions for advancing Shift-Left practices.

\section{Shift-Left Techniques in EDA: An Overview}
\label{key_concepts}
The goal of this paper is to formulate the Shift-Left problem in the EDA field and organize existing work according to the Shift-Left paradigm. Before reviewing prior researches, this section first defines the problem and introduces the key concepts and metrics used to evaluate the quality of Shift-Left approaches.
\begin{table}[t!]
\centering
\caption{Comparison of Shift Left, DTCO, HW/SW Co-Design, and Cross-Layer Design in EDA}
\footnotesize
\renewcommand{\arraystretch}{1.0}
\resizebox{\textwidth}{!}{
\begin{tabular}{|l|c|c|c|c|}
\hline
\textbf{Approach} & \textbf{Shift-Left} & \textbf{DTCO} & \textbf{HW/SW Co-Design} & \textbf{Cross-Layer Design} \\
\hline
\textbf{Main Axis} & Time (Early Dev $\leftarrow$) & Design $\leftrightarrow$ Tech & HW $\leftrightarrow$ SW & Multi-Abstraction Layers \\
\hline
\textbf{Scope} & Full Design Flow & Tech + Circuit/Layout & Embedded Systems, System on Chips (SoCs) & Full Stack (Device $\rightarrow$ App) \\
\hline
\textbf{Focus} & Early Validation, Testing & PPA via Tech-Aware Design & Functional Partitioning & Holistic Optimization \\
\hline
\textbf{Stakeholders} & Designers, EDA Vendors & Foundry + Designers & HW/SW Architects & Multi-disciplinary Teams \\
\hline
\textbf{Goal} & Early Feedback, Reduce Bugs & Tech-Driven PPA Gains & Joint HW/SW Efficiency & End-to-End System Efficiency \\
\hline
\textbf{Example} & RTL Power Estimation & FinFET Cell Co-Design & ML Accel on FPGA & Aging- or Thermal-Aware Logic \\
\hline
\end{tabular}
}
\label{tab:eda_comparison}
\end{table}

% \begin{table}[t]
% \centering
% \caption{Differences/Similarities between Shift Left, DTCO, HW/SW Codesign, and Cross-layer Design in EDA.}
% \resizebox{\textwidth}{!}{
% \begin{tabular}{|c|c|c|c|c|}
% \toprule \toprule
% \textbf{Aspect} & \textbf{Shift Left} & \textbf{DTCO} & \textbf{HW/SW Codesign} & \textbf{Cross-layer Design} \\ \hline
% \textbf{Focus} &
% Early tasks (verification, testing). &
% Design and tech co-opt. &
% HW and SW co-design. &
% Layered optimization. \\ \hline

% \textbf{Goal} &
% Catch issues early, save time/cost. &
% Align design with manufacturing. &
% Balance HW and SW. &
% Optimize across layers. \\ \hline

% \textbf{Use} &
% Logic synthesis, verification. &
% Layout, device design. &
% Embedded systems, SoC. &
% Low-power, high-perf systems. \\ \hline

% \textbf{Challenges} &
% Early accuracy, task balance. &
% Design-tech trade-offs. &
% HW/SW sync, trade-offs. &
% Layer coordination, complexity. \\ \hline

% \textbf{Benefits} &
% Faster convergence, better QoR. &
% Better PPA, manufacturability. &
% Balanced system, integration. &
% System-level performance. \\ \hline

% \textbf{Stage} &
% Early design stages. &
% Throughout design flow. &
% System level. &
% Across all layers. \\ \hline

% \textbf{Collaboration} &
% Verification teams. &
% Design/tech teams. &
% HW and SW teams. &
% HW, SW, and systems teams. \\ \bottomrule
% \end{tabular}}
% \label{tab:eda_comparison}
% \end{table}

\subsection{Formulate Shift-Left Problems in EDA}
As the EDA industry advances into the ``age of fusion,'' several key design methodologies have emerged to enhance design efficiency and quality. These include Shift-Left, Design-Technology Co-Optimization (DTCO), Hardware and Software (HW/SW) Co-design, and Cross-layer Design, each with its own distinct focus and goals, yet sharing commonalities. As summarized in Table ~\ref{tab:eda_comparison}, Shift-Left prioritizes moving tasks sequentially, such as verification and testing, earlier in the design cycle to reduce time, cost, and risk by identifying issues sooner. This approach has now been expanded to cover the entire design flow. On the horizontal stack, DTCO emphasizes the co-optimization of design and manufacturing processes, aligning design decisions with technological capabilities to improve manufacturability and overall PPA. HW/SW Co-design involves the simultaneous design and optimization of hardware and software, striving to balance and integrate these components for optimal system performance. Cross-layer Design takes a broader approach by optimizing across multiple abstraction layers, from hardware to software, to enhance system-level performance, reliability, power efficiency, and area. While each paradigm targets different stages and aspects of the design process, they all share the common goal of improving design quality by addressing challenges earlier and more holistically. Shift-Left and DTCO are primarily concerned with the timing of tasks and the integration of technology and design, respectively, whereas HW/SW Co-design and Cross-layer Design focus on collaboration and optimization across components and layers. Despite these differences, all four paradigms require close collaboration among various teams and tools (e.g., verification, design, technology, hardware, and software) to achieve the desired results. Together, they contribute to a more efficient and effective EDA process by ensuring that various aspects of design are addressed in a timely and integrated manner. In this paper, we primarily focus on the Shift-Left approach.

% Testing has been the earliest adopter for the Shift-Left approach. It was defined as an approach to software testing and system testing in which testing is performed earlier in the lifecycle (i.e. moved left on the project timeline) ~\cite{wikipedia}. Typically large waterfall V model’s types of testing are shifted left to become increments of the corresponding types of testing in the smaller incremental V models. While testing is one critical part of the whole design phase, there are many steps are typically run sequentially, which now needs to run concurrently to ensure the design efficiency and quality. 
If we take any design step in a long chip design cycle and deem it as $Flow_i$, the key Shift-Left problem can be formulated as:

\begin{quote}
\textit{How can we \textbf{predict}, \textbf{detect}, and \textbf{address} design issues systematically at \textbf{earlier} stages of the design flow, reducing reliance on late-stage fixes?}
\end{quote}

In a more abstract form, the Shift-Left problem can be described as:
\begin{itemize}
    \item \textbf{Given}:
    \begin{itemize}
        \item Partial or incomplete design data (e.g., behavioral models, early Register Transfer Level (RTL), netlists, floorplans, etc.)
        \item Predictive models for PPA, timing, reliability, security, manufacturability, etc.
    \end{itemize}

    \item \textbf{Find}:
    \begin{itemize}
        \item High-confidence detection of critical design issues
        \item Optimal design adjustments at minimal cost in early phases
    \end{itemize}

    \item \textbf{Objective}:
    \begin{align*}
    \text{\textit{Minimize}} \quad & \text{COST}_{\text{late-stage corrections}} + \text{Design Cycle Time} \\
    \text{\textit{s.t.}} \quad & \text{PPA, Reliability, Manufacturability Constraints}
    \end{align*}
\end{itemize}
where $\text{COST}_{\text{late-stage corrections}}$ includes the design effort required to fix violations and perform rework in the later stages of the design cycle. The ``Design Cycle Time'' refers to the total duration needed to achieve design closure. The Shift-Left strategy must ensure that design quality is not compromised by maintaining key metrics such as PPA, as well as reliability. Moreover, the design must remain clean and compliant with manufacturing rules and constraints.

%  semiconductor development flow, tasks once performed sequentially must now be done concurrently.

% Electronic Design Automation (EDA) refers to detecting and addressing design issues earlier in the semiconductor development lifecycle, thus improving time-to-market, reducing costs, and improving quality. Here are several formulated Shift Left problems within the EDA context:

%\renewcommand{\arraystretch}{1.42}
% Waterfall
% V-curve
%Shift Left in EDA has been presented in different formats. For example, the DTCO, the CoDesign and Cross-layer design all present similar property as shift left. The aim is to interconnect the different steps or aspects in a closer way to achieve a common best goal.

\subsection{Quantify Shift-Left in EDA}
Shift-Left strategies have emerged across various design steps and contexts within the EDA flow. As discussed in the previous section, it is crucial to develop clear metrics for characterizing the quality and effectiveness of these Shift-Left efforts. Defining anchor points in the design flow is essential for tracking and quantifying ``Shift-Left'' movements. These anchor points (AP) represent key, well-defined stages in a typical EDA flow. Table \ref{tab:anchor_points_ecoincluded} summarizes a set of anchor points, moving from early design to final signoff in a typical digital design flow.

\begin{table}[!t]
\centering
\footnotesize
\caption{Grouped Anchor Points in the EDA Design Flow, Including ECO and Tapeout.}
\resizebox{\textwidth}{!}{
\begin{tabular}{|c|l|p{8.8cm}|}
\hline
\textbf{Anchor ID
%\footnote{If one stage (e.g., synthesis, placement, route, etc.) has multiple detailed steps, these are viewed as one single anchor ID.}
} & \textbf{Name} & \textbf{Description} \\
\hline
\multicolumn{3}{|c|}{\textbf{Pre-RTL}} \\
\hline
AP0 & Specification & High-level functional and architectural intent; no HDL available yet. \\
AP1 & Electronic System Level (ESL) Modeling & System-level modeling (e.g., SystemC, C++); used for early functional and performance validation. \\
\hline
\multicolumn{3}{|c|}{\textbf{Front-End}} \\
\hline
AP2 & RTL Design (including High Level Synthesis (HLS)) & Register-transfer level description in Verilog/VHDL/High-level languages. \\
AP3 & Logic Synthesis & RTL is compiled into a gate-level netlist with early estimates for timing and power. \\
\hline
\multicolumn{3}{|c|}{\textbf{Back-End/Physical design}} \\
\hline
AP4 & Floorplanning & Chip-level layout planning, block shaping, and I/O placement. \\
AP5 & Placement & Macros/Standard cells placed with basic timing and congestion estimates. \\
AP6 & Clock Tree Synthesis (CTS) & Clock buffers inserted and routed to meet skew and latency targets. \\
AP7 & Routing & Interconnects routed; parasitics extracted for accurate analysis. \\
\hline
\multicolumn{3}{|c|}{\textbf{Signoff}} \\
\hline
AP8 & Signoff Analysis & Final verification of timing, power, Design Rule Chesk (DRC), Layout Versus Schematic (LVS), Signal Integrity (SI), and reliability. \\
\hline
\multicolumn{3}{|c|}{\textbf{ECO (Engineering Change Order)}} \\
\hline
AP9 & ECO Fixes & Late-stage functional/timing fixes with minimal physical impact. \\
\hline
\multicolumn{3}{|c|}{\textbf{Tapeout}} \\
\hline
AP10 & Tapeout & Final GDSII generation and data preparation for fabrication. \\
\hline
\end{tabular}
}
\label{tab:anchor_points_ecoincluded}
\end{table}

Based on this flow, we define the following taxonomy, which will be used consistently throughout this paper to describe and evaluate Shift-Left movements:

\begin{enumerate}
    \item \textbf{Starting Point:} The original stage in the design flow where a task or analysis traditionally occurs.
    \item \textbf{End Point:} The earlier stage to which the task or analysis is shifted.
    \item \textbf{Number of Steps:} The number of design stages traversed during the shift (i.e., how early the task has been moved).
    \item \textbf{Shift Runtime Overhead:} The additional computational or tool runtime introduced by performing the task earlier in the flow.
    \item \textbf{Shift QoR Gain:} The measurable improvement in Quality of Results (QoR), such as PPA, reliability, or manufacturability, achieved through the shift.
\end{enumerate}

It should be noted that, in parallel with the design stages summarized in Table~\ref{tab:anchor_points_ecoincluded}, test and verification activities are conducted throughout the entire flow, accompanying each anchor point. While these efforts are tightly integrated with the respective design phases, this paper will discuss them separately to highlight their distinct Shift-Left opportunities and methodologies.

% Define diffent ancor poins in the design flow
% Starting point
% End point
% Number of Steps
% Runtime Overhead
% QoR Gain

\subsection{Shift Approaches}
Based on our observations and a comprehensive survey of existing Shift-Left practices in both industry and academia within the EDA domain, we classify Shift-Left approaches into two main categories: \textbf{Virtual Prototypes (VPs)} and \textbf{Fused Actions (FAs)}. These acronyms will be used throughout the remainder of this paper. The details and distinguishing characteristics of each category are presented in the following sections.

\subsubsection{Virtual Prototypes (VPs)}

A \textit{Virtual Prototype} creates a digital twin that mimics the behavior of a downstream design stage by predicting either the design behavior or the corresponding optimization targets. This allows earlier stages to gain insight into the impact of their decisions on later stages without invoking full execution of the corresponding EDA tools. Typical strategies include predicting optimal design configurations, estimating PPA metrics ~\cite{lin2023hl}, or approximating them using intermediate proxies such as wirelength ~\cite{chen2022pros}, congestion ~\cite{gao2022congestion}, or timing slack ~\cite{guo2022timing}.

By integrating these predicted “twins,” the starting point acquires awareness of downstream constraints and outcomes, enabling early-stage guidance. The effectiveness of this strategy hinges on the accuracy of the prediction model. Discrepancies between predicted and actual outcomes may reduce the benefit of early action and, in some cases, lead to over- or under-optimization. Recent advances in ML, deep learning (DL), and LLMs have significantly improved the fidelity of such predictions through data-driven modeling. Additionally, reinforcement learning (RL) has been adopted in many works (e.g.,~\cite{huang2021machine}) to iteratively optimize design decisions, using virtual prototypes as calibrated feedback mechanisms for learning agents.

\subsubsection{Fused Actions (FAs)}
A second Shift-Left strategy is action fusion, in which later stages or sub-stages of the EDA flow are moved earlier and integrated with preceding steps. By aligning with early-stage operations and establishing a uniform interface among tools, fused actions help reduce turnaround time and accelerate PPA convergence. For instance, physically-aware synthesis can incorporate early global placement or even clock tree synthesis, allowing stages such as AP3 and AP6/AP7 to overlap, thereby improving QoR. Moreover, this approach reduces runtime by executing certain tasks earlier, which is especially beneficial for large-scale designs.

Fundamentally, this strategy addresses the limitations of the traditional divide-and-conquer paradigm in EDA, where iterative corrections are costly and tools typically optimize within narrow design space. As a consequence of this paradigm shift, many EDA vendors have started developing shared databases that enable seamless data exchange across tools. An example is the open-source toolchain OpenROAD ~\cite{openroad}, which integrates multiple standalone tools into a unified physical synthesis flow. It also leverages a data collection and storage infrastructure called METRICS 2.0 ~\cite{jung2021metrics2} to facilitate performance tracking and optimization. Overall, this fusion-based Shift-Left strategy has attracted growing attention in the industry, with several recent commercial EDA products following this trend ~\cite{cadence,synospsys,Siemens,keysight}.

\subsubsection{Comparisons of Two Shift Approaches}

\begin{table}[!t]
\centering
\caption{Comparison of Virtual Prototypes (VPs) and Fused Actions (FAs)}
\renewcommand{\arraystretch}{1.0}
\resizebox{0.8\textwidth}{!}{
\begin{tabular}{|c|l|l|}
\hline
\textbf{Shift Approach} & \textbf{Virtual Prototypes (VPs)} & \textbf{Fused Actions (FAs)} \\
\hline
\textbf{Core Idea} & Predict downstream stages & Merge downstream actions early \\
\hline
\textbf{Main Shift Goals} & Early-stage awareness & Speed up flow, improve QoR \\
\hline
\textbf{Approach} & ML/DL/RL-based prediction & Tool-stage integration \\
\hline
\textbf{Example} & Estimating PPA during synthesis & Placement-aware synthesis \\
\hline
\textbf{Advantages} & Fewer iterations, better early decisions & Faster runtime, tighter convergence \\
\hline
\textbf{Risks} & Prediction errors, overfitting & Tool coupling, reduced modularity, data consistency \\
\hline
\textbf{Usage} & ML-augmented EDA tools & Unified toolchains (e.g., OpenROAD) \\
\hline
\textbf{Viewpoint} & ``Look ahead'' & ``Pull forward'' \\
\hline
\end{tabular}
}
\label{tab:vp_vs_fa}
\end{table}

Table~\ref{tab:vp_vs_fa} presents a concise comparison between VPs and FAs, two Shift-Left strategies aimed at improving the efficiency and quality of modern EDA flows. VPs rely on predictive models—often driven by ML, DL, or RL to forecast downstream design metrics such as Power and Timing, enabling early-stage tools to make more informed decisions. In contrast, FAs focus on merging or overlapping traditionally sequential design stages, such as integrating placement information into synthesis, to reduce runtime and accelerate convergence. While VPs offer the advantage of early insight with minimal disruption to existing flow structure, they are susceptible to prediction inaccuracies. FAs, on the other hand, can achieve tighter tool integration and better runtime but may introduce complexity in tool interoperability and reduce modularity. These two strategies represent complementary directions in rethinking the traditional divide-and-conquer paradigm in EDA.

\begin{figure*}[!ht]
  \centering
\includegraphics[width=\textwidth]{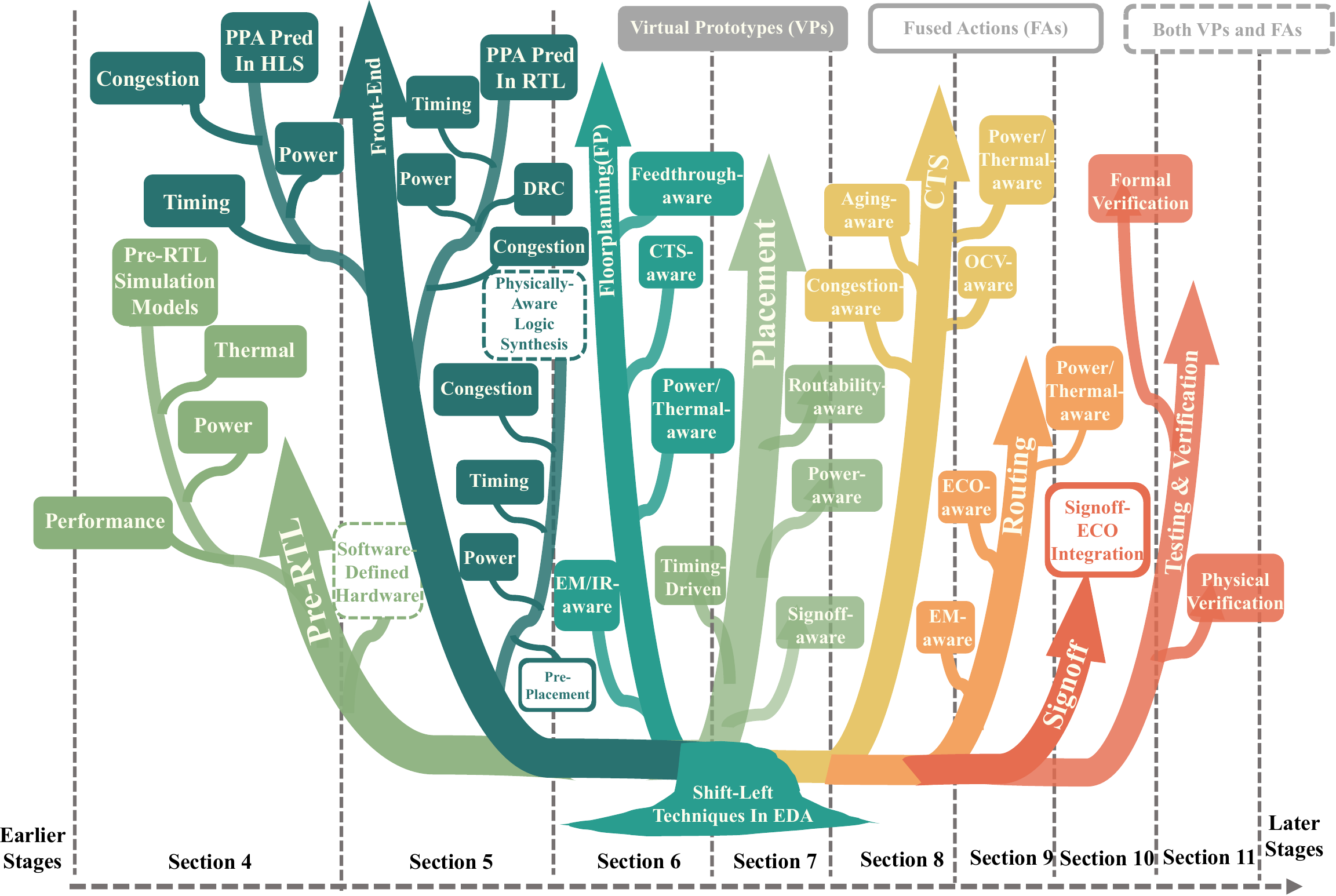}
%\vspace{-10pt}
  \caption{A taxonomy of Shift-Left techniques reviewed in this paper.}
  \label{fig:taxonomy}
\end{figure*}

In the following sections, we will discuss existing Shift-Left movements at various anchor points, along with their corresponding shift strategies as illustrated in Figure \ref{fig:taxonomy}, based on the classifications introduced in this section.

\section{Shift-Left in Pre-RTL Design Phase}
\label{sl-pre-rtl}
The Pre-RTL design phase focuses on collecting design specifications and developing behavioral and functional models in both software and hardware domains, typically involving extensive design space exploration (DSE) to identify optimal design configurations from a vast solution space. Although traditionally labor-intensive and experience-driven, the Pre-RTL phase now employs system-level modeling, virtual prototypes, AI-accelerated engines and thermal analysis to let designers explore architectures, optimize PPA and uncover flaws before RTL, using (1) abstract VP models to forecast metrics and (2) software-driven functional exploration that demands flexible hardware adapting to software-defined requirements.%It is often a labor-intensive process, requiring weeks of experimentation and heavily relying on past experience and domain expertise. With the advancement of pre-RTL simulation tools, this phase now enables designers to explore architectural trade-offs, optimize system-level performance, and detect potential design flaws well before RTL development begins. These tools incorporate system-level modeling, virtual prototyping, AI-driven simulation engines, and thermal-aware analyses to significantly reduce the time and cost traditionally associated with late-stage validation. Simulation strategies in this phase generally fall into two categories: (1) developing abstract VP models to predict PPA behaviors, and (2) enabling software-driven functional exploration, where the hardware must remain flexible to allow hardware adaptation to software-defined requirements.

\subsection{Pre-RTL Simulation Models}
\label{pre-rtl-model}
Pre-RTL models or simulators are different from several perspectives, including the level of simulation detail, the scope of the simulated target, and the input to the simulator ~\cite{akram2019survey}. 
% common features
One common feature of almost all the Pre-RTL simulators is that the modeling process is not as detailed as that utilized in commercial EDA tools. This is due to the simplified input abstractions and less complicated models applied. Unlike the input to commercial EDA tools, which is composed of RTL, gate, or transistor-level details, the input abstraction of Pre-RTL simulators typically only includes component-wise parameters (e.g., the number of CPU cores, cache size) and the interconnection specification. Performance models in Table ~\ref{tab:pre_rtl_models_grouped} list representative simulators that mainly generate system behaviors and performance metrics. %The nature of the Pre-RTL simulators degrades their modeling accuracy to some extent compared with that of industry-standard tools. However, Pre-RTL simulators excel in their short runtime and the capability to catch potentially better design choices or hazards in the early design stage.
% Hardware modeling at the Pre-RTL stage is typically coarse-grained, as it omits RTL-level details and focuses on high-level abstractions. Unlike formal design and verification flows that rely on industry-standard EDA tools and require transistor-level or gate-level representations of a finalized design, Pre-RTL simulation operates with abstract models that include major functional blocks and basic interconnections. Despite this abstraction, Pre-RTL models can yield valuable predictions of system behavior, power consumption, temperature profiles, and other relevant metrics.
%From the Shift-Left point of view, they effectively shift analysis and optimization tasks typically performed in the AP4+ stages (such as floorplanning and placement) to much earlier phases like AP0 or AP1, thus traversing at least four steps in the traditional flow. %By enabling early-stage simulation and design exploration, these models significantly improve design efficiency, accelerate DSE, and contribute to the development of more reliable and optimized final chips. As for the modeling accuracy, mature simulators can still yield predictions regarding performance, power, and thermal metrics with high quality and strike a good balance between runtime and accuracy.

\begin{table}[!t]
\centering
\caption{Summary of Representative Pre-RTL Models Grouped by Focus Areas}
\renewcommand{\arraystretch}{1.0}
\resizebox{0.9\textwidth}{!}{
\begin{tabular}{|l|p{11cm}|}
\hline
\textbf{Pre-RTL Model Type} & \textbf{Model Name} \\
\hline
\multirow{3}{*}{Performance Model}
& gem5 ~\cite{binkert2011gem5, lowe2020gem5}, Sniper ~\cite{carlson2011sniper}, ZSim ~\cite{sanchez2013zsim}, DRAMSim2 ~\cite{rosenfeld2011dramsim2}, DRAMSim3 ~\cite{li2020dramsim3}, TimeLoop ~\cite{parashar2019timeloop}, Aladdin ~\cite{shao2014aladdin}, SparseLoop ~\cite{wu2022sparseloop}, CiMLoop ~\cite{andrulis2024cimloop}, NeuroSim ~\cite{peng2019neurosim} \\
\hline
\multirow{1}{*}{Power Model}
& McPAT ~\cite{li2009mcpat}, CACTI ~\cite{muralimanohar2009cacti, balasubramonian2017cacti}, Accelergy ~\cite{wu2019accelergy}, Aladdin ~\cite{shao2014aladdin} \\
\hline
\multirow{1}{*}{Thermal Model}
& HotSpot ~\cite{huang2006hotspot, zhang2015hotspot6, han2022hotspot7}, 3D-ICE ~\cite{sridhar20103dice}\\
\hline
\multirow{1}{*}{Integrated Toolchain}
& HotSniper ~\cite{pathania2018hotsniper}, HotGauge ~\cite{hankin2021hotgauge}, CoMeT ~\cite{siddhu2022comet}, Cool-3D ~\cite{wang2023,wang2025cool3d}\\
\hline
% \multirow{2}{*}{Virtual Prototyping / Floorplanning}
% & Virtual Prototyping ~\cite{kogel2024} & Early software validation and platform simulation \\
% & ArchFP ~\cite{faust2012} & Pre-RTL floorplanning and spatial design \\
% \hline
% \multirow{3}{*}{AI-Driven PPA Estimation}
% & FastPASE ~\cite{levy2024} & AI-driven design space exploration and performance prediction \\
% & MasterRTL ~\cite{fang2023} & Fast pre-synthesis PPA estimation \\
% \hline
\end{tabular}
}
\label{tab:pre_rtl_models_grouped}
\end{table}

System behavior is abstracted by component-access traces. Processor granularity scales from cycle-accurate (slowest, most detailed) through event-driven to functional levels ~\cite{akram2019survey}. Many widely-used models are mixtures of them so as to strike a better balance between the modeling accuracy and the runtime. For example, gem5 ~\cite{binkert2011gem5, lowe2020gem5}, an event-driven simulator but was still able to track the process in a cycle-by-cycle manner ~\cite{akram2019survey}. Sniper ~\cite{carlson2011sniper}, another widely used processor-oriented simulator, utilized the interval simulation technique which is specially designed for efficient multi-core system simulation by ~\cite{genbrugge2010interval}, was not as detailed as the cycle-accurate model but possessed a short simulation time. ZSim ~\cite{sanchez2013zsim}, especially targeting large multi-core systems, used a timing model with dynamic binary translation for quick simulation. 
Despite these models simulating general-purpose processors, specialized simulators customized for specific components or even Application-Specific Integrated Circuits (ASICs) also exist. For example, for memory-related Pre-RTL modeling, the simulators tailored for dynamic random access memory (DRAM) such as DRAMSim2 ~\cite{rosenfeld2011dramsim2} and DRAMSim3 ~\cite{li2020dramsim3}, can be used. There are also architectural performance evaluation simulators for different specialized accelerators, such as TimeLoop ~\cite{parashar2019timeloop}/Aladdin ~\cite{shao2014aladdin} for accelerators, SparseLoop ~\cite{wu2022sparseloop} for sparsity-aware computing, and CiMLoop ~\cite{andrulis2024cimloop}/NeuroSim ~\cite{peng2019neurosim} for compute-in-memory (CiM) based accelerators. 
%%% TODO: consider to add gem5 extension for different hardware

%%% power model
For early power modeling, most tools can cooperate with the performance model to generate the power trace, use the basic block-wise access traces to preliminarily estimate the power consumption. For example, McPAT ~\cite{li2009mcpat} is a representative power model for processors that is usually coupled with gem5 or Sniper for early power prediction. For memory-related power modeling, CACTI ~\cite{muralimanohar2009cacti, balasubramonian2017cacti} provides more specialized details and can also work with those performance models. As for accelerator designs with higher design flexibility, special power or energy estimation support, such as Accelergy ~\cite{wu2019accelergy}, was proposed to work in assistance of accelerator-oriented performance models, such as TimeLoop ~\cite{parashar2019timeloop}, to provide complete PPA estimation. There are also combined models, with which both performance and power can be evaluated. For example, Aladdin ~\cite{shao2014aladdin} is an accelerator-oriented simulator featuring both performance and power modeling. 
% Similarly, ~\cite{jacobson2011} introduced a model that supports fast early-stage power modeling using abstract representations.

% Pre-RTL power modeling is crucial for enabling early-stage power and energy estimation, allowing designers to optimize complex systems before detailed RTL implementations are available. Early works like Aladdin ~\cite{shao2014aladdin} initiated fast power-performance modeling for accelerators, enabling broad design space exploration. VoltSpot ~\cite{zhang2015voltspot} contributed significant advancements in on-chip power delivery modeling at the architectural level. The introduction of machine learning significantly improved estimation accuracy, as seen in HL-Pow ~\cite{lin2020hlpow} and PRIMAL ~\cite{zhou2019primal}. Runtime modeling approaches like Simmani ~\cite{kim2019simmani} enhanced power modeling adaptability by automatically selecting critical signals. Domain-specific frameworks such as SimuNN ~\cite{cao2020simunn} and Gem5-Accel ~\cite{vieira2023gem5} extended pre-RTL power estimation to neural networks and accelerators. OldSpot ~\cite{guo2018oldspot} focused on aging and lifetime prediction, while Lumos+ ~\cite{wang2016lumos} provided rapid design space exploration in heterogeneous architectures. Collectively, these methodologies demonstrate a robust trajectory toward high-fidelity, scalable, and early power modeling in modern electronic systems.

%Need to check these references

%\subsection{Thermal}
% \href{https://semiengineering.com/the-3d-ic-multiphysics-challenge-dictates-a-shift-left-strategy/}{The 3D-IC Multiphysics Challenge Dictates A Shift-Left Strategy}
% ~\cite{hossam2023fast}

%%% Thermal model
Thermal estimation is another critical part of chip design. Early access to thermal behaviors can reduce related design costs regarding resolving thermal issues ~\cite{huang2006hotspot, zhang2015hotspot6,han2022hotspot7,sridhar20103dice}. %%% Toolchain
Beyond those individual models that generate the metrics within a limited coverage (e.g., thermal only), there are toolchains that integrate a series of models to offer a complete set of metric measurement and a better usage experience ~\cite{pathania2018hotsniper,hankin2021hotgauge,siddhu2022comet,wang2023,wang2025cool3d}. %The end-to-end feature enabled in these toolchains eases the designers from creating a flow for quick evaluation, enhancing the design efficiency for phase AP0 or AP1. 

\subsection{Software-defined Hardware (SDH)}
\label{software-defined-hardware}
SDH ~\cite{chakravarthi2025software} represents a transformative approach where hardware is described, configured, and optimized through high-level software-like abstractions instead of RTL, yielding modular, reconfigurable, and adaptable designs via high-level languages, virtual prototypes, and parameterized architectures. This paradigm streamlines HW/SW co-design, enabling early functional verification and system-level prototyping without mature RTL or silicon, and thus advances Pre-RTL shift-left by supporting rapid iterations, upfront power/performance analysis, and pre-silicon software development to compress schedules, cut trial-and-error costs, and accelerate time-to-market; representative models appear in Section ~\ref{pre-rtl-model}.
%represents a transformative approach where hardware is described, configured, and optimized using high-level, software-like abstractions rather than traditional RTL formats. By leveraging high-level languages, virtual prototyping, and parameterized architectures, SDH enables hardware that is modular, reconfigurable, and adaptable. 
%This paradigm enables an efficient HW/SW Co-design process and simplifies the collaboration between hardware and software designers, allowing early-design-phase functional verification and system-level prototyping without fully developed RTL or silicon.
%Thus, SDH plays a critical role in advancing the Shift-Left methodology of the pre-RTL stage by enabling faster design iterations, early performance and power analysis, and pre-silicon software development. 
% Its software-driven nature supports rapid design space exploration, continuous integration pipelines, and AI-based optimization at much earlier design stages. 
%By allowing HW/SW Co-design and early validation, SDH can effectively compress the development timeline, reduce trial-and-error costs, and accelerate time-to-market for complex systems.
%Some models described in Section ~\ref{pre-rtl-model} provide representatives of SDH. 
For example, gem5 ~\cite{binkert2011gem5, lowe2020gem5} can be programmed using C++ and Python to configure a specific processor. A software designer can test a newly developed algorithm using gem5 even without real hardware on hand. And for a hardware designer, the SDH modeled in these simulators can help to identify potential workload running issues and verify the target functionality on the real hardware in the early design phase.

 % ``Traditionally, software and hardware were developed separately and then integrated together. The extended consumer products upgrade cycles and the decade-long automotive electronics lifespan have definitely driven the adoption of the SDH approach. Software-defined hardware tackles that integration from outset of the design process. The software defines a system’s required functionality, and the silicon must be malleable enough for design teams to optimize the hardware for that functionality.'' -- https://semiengineering.com/toward-a-software-defined-hardware-world/
% Hardware/software co-design has been a goal for several decades,

\subsection{Workload Modeling}

Workload modeling is another critical component of the pre-RTL simulation process. The workload stimulates the under-test hardware and enables the process to catch the hardware behavior close to real life. Unlike a fabricated chip, onto which we can burn a program for testing, hardware designs in the pre-RTL stage need special modeling support for testing programs to run. There are diverse implementation strategies within different Pre-RTL models. There are simulators that support the direct running of a compiled program as that on real hardware, such as gem5 ~\cite{binkert2011gem5} and Sniper ~\cite{carlson2011sniper}. 
But this property relies on compiler support and the capability of the simulator to catch the instructions and system calls passed from the program. 
% Fully-customized/Specialized formats
% Nested-loop model for ML workloads
For simulators modeling specialized hardware or a certain component, fully customized formats of the workload are possible. For example, to estimate the layer-wise performance and power of a ML accelerator, a nested-loop abstraction is widely used to model the convolutional layer or fully connected layer in the ML algorithm. The nested-loop abstraction is used in tools such as TimeLoop ~\cite{parashar2019timeloop} and ZigZag ~\cite{mei2021zigzag}. It is advantageous to model the data mapping and computing parallelism for a hierarchy of hardware and a multi-dimensional workload. 

% Direct compiling: emulation based vs. instrumentation based

\subsection{DSE in Early Design Phase}

Pre-RTL models serve as early-design-phase DSE tools to identify optimal design options from large candidate pools. We categorize pre-RTL DSE into three groups: component selection, mapping scheme selection, and simultaneous execution of both. For component selection, the optimal hardware characterization is chosen based on specific workloads. For example, the framework developed in ~\cite{dai2024multichiplet} was designed to explore the optimal multi-chiplet system composed of CiM chips given an AI workload. 
For the second category, the hardware composition is defined by the designer while the design options to be explored are the workload mapping on the hardware. One representative example is TimeLoop ~\cite{parashar2019timeloop}, where given layers of an AI model, data mapping across the memory hierarchy and the processing elements (PEs) is to be explored for optimization. The workload mapping or distribution is also critical for heterogeneous processing systems. So for tools similar to AERO ~\cite{yang2022aero}, they provide the framework to find the most efficient way to distribute workload to different processing/computing cores. 
Since both the hardware component selection and workload mapping are critical design options in the early design phase, some tools automate the two processes as one ~\cite{mei2021zigzag}. %For example, ZigZag framework ~\cite{mei2021zigzag} provides support to generate the pareto-optimal accelerator architecture as well as the NN workload mapping on the hardware. 

\tikzset{
    double color fill/.code 2 args={
        \pgfdeclareverticalshading[%
            tikz@axis@top,tikz@axis@middle,tikz@axis@bottom%
        ]{diagonalfill}{100bp}{%
            color(0bp)=(tikz@axis@bottom);
            color(50bp)=(tikz@axis@bottom);
            color(50bp)=(tikz@axis@middle);
            color(50bp)=(tikz@axis@top);
            color(100bp)=(tikz@axis@top)
        }
        \tikzset{shade, left color=#1, right color=#2, shading=diagonalfill}
    }
}

\section{Shift-Left in Front-End Design Phase}
\label{shift-front-end}
The Shift-Left paradigm in the front-end design phase involves moving tasks and awareness from later stages into earlier design phases like HLS and RTL design stages. The primary strategies to achieve this are creating VP models, developing physically-aware algorithms, and combining VP and FA together to integrate downstream knowledge directly into HLS and RTL stages. Table ~\ref{tab:shiftleft_frontend} lists the representative works mentioned in this section.

\begin{table}[htbp]
  \centering
  \caption{Shift Left in Front-End Design}
  \resizebox{\textwidth}{!}{%
    \renewcommand{\arraystretch}{1.2}  
    \begin{tabular}{|>{\centering\arraybackslash}m{8em}|
                    >{\centering\arraybackslash}m{10em}|
                    >{\centering\arraybackslash}m{32em}|} 
    \hline
    \textbf{Type} & \textbf{Target} & \textbf{Representative Works / Applications} \\
    \hline
    \multirow{3}[6]{*}{\shortstack{Prediction in HLS}} 
        & Performance
        & Analytical models estimate latency/resources; ML refines accuracy and enables cross-platform prediction~\cite{zhao2017comba,wu2021ironman,o2018hlspredict}. \\
    \cline{2-3}          
        & Timing
        & From analytical area–delay models to ML frameworks for frequency and delay prediction~\cite{bilavarn2006design,choi2017hlscope+,makrani2019pyramid}. \\
    \cline{2-3}          
        & Power/Congestion
        & ML-based power estimation and congestion prediction from code/netlist features~\cite{lin2020hl,lee2015dynamic,zhao2019machine}. \\
    \hline
    \multirow{3}[6]{*}{\shortstack{Prediction in RTL}} 
        & Timing
        & CNN/LSTM for synthesis flows; GNN/Transformer/AST-based models for fine-grained slack prediction~\cite{10.1145/3195970.3196026,10.1145/3380446.3430638,9599868,wang2025bridging}. \\
    \cline{2-3}          
        & Power
        & Regression/GNN-based power models (GRANNITE, PRIMAL, Simmani) and commercial tools (PowerArtist)~\cite{9218643,10.1145/3316781.3317884,10.1145/3352460.3358322,ansys_powerartist_2022}. \\
    \cline{2-3}    
        & Congestion
        & GNN-based frameworks (CircuitGNN, CongestionNet) and LLM-based VeriLoC predict congestion/wirelength early~\cite{8920342,hemadri2025veriloclineofcodelevelprediction}. \\
    \hline
    \multirow{3}[6]{*}{\shortstack{Physically-Aware \\ Logic Synthesis}} 
        & Timing–Area
        & Fuse technology mapping with placement or interleave synthesis/placement; hybrid methods with predictive slack models~\cite{810630,742830,968622,zhu2023delay}. \\
    \cline{2-3}          
        & Congestion
        & Modify mapping cost for wirelength; predictive placement for congestion-aware resynthesis~\cite{1382637,garg2016study,10.1145/611817.611836}. \\
    \cline{2-3}          
        & PPA
        & Embed placement/routing engines (iSpatial) or use trial placement in hybrid flows (Encounter RTL Compiler)~\cite{Joshi2020iSpatial,Cadence2013Encounter}. \\
    \hline
    \multirow{2}[4]{*}{\shortstack{Physically-Aware \\ HLS}} 
        & FA Integration
        & Floorplanning during scheduling/allocation or resource binding~\cite{979797,10.1145/3649329.3658268}. \\
    \cline{2-3}          
        & VP-guided Hybrid
        & Fast placement/routing or delay/skew prediction models for routability-aware HLS~\cite{10.1145/2554688.2554775,9218718}. \\
    \hline
    \end{tabular}%
    }
  \label{tab:shiftleft_frontend}%
\end{table}

\subsection{Shift-Left Prediction in HLS}
\label{prediction-in-hls}
% Performance&Area
Performance and area co-estimation has been a primary focus, addressed by both analytical and ML-based VPs.
Early analytical VPs deconstruct high-level code to build mathematical hardware models. COMBA~\cite{zhao2017comba} analyzed LLVM Intermediate Representation to predict latency and resource usage. Similar static analysis methods guided transformations~\cite{prost2013fast}, explored design trade-offs~\cite{zhong2014design,liu2012compositional}, and estimated execution time for OpenCL kernels~\cite{wang2017flexcl,wang2016performance}. A visual-analytical approach even extended the Roofline model~\cite{williams2009roofline} to HLS for bottleneck identification~\cite{da2013performance}. MLs are also used to improve downstream prediction. ~\cite{wu2021ironman} and ~\cite{10.1145/3489517.3530408} operated on dataflow graphs to predict post-synthesis results, while other models like Gradient Boosting Machines~\cite{zhong2017design}, Rival Penalized Competitive Learning~\cite{liu2016efficient}, Random Forests~\cite{liu2013learning}, and other ensemble models~\cite{dai2018fast} leveraged source-code or HLS-report features to refine area and latency.
%TODO: need to decide if this part is necessary.
ML models are also effective for cross-platform prediction, such as translating CPU execution profiles into cycle counts~\cite{o2018hlspredict}, or using neural networks to predict speedup on different FPGAs~\cite{makrani2019xppe}. %A hybrid approach combined ML with pre-characterized templates~\cite{koeplinger2016automatic}.

% Timing
Timing and latency prediction, critical for performance, has also been a major Shift-Left target. Early efforts used analytical models to estimate area-delay trade-offs~\cite{bilavarn2006design} and memory latency~\cite{pham2015exploiting}. More advanced techniques used instrumented C-simulation or trace analysis for higher-fidelity predictions~\cite{choi2017hlscope+,zhong2016lin}. ML has dramatically improved accuracy. Some frameworks corrected the final frequency estimates from HLS tools~\cite{makrani2019pyramid}, while others bypassed HLS entirely to predict post-layout timing directly from source code and corresponding IR. At a finer granularity, GNNs have been used to virtualize the technology mapping stage by predicting the delay of individual arithmetic operations~\cite{ustun2020accurate}.

% Power&congestion
Predicting power and routing congestion early represents a critical Shift-Left toward physical design awareness. For power, ML-based VPs like HL-Pow~\cite{lin2020hl} and HLSPredict~\cite{o2018hlspredict} forecasted final design power from HLS reports or execution proxies. Another powerful VP technique involved training models on gate-level data and back-annotating power characteristics onto the HLS IR~\cite{lee2015dynamic}. To prevent routability issues,  some works used fast structural metrics extracted from the HLS-generated netlist to predict final interconnect congestion and wirelength~\cite{10.1145/2160916.2160952}. Pushing predictions even earlier, some operated directly on source code and used analytical techniques to identify congestion-prone structures like large multiplexers from graphs~\cite{8399766}, or used ML to predict routing congestion by analyzing features extracted from source code~\cite{zhao2019machine}. Complementary feedback mechanisms analyzed post-route reports and automatically traced congestion hotspots back to the source code~\cite{tariq2021high}.

\subsection{Shift-Left Prediction in RTL}
Several frameworks have leveraged ML to build PPA predictors. MasterRTL~\cite{10323951,10577671} introduced a novel graph representation SOG of RTL to train models that forecast the final PPA. SNS used deep learning to predict from HDL code~\cite{10.1145/3470496.3527444}. Techniques like learning-based recommender system from~\cite{10.1145/3316781.3323919}, Fist~\cite{9045201} and others~\cite{9643533} built models to guide the tuning of synthesis tool parameters, while ASAP~\cite{10299840} also captured the inherent trade-offs between metrics. 
Commercial tools like Synopsys's RTL Architect~\cite{synopsys_rtlarchitect_2021} enabled early PPA and congestion analysis directly by leveraging physical implementation and signoff engines to create a fast, predictive model. The most recent approaches employed foundation models to create more sophisticated VPs. CircuitFusion~\cite{fang2025circuitfusionmultimodalcircuitrepresentation} fused multiple modalities to align RTL representations with post-synthesis netlist data. VeriDistill~\cite{moravej2025graphsapprenticeteachingllm} used knowledge distillation to teach an LLM about physical properties, with a GNN trained on post-synthesis logic graphs. Another emerging trend is the use of large language models to directly automate RTL code optimization. For instance, RTLRewriter~\cite{10.1145/3676536.3676775} combined multi-modal analysis, retrieval-augmented generation for optimization knowledge, and a cost-aware search algorithm to directly generate optimized RTL.

A major focus of RTL-level prediction is forecasting timing. Early works integrated CNNs~\cite{10.1145/3195970.3196026} and LSTMs~\cite{10.1145/3380446.3430638} to model synthesis flows as a sequence of commands. More recent approaches directly estimated timing metrics from circuit structure. Similarly, word embedding and CNNs have been used to model logic paths as sentences to predict their timing criticality and guide optimization~\cite{9371553}. 
GNNs can also treat the RTL design as a graph to predict post-synthesis metrics like component delay, slew, and arrival times~\cite{10299859,10.1145/3508352.3561095}. Other works like LSTP~\cite{10473925} combined GNNs with Transformers to interpret both the circuit structure and the synthesis script for accurate timing prediction. ML models built on Abstract Syntax Trees~\cite{10.1145/3508352.3549375,10299879} and And-Inverter-Graph~\cite{9599868} provided fast feedback on metrics like Total Negative Slack (TNS). RTL-Timer~\cite{10.1145/3649329.3655671} and RTLDistil~\cite{wang2025bridging} provided fine-grained slack predictions directly on Verilog code.

Predicting power consumption early in the design flow is another critical Shift-Left objective. Frameworks like GRANNITE~\cite{9218643} and GRASPE~\cite{10181823} utilized GNNs to learn the complex relationships between anetlist and post-synthesis switching activity. Other approaches like APOLLO~\cite{10.1145/3466752.3480064}, PRIMAL~\cite{10.1145/3316781.3317884}, and Simmani~\cite{10.1145/3352460.3358322} built regression models that correlate RTL signal activity with detailed power data. ~\cite{7059105} designed an abstracted power model based on Singular Value Decomposition into the RTL. Besides academical researches, commercial EDA tools have long focused on this problem. Platforms such as Ansys PowerArtist~\cite{ansys_powerartist_2022}, Ennocad's EnFortius~\cite{ennocad_enfortius_2023}, Power Analysis tool from Synopsys~\cite{synopsys_rtlpower_2023} leveraged physically-aware modeling to provide accurate power estimation.

A crucial set of VPs aims to predict physical design challenges, primarily routing congestion and wirelength, directly from a pre-placement netlist. Frameworks like CircuitGNN~\cite{10.5555/3600270.3601747}, CongestionNet~\cite{8920342}, and others~\cite{9643446} applied GNNs to the logic graph, learning topological patterns that lead to downstream congestion hotspots. Customized Graph Attention Networks~\cite{9707500} have been developed to predict not just congestion but also individual net lengths and timing metrics before placement. Advanced methods like VeriLoC~\cite{hemadri2025veriloclineofcodelevelprediction} leveraged LLMs to provide line-of-code level predictions for both timing and congestion.
% \subsubsection{Shift Left with Power Emulation and Accurate RTL Power Analysis}
% Back in 1996, researchers started to look at methods for modeling power at a high level. ~\cite{landman1996high}
% ~\cite{raghunathan1999register} proposed design-for-low-power techniques for register-transfer level (RTL) controller/data path circuits.
\subsection{Physically-Aware Logic Synthesis}
% \label{physically-aware-ls}
To bridge the gap between logical synthesis and timing estimation, physically-aware synthesis methodologies shift physical implementation knowledge into the RTL-to-netlist stage. A dominant approach involves directly fusing physical design tasks with logical optimization. Numerous works co-optimized timing and area by fundamentally combining technology mapping with placement, making logic optimization decisions concurrently with physical layout choice~\cite{10.1145/277044.277072,810630,742830,5715609,406709,979696,10.5555/1509456.1509489,968622,185212}. Other techniques fused physical constraints into logic restructuring~\cite{742830}, interleaved synthesis with incremental placement~\cite{968622}, or integrated optimization directly into a comprehensive synthesis loop~\cite{4167765}. Hybrid methods combined these approaches, using a physical model to guide a fused and physically-aware local synthesis process~\cite{727128,Reis2018}. More advanced hybrid technique like DDPAS~\cite{zhu2023delay} introduced a lightweight NLP-based model to predict post-routing timing slack and fused it directly into the synthesis engine's search algorithm. This model was later applied in work~\cite{cai2025revisit} to develop novel early-stage multi-bit flip flop (MBFF) clustering right after logic synthesis.

Another critical target for Shift-Left methodologies is routing congestion. These FA methods modified the cost functions of synthesis operations like decomposition and technology mapping to directly include penalties for wirelength or layout, thereby inherently optimizing for routability~\cite{1382637, 968621,998371}, or performed physical design tasks directly and integrate automated macro placement into the synthesis flow to generate a congestion-aware initial floorplan~\cite{garg2016study}. More common hybrid models enabled a FA with a VP by generating an initial placement of the technology-independent netlist as a predictive model of the final layout. The physical information was then used to guide congestion-aware technology mapping or resynthesis, addressing routing challenges early in synthesis~\cite{998370,10.1145/611817.611836,410815,1190987}.

Many frameworks achieved comprehensive PPA optimization by embedding physical awareness. For example, the iSpatial flow ~\cite{Joshi2020iSpatial} integrated the Innovus placement and routing engine directly within the Cadence Genus synthesis tool to provide accurate PPA feedback. Other FAs performed placement-aware resynthesis~\cite{10.1145/3649329.3656243} or automated the complex parameter tuning of these physically-aware flows~\cite{7459483}. The predominant trend is also a hybrid methodology that leverages a VP to drive a FA. These systems, such as the Cadence Encounter RTL Compiler~\cite{Cadence2013Encounter}, created a prototype of the physical layout through trial placement or predictive models~\cite{10.1145/611817.611836,410815,geralla2018optimization,Clarke2011Eliminating}. %This prototype forecasts downstream PPA metrics, including power~\cite{10.1145/3676536.3676676,10186351}, timing and congestion~\cite{guo2024integration, Cadence2013Encounter}.

\subsection{Physically-Aware HLS}
Early prediction of timing in HLS is achieved by shifting physical awareness into early design stage. Some approaches employed FA by integrating physical design tasks directly into HLS by concurrently performing floorplanning during scheduling and allocation~\cite{979797}, combining placement with resource sharing~\cite{1167596}, or integrating a full 3D macro placement tool into the HLS resource binding stage~\cite{10.1145/3649329.3658268}. Other works used a hybrid approach where a VP guided the FA. These VPs can be early, fast placement-and-routing runs that provide realistic timing feedback~\cite{10.1145/2554688.2554775}, predictive models for net delays~\cite{9218718}, or clock skew estimation models that enable floorplan-aware HLS~\cite{7516905}. These hybrid techniques created tight and iterative feedback loops to correct for timing violations early~\cite{5733713}. Addressing wire congestion and routability is another key focus. Most common method is using a VP to enable FA. These methods performed a fast initial placement to estimate physical locations and congestion, and then used this as a direct cost function within HLS loop to guide scheduling and allocation decisions~\cite{10.1145/2744769.2744893,4756913}. Beside above, there are also Shift-Left strategies achieved multi-objective PPA improvements, whose dominant approach is FA~\cite{10.1145/337292.337769,10.1145/2744769.2744801,10.1145/2684746.2689063,742906,10.1109/ISVLSI.2006.8}.

%For overall PPA improvement, similar shift-left strategies are applied. FA is a dominant approach. Work~\cite{10.1145/337292.337769} unifies behavioral synthesis and physical design into a single, constructive process. Furthermore, awareness of LUT-based technology mapping is integrated directly into HLS pipeline scheduling to create more area-efficient designs~\cite{10.1145/2744769.2744801,10.1145/2684746.2689063}. Hybrid methods also combine a VP with an FA, using lightweight VPs like an incremental floorplanner to provide area estimates that guide HLS binding decisions~\cite{742906}, or employing probabilistic gain models to predict the physical impact of allocation choices~\cite{10.1109/ISVLSI.2006.8}. 

% \section{Shift Left Major Approaches}

% \section{Shift Left in Testing}

% \subsection{Duplicated Twins}

% \subsection{Predicted Twins}

% \subsection{Data Consistency}

% \subsection{Where to start?}
% Based on existing work, and plot the steps and starting point stats.
% \subsection{Where to end?}

\section{Shift-Left in Floorplanning}
\label{sl-floorplanning}
Early top-level floorplanning that jointly optimizes congestion, power, and thermal sign-off metrics shifts downstream risks left, eliminating costly iterations. To operationalize Shift-Left, floorplanning is extended into four anticipatory sub-tasks including feedthrough, CTS, power/thermal, and EM/IR-aware generation, each pre-empting a downstream sign-off constraint. Table ~\ref{tab:shiftleft_floorplan} enumerates the notable works referenced in this section.

\begin{table}[!ht]
  \centering
  \caption{Shift Left in Floorplanning}
  \resizebox{\textwidth}{!}{%
    \renewcommand{\arraystretch}{1.3}  
    \begin{tabular}{|>{\centering\arraybackslash}m{8em}|>{\centering\arraybackslash}m{10em}|>{\centering\arraybackslash}m{32em}|} 
    \hline
    \textbf{Type} & \textbf{Methodology} & \textbf{Representative Works / Applications} \\
    \hline
    \multirow{3}[6]{*}{\shortstack{Feedthrough-Aware}} 
            & SA-based
            & Use simulated annealing with embedded cost models \cite{3697728, xu2025stepbeyondfeedthrough} \\
        \cline{2-3}          
            & Channel Reservation
            & Reserve feedthrough channels early between hierarchical blocks to improve routability and timing closure~\cite{hong2022feedthrough} \\
        \cline{2-3}          
            & Explicit Feedthrough Planning
            & Insert feedthroughs during floorplanning for non–IO-limited SoCs~\cite{sakariya2025decongesting}.\\
        \hline
    \multirow{2}[4]{*}{\shortstack{CTS-Aware}} 
        & Virtual Prototypes
        & Embeds clock tree estimation into simulated annealing to guide floorplan generation under timing budgets~\cite{srinath2021investigation,1466067}; \newline{}
        Estimates clock distribution early to balance skew, latency, power, and IR-drop, improving timing predictability~\cite{6498346,yim1999floorplan}\\
    \cline{2-3}          
        & Fused Actions
        & Embeds gated-clock, wirelength, and clock-network models into RTL or architectural exploration for early power reduction~\cite{donno2004power,butt2007system}; \newline{}
        Generates variation-resilient spines and co-optimizes their placement with skew, latency, and power, achieving significant savings~\cite{seo2015spine,kim2017spine} \\
    \hline
    \multirow{2}[4]{*}{\shortstack{Power / Thermal \\Aware}} 
        & Thermal Minimization
        & Hotspot spreading, genetic floorplanning, 3D/2.5D optimization, and predictive models ~\cite{chu1998matrix,cong2004thermal,jang2014thermal,molter2023thermal,guan2023thermal,wang2024atplace2} \\
    \cline{2-3}          
        & Power Integrity Management
        & Noise-Direct di/dt profiling and voltage-island-aware floorplanning for IR-drop and multi-supply~\cite{lee2006voltage, mohamood2007noise} \\
    \hline
    \multirow{3}[6]{*}{\shortstack{EM / IR-Aware}} 
        & IR-drop Minimization
        & Embed block-level power estimation, fast IR-drop analysis, and guided incremental fixes into floorplanning~\cite{yim1999floorplan,liu2001power,chen2005ir,li2009guided}.\\
    \cline{2-3}          
        & EM Reliability
        & Decap allocation and TSV-based current distribution~\cite{zhao2002noise,falkenstern2010pg}\\
    \cline{2-3}          
        & Pad/TSV Co-design
        & Jointly optimize block placement with pad or bump allocation to ensure robust supply across voltage domains~\cite{chu2013multi,basha2022pg}\\
    \hline
    \end{tabular}%
    }
  \label{tab:shiftleft_floorplan}%
\end{table}

\subsection{Feedthrough-aware Floorplanning}
Feedthroughs, which occur when nets traverse through intermediate blocks in hierarchical SoCs, often degrade timing, increase routing congestion, and force late design changes if not planned early. Thus, feedthrough-aware floorplanning has attracted attention as a Shift-Left optimization strategy to mitigate these issues, which is mainly achieved through FAs. ~\cite{3697728} proposed FTAFP, an SCB-Tree-based, two-phase simulated-annealing floorplanner that co-optimizes module count and wirelength of feedthroughs via an embedded cost model. ~\cite{xu2025stepbeyondfeedthrough} presented Flora, a three-stage floorplanner that SA-co-optimized wirelength and feedthrough,  rectilinearly reshaped modules to absorb whitespace, and refined macro/cell placement.
~\cite{hong2022feedthrough} proposed a channel-based feedthrough-reservation scheme between hierarchical blocks enhances early routability and timing closure, eliminating late SoC iterations.
~\cite{zhang2012reclaiming} addressed the challenge of over-the-block routing resources, which directly relates to feedthrough utilization. They proposed a buffering-aware rectilinear Steiner minimum tree construction that reclaims routing resources over IP blocks.
~\cite{sakariya2025decongesting} explicitly planed and inserted feedthroughs during the floorplanning stage, targeting non–IO-limited channel-based SoC architectures.

\subsection{CTS-aware Floorplanning}
As the clock network often dominates both timing uncertainty and dynamic power, conventional CTS performed after placement may lead to late-stage surprises in skew, latency and power overhead. CTS-aware floorplanning therefore integrates clock tree considerations into early physical planning. Existing approaches can be broadly divided into VPs emulating clock trees during floorplanning to provide predictive guidance, and FAs merging CTS construction directly with the floorplanning process.

\subsubsection{VPs} This category focuses on embedding virtual models of clock trees into the floorplanning stage, enabling the prediction of skew, latency, and power long before detailed CTS by creating lightweight but sufficiently accurate approximations of clock distribution that steer macro placement toward CTS-friendly layouts.
~\cite{srinath2021investigation} proposed a simulated annealing (SA)-based MSV floorplanning methodology for the design of ICs within the timing budget.
~\cite{1466067} proposed a two-stage SA floorplanner based on sequence-pair representation. A clock tree estimation model is embedded into the floorplanner to guide the process of floorplan generation.
~\cite{6498346} provided a study to understand the relationship in terms of clock network latency, clock skew, and clock network power with respect to registers count, floor planning aspect ratio, design utilization and clock tree synthesis constraint.
~\cite{yim1999floorplan} proposed a floorplan-based planning methodology for power and clock distribution in ASICs, where the clock tree structure was estimated during the floorplanning stage to balance interconnect sizing and buffer insertion. By considering power and clock simultaneously, their method resolved global issues such as IR-drop and skew much earlier in the flow, improving the predictability of timing closure.

\subsubsection{FAs} While VPs mainly focus on prediction, FA methods integrate CTS synthesis directly into the floorplanning process, where the floorplanner’s action space includes decisions about clock distribution, resulting in joint optimization of skew, latency, power, and physical placement.
~\cite{donno2004power} introduced a RTL-driven, power-aware clock-tree synthesis that embedded gated-clock and pre-placement wirelength data into cost-driven tree generation, yielding measurable power reduction before CTS.
~\cite{butt2007system} proposed a system-level CTS that embeds clock-network models into pre-floorplan power estimators enables architectural exploration with quantified clock-power savings, pushing clock planning from physical to architectural abstraction.
~\cite{seo2015spine} developed an early-floorplan clock-spine-synthesis flow that generated variation-resilient parallel-trunk networks, delivering 38 \% lower skew and 11 \% lower power than conventional CTS.
~\cite{kim2017spine} extended this concept with an algorithm for synthesis and exploration of clock spines, embedding spine allocation/placement into floorplanning co-optimization of skew, latency and power, yielding 56 \% power saving versus industrial CTS and demonstrating the payoff of clock-aware floorplanning.

\subsection{Power/Thermal-aware Floorplanning}
Rising power density and hotspots now enforce thermal–power co-constraints at floorplanning, driving research on thermal minimization and power integrity.\par
%As power density increases and thermal hotspots emerge in advanced technologies, power and thermal effects have become critical design constraints. To ensure reliability and performance, these factors are increasingly incorporated into the floorplanning stage, with research focusing on thermal minimization and power integrity management. \par
For thermal minimization, 
early thermal-driven placement evolved from matrix-synthesis formulation for uniform power spreading ~\cite{chu1998matrix} to hotspot-coupled genetic floorplanner that concurrently shrinks area and peak temperature ~\cite{cong2004thermal}.
%~\cite{chu1998matrix} first formalized the thermal placement problem as a matrix synthesis task, aiming to evenly distribute power dissipation and avoid localized hotspots. 
%~\cite{hung2005thermal} integrated HotSpot thermal modeling into a genetic algorithm floorplanner, achieving reductions in both chip area and peak temperature by spreading out high-power modules. 
Based on that, several works extended thermal optimization to 3D-ICs ~\cite{cong2004thermal,jang2014thermal}. %~\cite{cong2004thermal} embedded thermal models into a simulated annealing floorplanner and reported over 50\% reduction in maximum temperature. ~\cite{jang2014thermal} combined thermal-driven floorplanning with min-cut die partitioning, balancing power density across dies while minimizing TSV count, thereby improving both thermal and routability metrics. 
At the micro-architectural level, research has progressed from power-aware unit distribution ~\cite{sankaranarayanan2005case} and temperature-dependent delay modeling ~\cite{winther2015thermal} to predictive thermal optimization, encompassing Bayesian surrogate-based SoC placement ~\cite{molter2023thermal}, end-to-end RL-driven 3D-IC floorplanning ~\cite{guan2023thermal}, and analytical 2.5D placement with embedded compact thermal models ~\cite{wang2024atplace2}.
%At the microarchitectural level, ~\cite{sankaranarayanan2005case} showed that distributing high-power functional units reduces CPU peak temperature with minimal performance impact. ~\cite{winther2015thermal} further improved timing realism by including temperature-dependent wire delay in floorplan evaluation. Recent research emphasizes predictive and learning-based methods for thermal optimization. ~\cite{molter2023thermal} developed a Bayesian optimization framework for SoC and multi-chip module placement, using surrogate models to accelerate exploration while reducing peak temperatures. ~\cite{guan2023thermal} proposed an end-to-end learning-based 3D-IC floorplanner that fuses clustering, MDP, and RL, outperforming heuristics in wirelength and thermal metrics. ~\cite{wang2024atplace2} extended analytical frameworks to 2.5D ICs with ATPlace2.5D, integrating compact thermal models into an analytical placer to improve scalability while significantly reducing chiplet temperatures.

For the power integrity management, ~\cite{mohamood2007noise} introduced Noise-Direct, a methodology that profiles current surges and clusters high di/dt modules near power pins. ~\cite{lee2006voltage} proposed a voltage-island-aware floorplanner for multi-supply-voltage SoCs. These studies exemplified how IR-drop and power delivery challenges can be proactively addressed during floorplanning to enhance power integrity and reduce late-stage design iterations.

Beyond academic research, ~\cite{cadence2019celsius} highlights the industrial adoption of shift-left thermal analysis. Cadence Celsius integrates electrothermal co-simulation into implementation platforms, enabling designers to evaluate and mitigate thermal issues as early as floorplanning and placement. This demonstrates the industry-wide recognition that early-stage power and thermal analysis is essential for reliable and efficient design closure.

\subsection{EM/IR-aware Floorplanning}
With lowered supply voltages and rising current densities, Voltage Drop (IR-drop) and electronmigration (EM) have become first-order constraints, driving floorplanners to early co-optimize IR minimization, EM reliability, and power pad/TSV distribution.

For the IR-drop minimization, early work by ~\cite{yim1999floorplan} incorporated block-level power estimation into floorplanning to proactively size power meshes and reduce voltage drop. ~\cite{liu2001power} further co-optimized floorplans with power supply networks using network flow formulations, ensuring high-power modules receive stable supply. ~\cite{chen2005ir} introduced fast block-level IR-drop estimation into a simulated annealing framework, directly minimizing peak and average voltage drop during optimization. ~\cite{li2009guided} improved efficiency with guided incremental moves to fix local IR-drop violations without global re-annealing. %Together, these works demonstrated that embedding IR-drop objectives into floorplanning significantly reduces late-stage power grid redesign.

For the EM reliability, since IR-drop primarily focuses on voltage stability, EM-aware floorplanning considers current density to avoid long-term reliability failures. ~\cite{zhao2002noise} combined block placement with decap allocation to reduce both voltage droop and current surges, indirectly mitigating EM risk. ~\cite{falkenstern2010pg} extended these ideas to 3D ICs, showing that distributing current paths vertically with TSVs reduces current density in long horizontal wires, thereby alleviating EM-induced reliability concerns.

For the Power Pad and TSV Co-Design, supply pad or TSV placement plays a key role in power integrity. ~\cite{chu2013multi} addressed multi-Vdd SoCs by integrating power pad assignment into floorplanning, ensuring each voltage island receives robust supply. ~\cite{basha2022pg} focused on flip-chip ICs, iteratively optimizing both block positions and C4 bump allocations to balance IR-drop across multiple domains. %These works highlight that co-optimizing block placement and physical supply entry points is critical for IR/EM security in modern packaging.

\section{Shift-Left in Placement}
\label{shift-left-placement}
Placement, the NP-hard positioning of millions of non-overlapping cells/macros under wirelength minimization, explodes to intractable state spaces (e.g., 1,000 clusters on a 1,000-cell grid)~\cite{nature_placement}. Placement quality gates downstream CTS and routing; excessive density incurs congestion and power, motivating early timing/congestion co-optimization. Beyond classic density/wirelength objectives, modern placers Shift-Left by integrating Static Timing Analysis (STA) net-weighting for timing, flop clustering for useful skew, clock-tree length control for power, and DRC-aware techniques for routability and manufacturability. Table ~\ref{tab:shiftleft_placement} provides a compilation of the key works that have been cited.

\begin{table}[!t]
  \centering
  \caption{Shift Left in Placement}
  \resizebox{\textwidth}{!}{%
    \renewcommand{\arraystretch}{1.3}  
    \begin{tabular}{|>{\centering\arraybackslash}m{8em}|>{\centering\arraybackslash}m{10em}|>{\centering\arraybackslash}m{32em}|} 
    \hline
    \textbf{Type} & \textbf{Methodology} & \textbf{Representative Works / Applications} \\
    \hline
    \multirow{2}[4]{*}{\shortstack{Timing-Aware}} 
        & Virtual Prototypes
        & RTL-level connectivity and extended dataflow models enabling placement and macro flipping~\cite{10.1145/3505170.3506731, 10372220, zhao2024standard, zhao2025incredflip, zhao2025mp}; \newline{}
        KD-tree diagnosis to push macros toward boundary~\cite{pu2024incremacro} \\
    \cline{2-3}          
        & Fused Actions
        & Encode slacks obtained from static timing analysis as net weights with dynamic updates during placement~\cite{liao2022dreamplace, lin2024timing, shi2025timing, fu2024hybrid}; \newline{}
        Directly minimize delays of critical paths, offering finer control but limited scalability~\cite{hamada1993prime, swartz1995timing}; \newline{}
        Use differentiable timing models to enable scalable, gradient-based optimization~\cite{guo2022differentiable}
        \\
    \hline
    \multirow{2}[4]{*}{\shortstack{Routability-Aware}} 
        & Global Routability Driven
        & Enhances congestion estimation and couples placement with congestion-aware routing models~\cite{he2011ripple,hsu2011routability,kim2011simplr,cheng2018replace,hsu2014ntuplace4h} \\
    \cline{2-3}
        & Detailed Routability Driven
        & Anticipates DRVs via ML-based hotspot prediction~\cite{xie2018routenet,baek2022pin,islam2025pgr} and embeds constraints into placement optimization~\cite{xie2018routenet,baek2022pin,islam2025pgr,huang2017ntuplace4dr,liu2022xplace}.\\    
    \hline
    \multirow{1}[2]{*}{\shortstack{Power-Aware}} 
        & Clock Tree-based
        & Constructs virtual CTS and optimizes clock net wirelength during placement to reduce dynamic power~\cite{lee2011obstacle,lu2015eplace,ding2023clock,kim2011simpl} \\
    \hline
    \multirow{1}[2]{*}{\shortstack{Signoff-Aware}} 
        & Embedded Sign-off Checks
        & Embeds timing, PPA, and fill checks into placement to eliminate post-sign-off iterations, exemplified by Calibre RealTime Digital~\cite{siemens_drc_placement}. \\
    \hline
    \end{tabular}%
    }
  \label{tab:shiftleft_placement}%
\end{table}

\subsection{Timing-Aware Placement}
At advanced nodes, wirelength minimization alone cannot close timing, necessitating Shift-Left Timing-driven placement (TDP) that optimizes Worst Negative Slack (WNS)/Total Negative Slack (TNS)~\cite{iccad2015contest} while resolving conflicts among timing, wirelength, density, routability, and legality, yet path improvements may destabilize others and lengthen iteration. TDP methods are generally classified into VPs and FAs, where FAs embed fast static timing analysis (STA) for analytical slack optimization, while VPs improve timing indirectly via timing-aware metrics, such as macro “push-boundary” operations.

\subsubsection{VPs}

\textbf{Dataflow-aware Placement}.
Dataflow, known as the data motion between macros and cells, critically sets timing/power as evidenced by designers’ routine flyline analysis and tight logic-team iteration. ~\cite{ma2007micro}.
Recent works addressed macro placement with dataflow awareness: statistical design of experiments for wire importance ranking  ~\cite{10.1145/1065579.1065731}, corner-stitching with simulated evolution ~\cite{9360844}, and multi-level strategies for memory-dominated designs ~\cite{9309318,vidal2019rtl}. RTL-MP ~\cite{10.1145/3505170.3506731} and Hier-RTLMP ~\cite{10372220} incorporated RTL-level connectivity, while ~\cite{zhao2024standard,zhao2025incredflip,zhao2025mp} extended this to multi-hop macro–macro and macro–cell interactions to support both placement and macro flipping.
% Dataflow defines the moving of data between the macros and standard cells and influences the timing and power consumption of the circuit.
% The importance of dataflow awareness is reflected in massive efforts spent by physical designers for flyline analysis and frequent interactions with logic designers to understand detailed dataflow information ~\cite{ma2007micro}.
% ~\cite{9309318,vidal2019rtl} proposes a novel multi-level approach for the macro placement problem of complex designs dominated by macro blocks, typically memories.
% ~\cite{10.1145/1065579.1065731} uses a statistical design of experiments strategy based on a multifactorial design, which intelligently uses a limited number of simulations to rank the importance of the wires.
% ~\cite{9360844} proposes a macro placement procedure based on the corner stitching data structure and then apply an efficient and effective simulated evolution algorithm to further refine placement results.
% ~\cite{10.1145/3505170.3506731} proposes RTL-MP, a novel macro placer which utilizes RTL information and tries to ``mimic'' the interaction between the frontend RTL designer and the backend physical design engineer to produce human-quality floorplans.
% And ~\cite{10372220} proposes Hier-RTLMP, an extension of RTL-MP, including one-hop connection between macro and cell.
% ~\cite{zhao2024standard} proposes dataflow-aware macro placer which involves the multi-hop connection of macro to macro, macro to cell cluster, and cell cluster to cell cluster.
\textbf{Push Boundary}.
Although expert designers place macros on the periphery to keep routing tracks open, mixed-size analytical placers often like DREAMPlace ~\cite{lin2019dreamplace} and RePlace ~\cite{cheng2018replace} embed them in the core to shorten wirelength; these centrally located macros partition the canvas into disjoint subregions, scattering cells of the same net and forcing long, detoured interconnects that inflate wirelength and congestion~\cite{pu2024incremacro}.
To mitigate such issues, ~\cite{pu2024incremacro} proposed a KD-tree-based macro diagnosis to detect poorly placed macros and analytically push them toward the boundary while minimizing wirelength perturbation. 

\subsubsection{FAs}
\textbf{Net-based} timing-driven placement methods leverage STA to convert critical paths or slacks into fixed net weights before placement, embedding timing guidance within the wirelength objective while sacrificing adaptability ~\cite{kong2002novel, chang2002net}.
Dynamic approaches, by contrast, iteratively update net weights during placement, achieving more accurate timing optimization.
~\cite{liao2022dreamplace} pioneered this framework by integrating timing analysis tools (e.g., OpenTimer ~\cite{huang2015opentimer}) into analytical placers and using momentum-based updates for net weights.
~\cite{lin2024timing} extended this with a pin-connectivity-aware clustering metric and historical weight averaging for stability.
Recent works ~\cite{shi2025timing, fu2024hybrid} further refined timing-driven placement using pin-pair attractiveness, hybrid pin- and path-weighting schemes, RC-tree models, and incremental timing calibration.
Comparative results on ICCAD 2015 benchmarks are shown in Figure ~\ref{fig:place-comparison}.
While \textbf{path-based} placement directly minimizes delay of critical paths via cell movement, yielding finer timing control than net-based methods but incurring scalability penalties from the exponential path count. ~\cite{hamada1993prime} formulated path-based timing-driven placement as a Lagrangian problem, solved via a primal–dual algorithm using piecewise-linear RC networks for the primal and Newton’s method for the dual, with distributed RC delays integrated into a nonlinear timing model.
~\cite{swartz1995timing} proposed a pin-pair algorithm that builds a timing graph, extracts the top-K critical paths via the Dreyfus algorithm, and embeds dynamically-weighted timing penalties into simulated annealing to balance wirelength and timing.
% approaches optimize placement by directly targeting critical paths, explicitly reducing their delays through cell movement.
% While they offer precise control and high-quality solutions compared to net-based methods, their scalability is limited due to the potentially large number of paths in complex designs.
% ~\cite{hamada1993prime} proposes a path-based timing-driven placement method that formulates the problem as a Lagrangian and solves it via a primal–dual algorithm, using a piecewise-linear RC network for the primal and Newton’s method for the dual, with distributed RC delays incorporated into a nonlinear timing model.
% ~\cite{swartz1995timing} introduces a pin-pair algorithm that automatically constructs a timing graph and extracts top-K critical paths using the Dreyfus algorithm, integrating timing penalties into the simulated annealing cost function with a dynamic weight to balance wirelength and timing optimization.
\textbf{Analytical STA}. Recent timing-driven placement methods ~\cite{guo2022differentiable} addressed the limitations of traditional TDP—indirect timing optimization, poor scalability on large circuits, and repeated STA calls—by introducing differentiable timing analysis that directly computes the gradient of timing metrics w.r.t. cell locations, enabling end-to-end timing optimization.

\definecolor{c1}{HTML}{FB4A47} % 红色
\definecolor{c2}{HTML}{FD9B23} % 橙色
\definecolor{c3}{HTML}{8BD1B4} % 浅绿
\definecolor{c4}{HTML}{436B3D} % 深绿

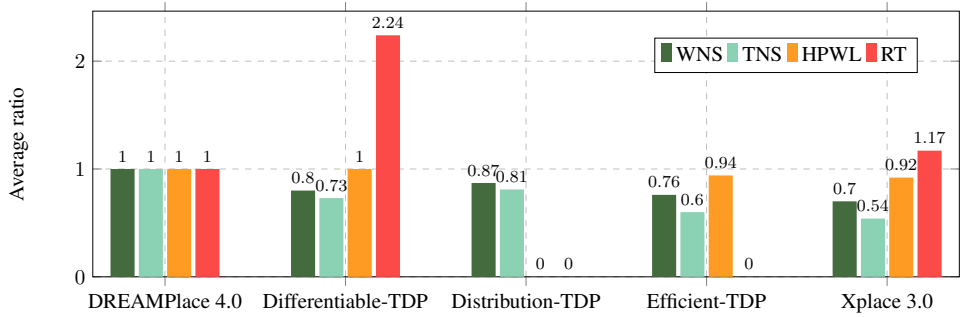
\begin{figure}[t]
\centering
\begin{tikzpicture}[scale=0.8]
\begin{axis}[
    ymin=0,
    ybar,
    grid=major,
    major grid style={dashed},
    ylabel=Average ratio,
    bar width=.4cm,
    width=14cm,
    height=6cm,
    symbolic x coords={
        DREAMPlace 4.0,
        Differentiable-TDP,
        Distribution-TDP,
        Efficient-TDP,
        Xplace 3.0
    },
    x=3cm,
    xtick=data,
    nodes near coords,
    nodes near coords style={font=\fontsize{8}{10}\selectfont, text=black},
    enlarge x limits=0.1,
    legend columns=4,
    legend style={at={(0.8, 0.9)}, anchor=north},
    every axis plot/.append style={
        legend image code/.code={
            \path[fill=#1] (0cm,-0.1cm) rectangle (0.18cm,0.2cm);
        }
    }
]

% --- WNS ---
\addplot+[draw=none, fill=c4]
  coordinates {(DREAMPlace 4.0, 1) (Differentiable-TDP, 0.8) (Distribution-TDP, 0.87) (Efficient-TDP, 0.76) (Xplace 3.0, 0.7)};

% --- TNS ---
\addplot+[draw=none, fill=c3]
  coordinates {(DREAMPlace 4.0, 1) (Differentiable-TDP, 0.73) (Distribution-TDP, 0.81) (Efficient-TDP, 0.6) (Xplace 3.0, 0.54)};

% --- HPWL ---
\addplot+[draw=none, fill=c2]
  coordinates {(DREAMPlace 4.0, 1) (Differentiable-TDP, 1.00) (Distribution-TDP, 0) (Efficient-TDP, 0.94) (Xplace 3.0, 0.92)};

% --- RT ---
\addplot+[draw=none, fill=c1]
  coordinates {(DREAMPlace 4.0, 1) (Differentiable-TDP, 2.24) (Distribution-TDP, 0) (Efficient-TDP, 0) (Xplace 3.0, 1.17)};

\legend{WNS, TNS, HPWL, RT};
\end{axis}
\end{tikzpicture}
\caption{Average ratios of WNS, TNS, HPWL, and RT across representative placers: DREAMPlace 4.0 \cite{liao2022dreamplace}, Differentiable-TDP \cite{guo2022differentiable}, Distribution-TDP \cite{lin2024timing}, Efficient-TDP \cite{shi2025timing}, and Xplace 3.0 \cite{fu2024hybrid}.}

\label{fig:place-comparison}
\end{figure}

% \subsubsection{Fused Actions}
% \subsubsection{Virtual}
% Log-sum-exp (LSE) wirelength model ~\cite{LSEwirelength}.

% Weighted average (WA) wirelength model ~\cite{WAwirelength}.

% Rectilinear Steiner minimum tree ~\cite{chu2007flute}.

% Timing driven placement (push boundary (virtual prototypes), STA (fused actions) ).

\subsection{Routability-Aware Placement} 
Routability has shifted from a secondary concern in wirelength-driven placement to a primary optimization objective due to technology scaling and increasing circuit complexity ~\cite{chow2016placement}.
Traditional wirelength-driven placement assumed that shorter wires reduce routing demand, but wirelength alone does not capture local congestion ~\cite{wang2000congestion}.
As technology scales below 45nm, limited routing resources and higher pin density make congestion more likely.
% In the field of VLSI physical design, routability has evolved from a derivative requirement of wirelength-driven placement in the early stages to a core objective of modern placement optimization.
% Its necessity stems from the rigid demands of technology scaling, increasing circuit complexity, and closed-loop design flows, making it a critical link in ensuring that circuit designs can be translated into manufacturable products ~\cite{chow2016placement}.
% Traditional wirelength-driven placement takes "minimizing the total wire length as its core objective.
% They assume that shorter wirelength can indirectly reduce the consumption of routing resources
% However, minimizing wirelength does not equate to minimizing congestion, as wirelength fails to reflect the local distribution of routing resources ~\cite{wang2000congestion}.
% As semiconductor technology nodes advance to 65nm and even below 28nm, physical constraints have become increasingly complex.
% While the number of pins per unit area has increased, the growth of routing resources has been limited, making local congestion highly likely.
% Routability not only determines the feasibility of routing but also directly impacts the design cycle and the final PPA (Power, Performance, Area) of the circuit.
% Thus, it must be treated as an independent optimization objective during the placement.
Routability-driven placement has evolved through two main stages: the global routability, highlighted by contests~\cite{viswanathan2011ispd,viswanathan2012iccad,viswanathan2012dac}, and the subsequent detailed routability, exemplified by ~\cite{yutsis2014ispd,bustany2015ispd}.
\subsubsection{Global Routability Driven Placement}
The global stage emphasizes feasibility of global routing, with optimization objectives such as Total Overflow (TOF) and Average Congestion of G-cell Edges (ACE) ~\cite{viswanathan2011ispd,viswanathan2012dac}. 
To estimate congestion more accurately, Ripple ~\cite{he2011ripple} improved RUDY-based estimation by decomposing multi-pin nets via Rectilinear Minimum Spanning Trees, while NTUplace4 ~\cite{hsu2011routability} employed an L-shaped probabilistic routing model with FLUTE-based two-pin subnet decomposition. 
At the placement–routing integration level, SimPLR ~\cite{kim2011simplr} coupled the BFG-R global router with the analytical placer SimPL ~\cite{kim2011simpl}, leveraging lookahead routing, dynamic cell bloating, and congestion-aware detailed placement for real-time coordination. 
Ropt ~\cite{liu2013optimization} further optimizes routability through a global routing model with local awareness, combining global re-placement, legalization, and local detailed placement. 
Analytical placement methods extend this line of work: Li ~\cite{li2014analytical} replaced cell density constraints with pin density constraints and introduces spreading and relocation strategies to alleviate macro blockages, while RePlace ~\cite{cheng2018replace} achieved fine-grained bin-level density control and integrates NCTU-GR ~\cite{liu2013nctu} for congestion optimization. 

Another important direction is incorporating design hierarchy. 
ICCAD 2012 ~\cite{viswanathan2012iccad} highlighted the role of hierarchy in improving both wirelength and routability. 
NTUplace4h ~\cite{hsu2014ntuplace4h} built on this by treating hierarchy as a core constraint, thereby addressing the disconnect between logical design and physical placement.

\subsubsection{Detailed Routability Driven Placement}
While global routability-driven placement focuses on overall routability, it typically ignores detailed design rules, so a placement feasible at the global routing stage may become non-routable during detailed routing ~\cite{alpert2010makes, liu2013case}.
Thus, potential detailed routing violations (DRVs) must be anticipated and incorporated into placement, modeled by early works such as ISPD 2014 ~\cite{yutsis2014ispd}. With the progress of ML, DRV and DRC hotspot prediction has become an important direction. 
~\cite{liu2021global} employed a fully convolutional network (FCN) to predict congestion maps and integrated this predictor into DREAMPlace ~\cite{lin2019dreamplace} with gradient descent optimization. 
RouteNet ~\cite{xie2018routenet} further demonstrated that CNNs and FCNs can predict DRC hotspots from placement features such as macro regions and RUDY pin density. 
PGNN~\cite{baek2022pin} combined GNNs and U-Net to capture pin accessibility and routing congestion simultaneously, while  ~\cite{lin2025early} improved efficiency by using GNNs with adaptive adjacency matrices. 
More recently, PGR-DRC ~\cite{islam2025pgr} introduced an unsupervised Gaussian-based framework with iterative thresholding to identify DRC hotspots without labeled data, highlighting the potential of label-free prediction.

Beyond identification, researchers have embedded routability constraints directly into placement optimization as a kind of FA-based methods. 
~\cite{huang2015detailed} proposed a full-flow, detailed-routing-driven analytical placement algorithm that integrates congestion into the wirelength model while explicitly considering DRVs during placement. 
ISPD 2015 ~\cite{bustany2015ispd} extended this by introducing fence region constraints, local density limits, and blockages to better mimic industrial congestion scenarios. 
NTUplace4dr ~\cite{huang2017ntuplace4dr} targeted mixed-size designs, deeply integrating technology rules and fence regions to reduce macro blockages. 
Finally, building on Xplace ~\cite{liu2022xplace}, ~\cite{dac25congestion} proposed differentiable congestion functions, virtual cell-guided netlist movement, and momentum-based cell inflation, enabling global and local optimization for routability with explicit DRV reduction.

\subsection{Power-Aware Placement}

% Placement directly impacts power distribution and integrity ~\cite{rabaey2002digital, alpert2008handbook}, as it determines signal wirelength and load capacitance, affecting dynamic power. Uneven cell density can also create thermal hotspots, increasing static power, making early-stage power optimization essential.
% In modern SoCs, the clock network can account for 30–50\% of dynamic power, since the clock drives many flip-flops at the highest frequency ~\cite{lee2011obstacle}.
% Longer clock trees increase capacitance, skew, and slew, complicating timing closure and necessitating additional buffers, which further raise power.
% Placing registers closer together reduces total clock tree wirelength, lowering power. Consequently, power-aware placement techniques focus on compacting the clock network to minimize both wirelength and buffer insertion.

Placement directly affects power distribution and integrity ~\cite{rabaey2002digital, alpert2008handbook} by determining wirelength and load capacitance, which influence dynamic power. 
Uneven cell density can create thermal hotspots, increasing static power, so early power optimization is essential. 
In modern SoCs, the clock network can consume 30–50\% of dynamic power ~\cite{lee2011obstacle}. Power-aware placement techniques aim to compact the clock network and minimize both wirelength and buffer insertion due to longer clock trees can increase capacitance, skew, and slew, complicating timing closure and requiring more buffers, which further raise power.  ~\cite{ding2023clock} built upon the ePlace ~\cite{lu2015eplace} and proposed a synchronized co-optimization framework for clock tree and placement that replaces costly Delaunay triangulation with a multi-grid search to accelerate tree construction and enable efficient per-iteration clock updates. In addition, gradient descent is employed to optimize the clock tree nets.
Lopper ~\cite{lee2011obstacle}, developed on top of the force-directed placement framework SimPL ~\cite{kim2011simpl}, in which the clock tree was decomposed into virtual nodes, and an Arboreal Contraction Force was formulated to optimize the clock tree nets.

\subsection{Signoff-Aware Placement}
Sign-off-aware placement embeds final-quality timing, PPA and fill checks into the placer itself, closing the correlation gap early and eliminating the native-to-sign-off iterations that traditionally follow. Calibre RealTime Digital ~\cite{siemens_drc_placement} developed by Siemens EDA enables this capability by integrating Calibre sign-off DRC and SmartFill directly into the placement environment, guaranteeing accuracy, boosting efficiency, and compressing the design cycle, accelerating DRC convergence by 2–4$\times$.

\section{Shift-Left in CTS}\label{overview:shift-left-in-cts}
CTS delivers clock with minimal skew, latency, slew and power; shift-left CTS embeds aging, variation and power awareness early, sustaining sub-ps skew and low slew/power under degradation while respecting congestion, eliminating late timing or routing iterations. The seminal works discussed in this section are cataloged in Table ~\ref{tab:shiftleft_CTS}.

\begin{table}[!ht]
  \centering
  \caption{Shift Left in CTS}
  \resizebox{\textwidth}{!}{%
    \renewcommand{\arraystretch}{1.3}  
    \begin{tabular}{|>{\centering\arraybackslash}m{8em}|>{\centering\arraybackslash}m{10em}|>{\centering\arraybackslash}m{32em}|} 
    \hline
    \textbf{Type} & \textbf{Methodology} & \textbf{Representative Works / Applications} \\
    \hline
    \multirow{4}[8]{*}{\shortstack{Aging-Aware}} 
            & Buffer Voltage Assignment
            & Assigning different voltages to buffers using linear programming \cite{choi2024bti, oh2020symmetrical} \\
        \cline{2-3}          
            & Voltage Threshold Replacement
            & Replacing standard-Vth buffers with high-Vth buffers \cite{chen2013novel} \\
        \cline{2-3}          
            & Clock Gating Ratio Balancing
            & Selective output implementation with NAND/NOR clock gates \cite{chakraborty2010skew}; \newline{} NAND-type matched clock tree with an ILP optimization model \cite{huang2013low, huang2010critical} \\
        \cline{2-3}          
            & Self-Heating-Aware Buffer Placement 
            & Buffer placement considering local thermal effects \cite{lin2014buffered}  \\
        \hline
    \multirow{2}[4]{*}{\shortstack{Congestion-Aware}} 
        & Structural and Device Optimization 
        & Fishbone finger sharing reduces routing demand for uneven loads \cite{vishnu2019clock}; \newline{} Replacing single-bit flip-flops with multi-bit flip-flops \cite{patel2013novel} \\
    \cline{2-3}
        & Interconnect and Routing Optimization
        & Hybrid RF/metal clock network reduces routing pressure on metal layers \cite{mohammadi2011hybrid} ; \newline{} Using probabilistic routing demand estimation \cite{farhangi2010pattern}\\    
    \hline
    \multirow{2}[4]{*}{\shortstack{Thermal / Power\\Aware}} 
        & Clock Gating Optimization
        & Simultaneous buffer and clock gate insertion for low dynamic power~\cite{lu2011fast};\newline{} Intelligent clock gates insertion when balancing stress workload~\cite{huang2013low} \\
    \cline{2-3}          
        & Thermal-Aware Buffer Insertion
        & Establish thermal-aware embedding with routing and buffer insertion with temperature profiles.  \cite{oh2019thermal} \\
    \hline
    \multirow{3}[6]{*}{\shortstack{OCV-Aware}} 
        & DME-Based
        & Buffer reallocation to reduce non-common path delay differences~\cite{rajaram2011robust} \newline{} Applied DME to estimate skew and guide routing for timing balance~\cite{lu2003process} \\
    \cline{2-3}          
        & Path-Level Adjustment
        & Path optimization with programmable skew delivery to flip-flops \cite{sivaswamy2008statistical} \newline{} Top-level clock buffer insertion to reduce slack across corners \cite{chan2014ocv} \\
    \cline{2-3}          
        & Cross-Link Insertion
        & Insert cross-links to reduce skew and clock period variation~\cite{10.1145/1960397.1960407, yang2011robust} \\
    \hline
    \end{tabular}%
    }
  \label{tab:shiftleft_CTS}%
\end{table}

\subsection{Aging-Aware CTS}
% Aging intro
Aging in ICs, which stems from physical mechanisms that degrade transistors and interconnects, is becoming increasingly critical with continued technology scaling. In transistors, negative/positive bias temperature instability (NBTI/PBTI) and hot carrier injection (HCI) generate traps that raise threshold voltage ($V_{th}$) and reduce carrier mobility, leading to delay increases at cell and path levels~\cite{singh2023reliability, amrouch2016reliability, guo2017towards}. For interconnects, EM displaces metal atoms under high current densities, while time-dependent dielectric breakdown (TDDB) accelerates dielectric wear-out~\cite{ho1989electromigration, lienig2018fundamentals, kim2018systematic}. These degradations strongly affect CTS, manifesting as asymmetric aging in buffers, inverters, and wires, causing non-zero skew, degraded slews, and reduced drive strength~\cite{ramadan2024impact}. This motivates incorporating aging prediction as VPs directly into the CTS stage~\cite{chakraborty2009analysis}. 

%Chakraborty et al.~\cite{chakraborty2009analysis} developed a mathematical framework to estimate NBTI-induced $V_{th}$ degradation under workload and temperature variations, and proposed circuit-level techniques to balance aging across clock devices. 

Beyond VPs, FA methods are also used to do mitigation focusing on reducing asymmetric degradation. Supply voltage alignment is a key technique to assign different voltages to buffers based on signal probabilities using linear programming, balancing stress and improving reliability~\cite{choi2024bti, oh2020symmetrical,chen2013novel}. Also, asymmetric aging is exacerbated by clock gating due to variations in stress probabilities across paths. Increasing the activity of low-activity gates can balance degradation but incurs high power cost, motivating low-power solutions ~\cite{huang2010critical,huang2013low,chakraborty2010skew}. Other aging challenges have been incorporated into CTS phase, such as EM ~\cite{lu2015electromigration} or thermal effects detailed in Section ~\ref{sec:thermal_CTS}.

\subsection{Congestion-Aware CTS}
Routing resources are significantly consumed by clock nets despite reserving most of the bottom two metal layers for standard cell internal wiring, impacting subsequent routing capability~\cite{westra2009congestion}. Consequently, a Shift-Left of congestion evaluation from the routing phase to CTS  becomes essential.

Several architectural and structural techniques have been proposed to reduce congestion in clock trees. ~\cite{vishnu2019clock} observed that when loads are slightly offset from clock straps and not uniformly distributed, employing fishbone finger sharing can save significant routing resources. On the other hand,~\cite{patel2013novel} explored the use of logically equivalent multi-bit flip-flops to replace single-bit flip-flops, which reduces the number of clock sinks and registers,  alleviating the routing resource demand during CTS while also saving the power. Other approaches leverage alternative interconnect strategies to reduce routing congestion ~\cite{farhangi2010pattern}. 
%focused on pattern-aware routing in the early CTS stage using a probabilistic routing demand estimation method to integrate expected routing demand into clock tree optimization metrics, thereby reducing the number of vias required and significantly mitigating congestion.

%By incorporating these techniques at the CTS stage, designers can proactively manage routing resource utilization, improve timing robustness, and reduce the likelihood of congestion-related performance degradation in the subsequent global routing and detailed routing stages.

% Collectively, these congestion-aware CTS strategies demonstrate a spectrum of solutions, from architectural-level optimizations like flip-flop aggregation and fishbone sharing, to interconnect-level innovations such as hybrid RF/metal networks and pattern-aware probabilistic routing. 

\subsection{Power/Thermal-Aware CTS}
\label{sec:thermal_CTS}
%Clock networks' power consumption demands early power-aware CTS to preempt post-layout verification. Buffer placement and switching activity cause IR-drop hotspots, impacting power integrity and thermal distribution, which can worsen transistor aging and reliability. Power- and thermal-aware CTS ensures robust timing and power while mitigating long-term reliability issues.

%Clock networks can consume a substantial portion of a chip's dynamic power, which makes early consideration of power effects critical during CTS, representing a shift-left approach that moves power analysis from post-layout verification of power consumption and quality to the early CTS stage. In particular, buffer placement and clock switching activity can create localized IR-drop hotspots and affect overall power integrity, which in turn influences thermal distribution across the chip. These thermal variations can lead to secondary effects such as exacerbated transistor aging and reduced reliability. Therefore, power- and thermal-aware CTS techniques are necessary to ensure both timing and power robustness while mitigating long-term reliability issues.

Power-aware CTS is essential to preempt post-layout verification for clock networks. Buffer placement and switching activity cause IR-drop hotspots, impacting power integrity, thermal distribution, and reliability. Power/Thermal-aware CTS therefore ensures robust timing and mitigates long-term reliability issues.~\cite{lin2014buffered} proposed a self heating-aware CTS methodology to acount for local thermal effects within buffers and mitigate them.~\cite{lu2011fast} proposed a CTS methodology with simultaneous buffer/gate insertion to achieve low switch capacitances and lower dynamic power.~\cite{huang2013low} optimized clock paths to address aging and power simultaneously, avoiding heuristic techniques that elevate activity in low-probability gates, leading to a more power-efficient CTS while reducing asymmetric aging.~\cite{oh2019thermal} proposed a thermal-aware 3D buffered CTS methodology to mitigate skew caused by thermal gradients in TSV-based 3D-ICs. %Their approach constructs a power-similarity-based topology for uniform TSV distribution, then performs thermal-aware embedding with routing and buffer insertion using grid-based temperature profiles.

% thermal
% ~\cite{donno2004power} proposes Power-aware clock tree planning
% \subsection{Early Routing}
\subsection{On-Chip Variation (OCV)-Aware CTS}
On-Chip Variation (OCV) refers to localized differences in transistor and interconnect properties caused by process non-uniformities. These variations introduce timing uncertainties at nominal conditions, closely tied to PVT variations, which degrade timing robustness if not addressed before physical implementation~\cite{9215244}. Deferred-merge embedding (DME) algorithms construct and merge clock subtrees while preserving common paths to reduce skew and variation sensitivity, thus are widely adopted across several studies~\cite{rajaram2011robust,velenis2003reduced,lu2003process}. Beyond structural subtree merging, path-level modifications also serve as an incremental technique for aging mitigation~\cite{sivaswamy2008statistical,chan2014ocv}. Cross-link insertion is also an investigated approach to mitigate variation effects in clock trees~\cite{10.1145/1960397.1960407,yang2011robust}. 

\section{Shift-Left in Routing}
\label{shift-left-routing}
Routing realizes inter-pin connections under scarce metal layers by trading off wirelength, congestion, crosstalk and signal integrity; it first globally plans paths and resources, then detail-routes geometries under design rules. Early heuristics were simple but runtime-prohibitive for complex designs, prompting Shift-Left metrics (wirelength, via count, speed) to be embedded during routing. Recently, AI/ML integration has shown significant promise in advancing these algorithms. Table ~\ref{tab:shiftleft_routing} showcases the highlighted works that have been brought up in this section.

%During routing phase, connections between the pins of various physical units are realized. Routers should strike a balance among multiple optimization objectives, including but not limited to minimizing wire length (which typically indicates better timing performance), alleviating congestion, reducing crosstalk, and ensuring signal integrity within the constraints of limited metal layer resources. Routing process is divided in to two steps, the former one, global routing plans high-level paths and allocates resources, then detailed routing precisely defines wire geometries to comply with design rules. Early automatic routing algorithms primarily relied on heuristic rules and graph theory algorithms, such as Maze routing, Channel Routing and Switchbox Routing. These algorithms were simple but inefficient with long runtimes, making them far not sufficient to meet the demands of complex designs. With the increasing complexity of integrated circuits, an increasing number of routing algorithms have begun to consider evaluation metrics of subsequent stages during the routing process, effectively shifting the design flow to the left. Significant progress has been made in reducing wire length, minimizing via count and enhancing routing speed. In recent years, the advent of artificial intelligence and machine learning has led to the incorporation of these advanced technologies into routing algorithms, demonstrating substantial research potential.

\subsection{ECO-Aware Routing}
ECO-aware routing iteratively refines paths under fixed area, layer budgets and utilization to co-optimize resource usage, wirelength, obstacle clearance, pin access, timing and IR-drop, halting when rules or iteration limits are met, which is mainly FA-based. The single-objective ECO-aware routing problem is the fundamental form of such tasks ~\cite{tseng2014power,cheng2021core}. Furthermore, a more common scenario in practical applications is balancing diverse design requirements to achieve a globally superior solution, thus relevant works often employed complex optimization frameworks~\cite{cong2001multilevel,wei2012eco}. With the introduction of advanced chip forms, ECO-aware routing based on specific architectures also has significant potential for development ~\cite{zhu2024timing}. %~\cite{zhu2024timing} proposed a timing-driven Steiner minimum tree algorithm that considers the X-architecture, obstacle avoidance, and timing slack constraints. In the future, ECO-aware routing may also have a larger impact on new chip types such as 3DIC, Chiplet and quantum computing chips.

\subsection{Power/Thermal-Aware Routing}
%Shift-left power-aware routing pre-optimizes dynamic and leakage power by lowering switching activity and buffer-induced capacitance on fixed placements, complementing placement-stage power reduction that acts through inter-component distance tuning.

%During the routing stage, the dynamic and static power performance of the chip can also be optimized in advance through Shift Left. Such optimization is usually achieved by reducing the switching activity during signal transmission and minimizing the impact of leakage current, which is similar to the power-aware placement mentioned earlier. However, the difference lies in the specific optimization methods. Power-aware routing mainly focuses on optimizing routing paths and buffer insertion based on a given layout, while power-aware placement mainly focuses on adjusting the distances between components during the early stages of design.

Power-aware routing pre-optimizes dynamic and leakage power by lowering switching activity and buffer-induced capacitance on fixed placements. Low-power paths are typically characterized by shorter lengths, lower capacitance along the path. Some created VPs, like ~\cite{bhardwaj2002quantifying} introduced a method to quantify power awareness via a "perfectly power-aware system" concept. There are also FA methods of heuristics ~\cite{dwibedi2024hybrid}, optimizing the position and size of buffers is also an effective routing strategy for improving power performance ~\cite{youssef2005pomr}. Power-aware routing can also consider different process and related reliability factors, like routing strategies under different working scenarios~\cite{bhardwaj2002quantifying} or thermal effect~\cite{cong2005thermal}, to achieve the power-saving aim. 

%When discussing power-aware routing, ~\cite{bhardwaj2002quantifying} emphasized the importance of dynamically adjusting routing strategies under different working scenarios to adapt to changing power demands, while also considering the impact of process parameter variations on routing power consumption. And in ~\cite{cong2005thermal}, although it focused primarily on solving the thermal-aware routing problem in 3D ICs, the approach not only reduced the maximum chip temperature but also to some extent considered power-oriented routing length optimization.

\subsection{EM-Aware Routing}

To restore convergence lost at advanced nodes, EM constraints are promoted from post-route sign-off to in-route design rules. At the net-level granularity, EM-aware algorithms confine constraints to entire nets or global wire segments, predominantly relying on graph-search or ILP techniques with a moderate size of problem instances~\cite{pak2012electromigration, jia2018electromigration}. At the subnet or segment level, the granularity of constraints is refined down to the width of each individual metal segment.~\cite{zhou2019aware} enforced “EM immortality” and introduced the concept of reservoir branches that allow selected segments to fail within the target lifetime while still guaranteeing global EM compliance.~\cite{axelou2024electromigration} encapsulated EM/IR constraints into a differentiable objective function with PSO as its solver. At the grid-level, constraints are resolved to layout-grid or Steiner-point/layer resolution through a solution process that is tightly coupled with on-die geometry while simultaneously accounting for parasitics, symmetry, and design-rule requirements~\cite{bigalke2018increasing}.

%At the grid-level, constraints are resolved down to the resolution of layout grids or Steiner points and layers, and the solution process is tightly coupled with on-die geometry while simultaneously accounting for parasitics, symmetry, and design-rule requirements. ~\cite{bigalke2018increasing} perform mesh discretization and boundary condition assignment directly inside the placement and routing engine, invoke external FEM/FIT solvers for rapid evaluation, and iteratively relocate cells and reassign layers guided by dual temperature-stress metrics to achieve early-stage EM robustness.

\begin{table}[htbp]
  \centering
  \caption{Shift-Left in Routing}
  \resizebox{\textwidth}{!}{%
    \renewcommand{\arraystretch}{1.3}  
    \begin{tabular}{|>{\centering\arraybackslash}m{7em}|>{\centering\arraybackslash}m{10em}|>{\centering\arraybackslash}m{32em}|} 
    \hline
    \textbf{Type} & \textbf{Methodology} & \textbf{Representative Works / Applications} \\
    \hline
    \multirow{3}[6]{*}{\shortstack{ECO-Aware}} 
        & Single-Objective 
        & Greedy pareto optimization ~\cite{tseng2014power}; \newline{} Combination of detailed routing and placement ~\cite{cheng2021core} \\
    \cline{2-3}          
        & Multiple-Objectives 
        & ECO timing framework using spare-cell replacement and rerouting ~\cite{wei2012eco}; \newline{} Multi-level gridless full-chip routing optimization ~\cite{wei2012eco} \\
    \cline{2-3}          
        & Specific Architectures 
        & PSO-based timing-driven Steiner tree for X-architecture ~\cite{zhu2024timing} \\
    \hline
    \multirow{2}[4]{*}{\shortstack{Thermal / Power\\Aware}} 
        & Low-Power Paths 
        & Fully power-aware systems with optimized routing and layout ~\cite{bhardwaj2002quantifying}; \newline{} Hybrid GA-PSO with SA for routing, reducing dynamic and static power ~\cite{dwibedi2024hybrid} \\
    \cline{2-3}          
        & Buffer Based 
        & Power optimization by buffer insertion and resizing ~\cite{youssef2005pomr} \\
    \hline
    \multirow{3}[6]{*}{\shortstack{EM-Aware}} 
        & Net-Level  
        & TSV stress gradients in directional EM for A*-based maze routing ~\cite{pak2012electromigration}; \newline{} Wirelength–current thresholds with maze pruning and ILP repair ~\cite{jia2018electromigration} \\
    \cline{2-3}          
        & Subnet-Level 
        & Two-stage LP with reservoir branches for EM reliability and IR-drop ~\cite{zhou2019aware}; \newline{} PSO optimizes wire widths, improving EM lifetime and reducing area ~\cite{axelou2024electromigration} \\
    \cline{2-3}          
        & Grid-Level  
        & Grid-level EM via FEM/FIT with iterative cell and layer adjustment ~\cite{bigalke2018increasing} \\
    \hline
    \end{tabular}%
    }
  \label{tab:shiftleft_routing}%
\end{table}

\section{Shift-Left in Signoff}
\label{shift-left-signoff}
Signoff, the final pre-tapeout verification stage, ensures the design meets all requirements through STA, EM, IR drop, and physical verification (DRC, LVS, DFM Check, etc.). New violations may arise due to unaccounted scenarios in earlier stages, requiring ECOs to resolve them without introducing new issues. This process often involves multiple iterations due to the separation of signoff analysis and ECO implementation. Recent works have explored tighter integration between these phases using GNNs. In ~\cite{zhu2024one,guo2024harnessing,zhu2023rc}, a learning-based framework was proposed for accurate and efficient cross-corner timing prediction, integrating learning-based reference corner selection and topology-aware interconnect timing prediction into the broader timing signoff and ECO flow. In~\cite{lu2023eco}, GNNs were leveraged to perform commercial-quality signoff power optimization, while~\cite{wang2022graph} introduced a directed GNN-based method that learns information from different neighbors for fast ECO leakage power optimization. On the IR-drop side,~\cite{fang2018machine} trained a machine learning model based on pre-ECO data to predict IR drop after ECO. Similarly,~\cite{lin2025graph} proposed a GNN-based model to estimate glitch rates, which, when integrated with a commercial power analysis tool, enables rapid identification of glitch hotspots and accelerates power optimization during signoff. In summary, most shift-left actions in the signoff stages involve adding VPs through GNN-based prediction models to accelerate analysis steps and interlock with ECO actions.

% \subsection{Interlocked Signoff and ECO Phases}
% The RTL-Level SDC Timing
% Exception Verification Ecosystem. \url{https://iccircle.com/static/upload/img20250107184558.pdf}

% In-Situ Timing-Error Prediction, ECO fix of targeted DFFs. \url{https://www.jstage.jst.go.jp/article/elex/advpub/0/advpub_20.20230145/_pdf}

% Pre-RTL Testbenches.  M. Orenes-Vera, M. Martonosi, and D. Wentzlaff, “From RTL to SVA:
% LLM-assisted generation of formal verification testbenches,” arXiv
% preprint arXiv:2309.09437, 2023.

% \url{}
\section{Shift-Left for Verification and Testing}
\label{shift-left-test-veri}

\begin{table}[!t]
  \centering
  \caption{Shift Left for Verification and Testing}
  \resizebox{\textwidth}{!}{%
    \renewcommand{\arraystretch}{1.3}  
    \begin{tabular}{|>{\centering\arraybackslash}m{6em}|>{\centering\arraybackslash}m{12em}|>{\centering\arraybackslash}m{30em}|} 
    \hline
    \textbf{Type} & \textbf{Methodology} & \textbf{Representative Works / Applications} \\
    \hline
    \multirow{6}[8]{*}{\shortstack{Formal\\Verification}} 
        & Functional Correctness Verification 
        & RTL-stage verification of SoCs and AI accelerators ~\cite{oye2024deep, anton2024vericheri}; \newline{} Correctness of HLS ~\cite{herklotz2021formal}; Verification of memory models ~\cite{herklotz2021formal} \\
    \cline{2-3}          
        & Equivalence Checking 
        & Pre-synthesis vs. post-synthesis equivalence ~\cite{lee2013equivalence}; \newline{} Cross-abstraction hierarchical verification ~\cite{lu2025hierarchical}; \newline{}
        RTL vs. high-level models ~\cite{marquez2013formal, raia2024case} \\
    \cline{2-3}          
        & Security Verification 
        & IP-level security assurance ~\cite{tan2025rtl}; Security-critical systems verification ~\cite{grimm2018survey} \\
    \cline{2-3}          
        & LLM-based Formal Verification 
        & Assertion generation and correction ~\cite{bai2025assertionforge, maddala2024laag, huang2024towards}; \newline{} RTL code generation ~\cite{liu2024openllm, liu2024rtlcoder}; \newline{} Security verification ~\cite{saha2025sv}; Benchmarking datasets/tools like FVeval ~\cite{kang2025fveval} \\
    \hline
    \multirow{4}[8]{*}{\shortstack{Physical\\Verification}} 
        & DRC/LVS Verification 
        & Synopsys IC Validator Explorer: early LVS at IP level ~\cite{SynopsysExplorerLVS}; \newline{} Siemens Calibre nmDRC Recon \& LVS Recon: early targeted checks ~\cite{SiemensDesignStageVerification} \\
    \cline{2-3}          
        & Parasitic Extraction (PEX) Verification 
        & Siemens Calibre RealTime Digital: real-time DRC feedback ~\cite{SiemensDesignStageVerification} \\
    \cline{2-3}          
        & Reliability Verification 
        & Cadence Innovus + Tempus (timing) + Voltus (power integrity) ~\cite{CadenceInnovus} \\
    \cline{2-3}          
        & Manufacturability / Yield Verification 
        & Synopsys IC Validator Explorer DFM capabilities ~\cite{SynopsysExplorerLVS}; \newline{} Siemens Calibre nmPlatform for lithography/DFM analysis ~\cite{SiemensDesignStageVerification} \\
    \hline
    \end{tabular}%
    }
  \label{tab:shiftleft}
\end{table}

% \textbf{General Overview of the motivation of shift left for verification and testing}
Shift-Left verification/testing migrates DRC/LVS, DFT and software validation from post-layout sign-off to incremental checks during RTL-to-layout evolution, catching SoC bugs when fixes are cheap and preventing the costly rework cascade that late-stage defects once triggered. Table~\ref{tab:shiftleft} presents a roster of the featured works that have been discussed.

%Traditionally, verification and testing in chip design were late-stage processes. Verification, including functional verification and physical verification including DRC and LVS, was a signoff activity after layout was complete. Testing focused on Design-for-Test (DFT) for manufacturing defect detection, while software testing began only after a physical chip was available. Bugs found late in these stages were expensive and time consuming to fix, and severely impacts productivity. A change to fix one bug can create others, leading to a cycle of rework and delays. The shift left concept for verification and testing is a progressive solution that moves these quality assurance activities to earlier stages, whose core motivation is to address the growing complexity of SoC designs and reduce time to market (TTM). By embedding continuous checks throughout the development cycle, issues are caught in early design stages, when they are simpler and cheaper to resolve, significantly reducing the risk of costly rework.

% \textbf{RTL Formal Verification}
\subsection{Formal Verification}
Formal verification at RTL and post synthesis shifts correctness proofs via static mathematical analysis left of dynamic simulation, exposing bugs before costly iterative runs ~\cite{sanghavi2010formal}. Unlike dynamic simulation, which can only confirm behavior for a finite set of test cases, formal methods can exhaustively prove or disprove certain properties of a design ~\cite{brinkmann2017formal}. For modern VLSI design, this RTL-stage verification is essential to ensure functional correctness and security at the register-transfer level, especially for complex designs like custom AI accelerators ~\cite{oye2024deep, anton2024vericheri}. These techniques are particularly critical for ensuring the security of IP ~\cite{tan2025rtl}, the correctness of high-level synthesis ~\cite{herklotz2021formal}, and the functional equivalence between pre-synthesis and post-synthesis programs ~\cite{lee2013equivalence}. This hierarchical approach allows engineers to verify hardware at different levels of abstraction ~\cite{lu2025hierarchical}, from high-level models to RTL designs ~\cite{marquez2013formal, raia2024case}, and even for safety-critical systems ~\cite{grimm2018survey}. Furthermore, these methods facilitate the verification of memory model implementations ~\cite{hsiao2021synthesizing} and offer a formal methodology for industrial setups ~\cite{devarajegowda2019formal, hasan2015formal}. In recent years, the integration of LLMs has also begun to revolutionize formal verification and related tasks~\cite{bai2025assertionforge, maddala2024laag, huang2024towards, liu2024openllm, liu2024rtlcoder, saha2025sv, jha2025large, kang2025fveval}. %By leveraging their ability to understand natural language specifications, LLMs can automatically generate and correct formal properties, significantly improving verification efficiency ~\cite{huang2024towards}. The application of LLMs extends to various other aspects of the design flow, including RTL code generation ~\cite{liu2024openllm, liu2024rtlcoder}, security verification ~\cite{saha2025sv}, and broader verification and testing tasks ~\cite{jha2025large}. Tools and datasets like FVeval are being developed to benchmark the capabilities of LLMs in formal verification of digital hardware ~\cite{kang2025fveval}. This new generation of AI-powered tools is designed to streamline the verification process, reduce manual effort, and help designers achieve faster convergence and higher confidence in their designs.

% ~\cite{guo2016automatic} proposes an automatic VHDL-to-Formal-HDL (Coq-compatible) converter to reduce Proof-Carrying Hardware (PCH)’s high proof cost from manual RTL conversion, aiding formal detection of malicious logic in untrusted IP cores (functional testing fails this) and demonstrating via AES IP.

% ~\cite{orenes2021autosva} proposes AutoSVA, a framework to automatically generate formal verification testbenches that verify liveness and safety of control logic involved in module interactions, addressing manual SVA's inefficiencies.

% ~\cite{orenes2023using} further extends AutoSVA, and explores using llms to generate correct SystemVerilog Assertions (SVA) for Formal property verification (FPV), via an FPV evaluation framework, AutoSVA extension, and tests on RTL to explore the capability of llms to capture RTL behavior and generate correct SVA properties.

\subsection{Physical Verification}
%Mirroring formal verification’s left shift, physical verification now migrates DRC, LVS and DFM checks from final sign-off into incremental steps during layout assembly, intercepting manufacturing violations while layouts are still fluid, cheap to repair and unable to cascade into respins or missed market windows.

%Building on the concept of shifting formal verification left to catch logical bugs early, the same proactive approach is now crucial for physical verification. Physical verification is a critical part of chip design, encompassing a suite of checks to ensure a design is ready for manufacturing, including DRC, LVS, and DFM analysis. Traditionally, these were run as a final signoff step, but with growing design complexity and shrinking timelines, waiting until a full-chip layout is complete leads to significant delays and risks. The core motivation for shifting physical verification left is to address these challenges by catching and fixing issues earlier in the design cycle, when they are easier and more cost-effective to resolve. This prevents a domino effect of late-stage problems that can result in costly re-spins and missed market windows.

Mirroring formal verification’s left shift, physical verification now migrates DRC, LVS and DFM checks from final sign-off into incremental steps during layout assembly, intercepting manufacturing violations while layouts are still fluid, cheap to repair and unable to cascade into respins or missed market windows. EDA companies have developed advanced tool suites to enable this Shift-Left methodology~\cite{siemens_shift_left_calibre, SynopsysExplorerLVS, CadenceInnovus, SiemensDesignStageVerification}. Synopsys's IC Validator Explorer provides an innovative approach to accelerate verification during full-chip integration by running early LVS checks on blocks and IPs before final signoff~\cite{SynopsysExplorerLVS}. Cadence's Innovus embeds signoff-grade Tempus timing and Voltus power engines to model electrical effects early and accelerate performance convergence~\cite{CadenceInnovus}. Siemens’ Calibre nmPlatform enables Shift-Left signoff by providing IP/block teams with early signoff-quality checks through nmDRC Recon and LVS Recon~\cite{SiemensDesignStageVerification}.%These solutions from EDA companies provide a comprehensive strategy for integrating physical verification into every stage of the design process, ensuring a more efficient and reliable path to design closure.
% \textbf{Physical Verification}

% Software Validation Using a Progressive Chip Design Process
% \url{https://www.lifecycleinsights.com/shift-left-with-software-validation-a-progressive-approach-to-chip-design/}

% \url{https://www.redline13.com/blog/2017/02/shift-left-testing-about/}

% \section{Shift Left with Pre-RTL Simulation}

% \section{Shift Left for Testing}

% \section{Hardware/Software Codesign: Shift Up}

% Cross-layer design

% \section{Fusion}

% \section{Shift Left in Emerging Technologies}
% 3DIC
% Chiplet
% PIM
% \section{Generating a Chip with AI}

% \section{Data Consistency}

% \section{Incremental Shift Left}

% ~\cite{kahng2022rosettastone}
\iffalse
\begin{figure*}[htbp]
  \centering
\includegraphics[width=0.9\textwidth]{./figures/3pictures_ver0913.pdf}
  \caption{Shift-Left comparison among different design stages.}
  \label{fig:spictures}
\end{figure*}
\fi

\section{Shift-Left Comparisons Among Different Stages}
\begin{figure*}[htbp]
  \centering
\includegraphics[width=\textwidth]{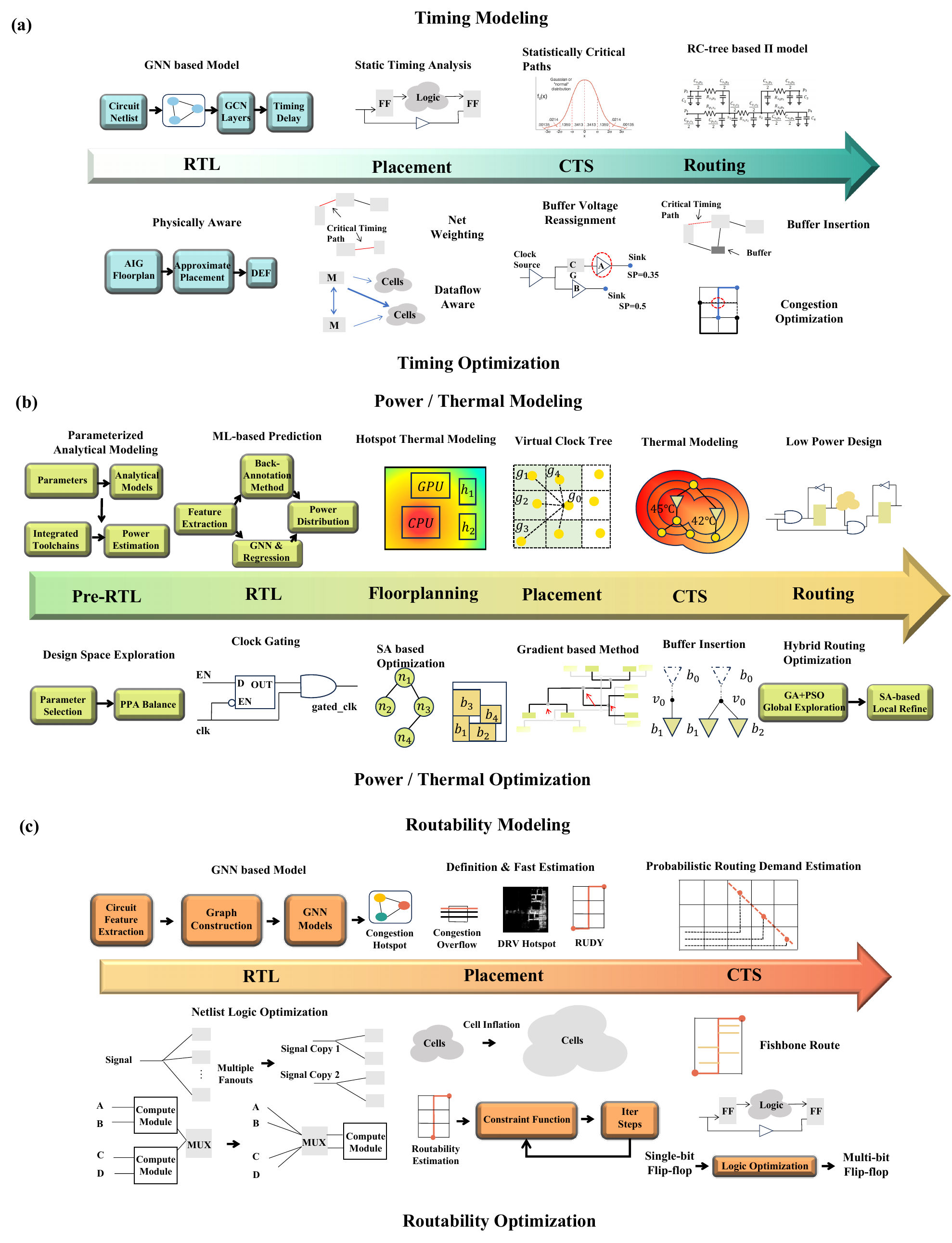}
  \caption{Comparison of three main optimization concerns among the design flow. (a) Timing. (b) Power/Thermal. (c) Routability.}
  \label{fig:spictures}
\end{figure*}

In the sections above, we present a stage-specific taxonomy of Shift-Left concept deployment; however, certain pervasive optimization objectives, including timing, power, and routability, are frequently addressed across virtually every design phase through divergent concrete techniques. To complement the stage-oriented exposition, we therefore provide a comparative analyzation that juxtaposes these universal considerations across the successive phases as Fig ~\ref{fig:spictures} shows.

Timing modeling and optimization strategies are shown in Fig ~\ref{fig:spictures}(a), the consideration of the which is primarily concentrated in RTL, placement, CTS and routing. From the viewpoint of VP, the modeling of timing becomes progressively more explicit and detailed as the design stage deepens, due to the multifaceted factors that influence timing at each stage. At the RTL level, the large gap to the final stage limits the availability of discernible patterns, and existing methods predominantly rely on machine learning techniques like GNNs to predict delays in the corresponding netlists. During the placement stage, the classical STA model becomes applicable. In CTS, the primary influencing factor is critical paths, and statistical methods are commonly used to identify them. While in routing, with the specific wiring already defined, RC-tree-based models can estimate the delay caused by metals from a physical rather than a signal perspective. From the perspective of FA, since the later phases have a greater impact on timing, effective and specific optimization measures for timing are still concentrated in the latter three stages, with the scope of considerations expanding progressively. In the RTL stage, the predicted information can be used to guide the initial floorplan. In placement, the objective can be systematically optimized through net weighting and dataflow. Then path timing can be adjusted by redistributing buffer voltages through CTS stage and buffer insertion adjustment and routing congestion can be optimized to avoid increased delay due to long detours in the routing stage.

Fig ~\ref{fig:spictures}(b) illustrates the construction of power or thermal models and their optimization strategies within the integrated circuit design process. For VPs, the design flow commences with parameterized analytical modeling, employing integrated toolchains and analytical models to estimate power consumption, followed by a machine learning-based prediction phase that utilizes feature extraction and GNN regression to forecast power distribution. Subsequently, hotspot thermal modeling is applied to analyze the thermal distribution of various modules while the virtual clock tree model optimizes the distribution of clock signals. The thermal modeling phase identifies critical paths through thermal distribution maps for further optimization, ultimately leading to low-power design techniques to reduce overall power consumption. In the optimization phase, design space exploration is conducted through parameter selection and PPA balancing. Clock gating technology reduces power consumption by controlling clock signals, and optimization based on SA employs the SA algorithm to seek optimal solutions. Gradient-based methods guide design optimization by calculating gradients, and buffer insertion improves signal integrity and reduces power consumption by inserting buffers on critical paths. Finally, hybrid routing optimization combines global exploration and local refinement strategies to optimize routing, thereby minimizing metal-induced delays and avoiding increased latency due to excessively long detours. This entire process underscores the significance and complexity of power and thermal management throughout the early design to the later implementation stages of integrated circuit design.

As for routability shown in Fig ~\ref{fig:spictures}(c), about the VP methods, the process begins with circuit feature extraction, graph construction, and the application of GNN-based models to identify congestion hotspots. This is followed by the definition and rapid estimation of key parameters such as congestion overflow, DRV hotspot, and RUDY, alongside probabilistic routing demand estimation, which provides a foundation for subsequent routing decisions. For FAs, starting from the RTL, netlist logic optimization is performed using multiple fanout signal copying techniques to reduce congestion. During the placement phase, cell inflation and routability estimation are utilized to optimize cell layout. Proceeding to the CTS phase, the fishbone routing strategy is employed to further refine the routing. Additionally, logic optimization is conducted between single-bit and multi-bit flip-flops to enhance the overall routability and performance of the design. This entire process underscores the significance and complexity of routability in the integrated circuit design, from early-stage design to later implementation, and the necessity of a series of optimization measures to improve design quality and efficiency.

The aforementioned processes leverage a variety of techniques including parametric modeling, physical application and MLs, thereby integrating the Shift-Left concept into EDA algorithms and tools. Concurrently, future challenges for Shift-Left may encompass enhancing the predictive accuracy of models, optimizing the computational efficiency of algorithms, improving generalizability, and more effectively integrating AI technologies for convenience. As technology evolves, it is also necessary to adapt to emerging technologies and consider new requirements such as environmental sustainability.

\section{Conclusions}
To conclude, the survey proposes a taxonomy with detailed methodology introduced about Shift-Left techniques in different design phases of EDA flow, as well as potential directions and future challenges. This study offers a detailed reference and practical guidance for future systematic investigations of Shift-Left methodologies, ultimately contributing to enhanced chip-design efficiency across the industry. To facilitate dynamic tracking of recent Shift-Left research in EDA, we also maintain a GitHub repository: \url{https://github.com/iCAS-SJTU/Shift-Left-EDA-Papers}.

\section*{Acknowledgment}
This work was supported in part by the National Science and Technology Major Project (Grant No. 2021ZD0114701), the State Key Laboratory of Integrated Chips and Systems (SKLICS) open fund, and the SJTU Explore-X fund.
\bibliographystyle{unsrt}
\bibliography{reference}

\begin{thebibliography}{100}

\bibitem{smith2001shift}
Larry Smith.
\newblock Shift-left testing.
\newblock {\em Dr. Dobb's Journal: Software Tools for the Professional Programmer}, 26(9):56--58, 2001.

\bibitem{hutchison2013shift}
Steven~J Hutchison and DEFENSE ACQUISITION UNIV FT~BELVOIR VA.
\newblock Shift left! test earlier in the life cycle.
\newblock {\em Defense AT\&L Magazine}, 2013.

\bibitem{phan2023challenges}
Quoc-Sang Phan, Kim-Hao Nguyen, and ThanhVu Nguyen.
\newblock The challenges of shift left static analysis.
\newblock In {\em 2023 IEEE/ACM 45th International Conference on Software Engineering: Software Engineering in Practice (ICSE-SEIP)}, pages 340--342. IEEE, 2023.

\bibitem{shiftleft}
Yun Liang, Cheng Zhuo, and Yongfu Li.
\newblock The shift-left design paradigm of eda: Progress and challenges.
\newblock {\em SCIENTIA SINICA Informationis}, 2023.

\bibitem{bhardwaj2021shift}
Vivek Bhardwaj.
\newblock Shift left trends for design convergence in soc: An eda perspective.
\newblock {\em Vivek Bhardwaj. Shift Left Trends for Design Convergence in SOC: An EDA Perspective. International Journal of Computer Applications}, 174(16):22--27, 2021.

\bibitem{synospsys}
Synopsys.
\newblock Fusion compiler.
\newblock \url{https://www.synopsys.com/implementation-and-signoff/physical-implementation/fusion-compiler.html}.

\bibitem{cadence}
Cadence.
\newblock ispatial: Next-generation common physical optimization flow.
\newblock \url{https://community.cadence.com/cadence_blogs_8/b/di/posts/ispatial-next-gen-flow}.

\bibitem{keysight}
Keysight.
\newblock Integrated software tools shift left design cycles to increase engineering productivity.
\newblock \url{https://www.keysight.com/us/en/about/newsroom/news-releases/2023/0926-pr24-126-keysight-eda-2024-integrated-software-tools-shift-.html}.

\bibitem{Siemens}
Siemens.
\newblock Shift left with calibre solutions.
\newblock \url{https://eda.sw.siemens.com/en-US/ic/calibre-design/shift-left-calibre/}.

\bibitem{10531906}
Xinfei Guo, Xiaotian Zhao, and Linyu Zhu.
\newblock Integration of shift-left updates into logic synthesis and macro placement.
\newblock In {\em 2024 Conference of Science and Technology for Integrated Circuits (CSTIC)}, pages 1--3, 2024.

\bibitem{kahng2022machine}
Andrew~B Kahng.
\newblock Machine learning for cad/eda: the road ahead.
\newblock {\em IEEE Design \& Test}, 40(1):8--16, 2022.

\bibitem{huang2021machine}
Guyue Huang, Jingbo Hu, Yifan He, Jialong Liu, Mingyuan Ma, Zhaoyang Shen, Juejian Wu, Yuanfan Xu, Hengrui Zhang, Kai Zhong, et~al.
\newblock Machine learning for electronic design automation: A survey.
\newblock {\em ACM Transactions on Design Automation of Electronic Systems (TODAES)}, 26(5):1--46, 2021.

\bibitem{fallon2020machine}
Elias Fallon.
\newblock Machine learning in eda: Opportunities and challenges.
\newblock In {\em 2020 ACM/IEEE 2nd Workshop on Machine Learning for CAD (MLCAD)}, pages 103--103. IEEE, 2020.

\bibitem{ren2022machine}
Haoxing Ren, Brucek Khailany, Matthew Fojtik, and Yanqing Zhang.
\newblock Machine learning and algorithms: Let us team up for eda.
\newblock {\em IEEE Design \& Test}, 40(1):70--76, 2022.

\bibitem{lopera2021survey}
Daniela~S{\'a}nchez Lopera, Lorenzo Servadei, Gamze~Naz Kiprit, Souvik Hazra, Robert Wille, and Wolfgang Ecker.
\newblock A survey of graph neural networks for electronic design automation.
\newblock In {\em 2021 ACM/IEEE 3rd Workshop on Machine Learning for CAD (MLCAD)}, pages 1--6. IEEE, 2021.

\bibitem{sanchez2023comprehensive}
Daniela S{\'a}nchez, Lorenzo Servadei, Gamze~Naz Kiprit, Robert Wille, and Wolfgang Ecker.
\newblock A comprehensive survey on electronic design automation and graph neural networks: Theory and applications.
\newblock {\em ACM Transactions on Design Automation of Electronic Systems}, 28(2):1--27, 2023.

\bibitem{xu2025large}
Kangwei Xu, Denis Schwachhofer, Jason Blocklove, Ilia Polian, Peter Domanski, Dirk Pfl{\"u}ger, Siddharth Garg, Ramesh Karri, Ozgur Sinanoglu, Johann Knechtel, et~al.
\newblock Large language models (llms) for electronic design automation (eda).
\newblock {\em arXiv preprint arXiv:2508.20030}, 2025.

\bibitem{zhong2023llm4eda}
Ruizhe Zhong, Xingbo Du, Shixiong Kai, Zhentao Tang, Siyuan Xu, Hui-Ling Zhen, Jianye Hao, Qiang Xu, Mingxuan Yuan, and Junchi Yan.
\newblock Llm4eda: Emerging progress in large language models for electronic design automation.
\newblock {\em arXiv preprint arXiv:2401.12224}, 2023.

\bibitem{chen2024dawn}
Lei Chen, Yiqi Chen, Zhufei Chu, Wenji Fang, Tsung-Yi Ho, Yu~Huang, Sadaf Khan, Min Li, Xingquan Li, Yun Liang, et~al.
\newblock The dawn of ai-native eda: Promises and challenges of large circuit models.
\newblock {\em arXiv preprint arXiv:2403.07257}, 2024.

\bibitem{he2024large}
Zhuolun He and Bei Yu.
\newblock Large language models for eda: Future or mirage?
\newblock In {\em Proceedings of the 2024 International Symposium on Physical Design}, pages 65--66, 2024.

\bibitem{alsaqer2024potential}
Shadan Alsaqer, Sarah Alajmi, Imtiaz Ahmad, and Mohammad Alfailakawi.
\newblock The potential of llms in hardware design.
\newblock {\em Journal of Engineering Research}, 2024.

\bibitem{fang2025survey}
Wenji Fang, Jing Wang, Yao Lu, Shang Liu, Yuchao Wu, Yuzhe Ma, and Zhiyao Xie.
\newblock A survey of circuit foundation model: Foundation ai models for vlsi circuit design and eda.
\newblock {\em arXiv preprint arXiv:2504.03711}, 2025.

\bibitem{pan2025survey}
Jingyu Pan, Guanglei Zhou, Chen-Chia Chang, Isaac Jacobson, Jiang Hu, and Yiran Chen.
\newblock A survey of research in large language models for electronic design automation.
\newblock {\em ACM Transactions on Design Automation of Electronic Systems}, 2025.

\bibitem{lin2023hl}
Zhe Lin, Tingyuan Liang, Jieru Zhao, Sharad Sinha, and Wei Zhang.
\newblock Hl-pow: learning-assisted pre-rtl power modeling and optimization for fpga hls.
\newblock {\em IEEE Transactions on Computer-Aided Design of Integrated Circuits and Systems}, 42(11):3925--3938, 2023.

\bibitem{chen2022pros}
Jingsong Chen, Jian Kuang, Guowei Zhao, Dennis J-H Huang, and Evangeline~FY Young.
\newblock Pros 2.0: A plug-in for routability optimization and routed wirelength estimation using deep learning.
\newblock {\em IEEE Transactions on Computer-Aided Design of Integrated Circuits and Systems}, 42(1):164--177, 2022.

\bibitem{gao2022congestion}
Xiang Gao, Yi-Min Jiang, Lixin Shao, Pedja Raspopovic, Menno~E Verbeek, Manish Sharma, Vineet Rashingkar, and Amit Jalota.
\newblock Congestion and timing aware macro placement using machine learning predictions from different data sources: Cross-design model applicability and the discerning ensemble.
\newblock In {\em Proceedings of the 2022 International Symposium on Physical Design}, pages 195--202, 2022.

\bibitem{guo2022timing}
Zizheng Guo, Mingjie Liu, Jiaqi Gu, Shuhan Zhang, David~Z Pan, and Yibo Lin.
\newblock A timing engine inspired graph neural network model for pre-routing slack prediction.
\newblock In {\em Proceedings of the 59th ACM/IEEE Design Automation Conference}, pages 1207--1212, 2022.

\bibitem{openroad}
Openroad-flow-script.
\newblock \url{https://github.com/The-OpenROAD-Project/OpenROAD-flow-scripts}.

\bibitem{jung2021metrics2}
Jinwook Jung, Andrew~B Kahng, Seungwon Kim, and Ravi Varadarajan.
\newblock Metrics2. 1 and flow tuning in the ieee ceda robust design flow and openroad iccad special session paper.
\newblock In {\em 2021 IEEE/ACM International Conference On Computer Aided Design (ICCAD)}, pages 1--9. IEEE, 2021.

\bibitem{akram2019survey}
Ayaz Akram and Lina Sawalha.
\newblock A survey of computer architecture simulation techniques and tools.
\newblock {\em Ieee Access}, 7:78120--78145, 2019.

\bibitem{binkert2011gem5}
Nathan Binkert, Bradford Beckmann, Gabriel Black, Steven~K Reinhardt, Ali Saidi, Arkaprava Basu, Joel Hestness, Derek~R Hower, Tushar Krishna, Somayeh Sardashti, et~al.
\newblock The gem5 simulator.
\newblock {\em ACM SIGARCH computer architecture news}, 39(2):1--7, 2011.

\bibitem{lowe2020gem5}
Jason Lowe-Power, Abdul~Mutaal Ahmad, Ayaz Akram, Mohammad Alian, Rico Amslinger, Matteo Andreozzi, Adri{\`a} Armejach, Nils Asmussen, Brad Beckmann, Srikant Bharadwaj, et~al.
\newblock The gem5 simulator: Version 20.0+.
\newblock {\em arXiv preprint arXiv:2007.03152}, 2020.

\bibitem{carlson2011sniper}
Trevor~E Carlson, Wim Heirman, and Lieven Eeckhout.
\newblock Sniper: Exploring the level of abstraction for scalable and accurate parallel multi-core simulation.
\newblock In {\em Proceedings of 2011 International Conference for High Performance Computing, Networking, Storage and Analysis}, pages 1--12, 2011.

\bibitem{sanchez2013zsim}
Daniel Sanchez and Christos Kozyrakis.
\newblock Zsim: Fast and accurate microarchitectural simulation of thousand-core systems.
\newblock {\em ACM SIGARCH Computer architecture news}, 41(3):475--486, 2013.

\bibitem{rosenfeld2011dramsim2}
Paul Rosenfeld, Elliott Cooper-Balis, and Bruce Jacob.
\newblock Dramsim2: A cycle accurate memory system simulator.
\newblock {\em IEEE computer architecture letters}, 10(1):16--19, 2011.

\bibitem{li2020dramsim3}
Shang Li, Zhiyuan Yang, Dhiraj Reddy, Ankur Srivastava, and Bruce Jacob.
\newblock Dramsim3: A cycle-accurate, thermal-capable dram simulator.
\newblock {\em IEEE Computer Architecture Letters}, 19(2):106--109, 2020.

\bibitem{parashar2019timeloop}
Angshuman Parashar, Priyanka Raina, Yakun~Sophia Shao, Yu-Hsin Chen, Victor~A Ying, Anurag Mukkara, Rangharajan Venkatesan, Brucek Khailany, Stephen~W Keckler, and Joel Emer.
\newblock Timeloop: A systematic approach to dnn accelerator evaluation.
\newblock In {\em 2019 IEEE international symposium on performance analysis of systems and software (ISPASS)}, pages 304--315. IEEE, 2019.

\bibitem{shao2014aladdin}
Yaqi~S. Shao, Brandon Reagen, Gu-Yeon Wei, and David Brooks.
\newblock Aladdin: A pre-rtl, power-performance accelerator simulator enabling large design space exploration of customized architectures.
\newblock In {\em ACM SIGARCH Computer Architecture News}, 2014.

\bibitem{wu2022sparseloop}
Yannan~Nellie Wu, Po-An Tsai, Angshuman Parashar, Vivienne Sze, and Joel~S Emer.
\newblock Sparseloop: An analytical approach to sparse tensor accelerator modeling.
\newblock In {\em 2022 55th IEEE/ACM International Symposium on Microarchitecture (MICRO)}, pages 1377--1395. IEEE, 2022.

\bibitem{andrulis2024cimloop}
Tanner Andrulis, Joel~S Emer, and Vivienne Sze.
\newblock Cimloop: A flexible, accurate, and fast compute-in-memory modeling tool.
\newblock In {\em 2024 IEEE International Symposium on Performance Analysis of Systems and Software (ISPASS)}, pages 10--23. IEEE, 2024.

\bibitem{peng2019neurosim}
Xiaochen Peng, Shanshi Huang, Yandong Luo, Xiaoyu Sun, and Shimeng Yu.
\newblock Dnn+ neurosim: An end-to-end benchmarking framework for compute-in-memory accelerators with versatile device technologies.
\newblock In {\em 2019 IEEE international electron devices meeting (IEDM)}, pages 32--5. IEEE, 2019.

\bibitem{li2009mcpat}
Sheng Li, Jung~Ho Ahn, Richard~D Strong, Jay~B Brockman, Dean~M Tullsen, and Norman~P Jouppi.
\newblock Mcpat: An integrated power, area, and timing modeling framework for multicore and manycore architectures.
\newblock In {\em Proceedings of the 42nd annual ieee/acm international symposium on microarchitecture}, pages 469--480, 2009.

\bibitem{muralimanohar2009cacti}
Naveen Muralimanohar, Rajeev Balasubramonian, and Norman~P Jouppi.
\newblock Cacti 6.0: A tool to model large caches.
\newblock {\em HP laboratories}, 27:28, 2009.

\bibitem{balasubramonian2017cacti}
Rajeev Balasubramonian, Andrew~B Kahng, Naveen Muralimanohar, Ali Shafiee, and Vaishnav Srinivas.
\newblock Cacti 7: New tools for interconnect exploration in innovative off-chip memories.
\newblock {\em ACM Transactions on Architecture and Code Optimization (TACO)}, 14(2):1--25, 2017.

\bibitem{wu2019accelergy}
Yannan~Nellie Wu, Joel~S Emer, and Vivienne Sze.
\newblock Accelergy: An architecture-level energy estimation methodology for accelerator designs.
\newblock In {\em 2019 IEEE/ACM International Conference on Computer-Aided Design (ICCAD)}, pages 1--8. IEEE, 2019.

\bibitem{huang2006hotspot}
Wei Huang, Shougata Ghosh, Sivakumar Velusamy, Karthik Sankaranarayanan, Kevin Skadron, and Mircea~R Stan.
\newblock Hotspot: A compact thermal modeling methodology for early-stage vlsi design.
\newblock {\em IEEE Transactions on very large scale integration (VLSI) systems}, 14(5):501--513, 2006.

\bibitem{zhang2015hotspot6}
Runjie Zhang, Mircea~R Stan, and Kevin Skadron.
\newblock Hotspot 6.0: Validation, acceleration and extension.
\newblock {\em University of Virginia, Tech. Rep}, 2015.

\bibitem{han2022hotspot7}
Jun-Han Han, Xinfei Guo, Kevin Skadron, and Mircea~R Stan.
\newblock From 2.5 d to 3d chiplet systems: Investigation of thermal implications with hotspot 7.0.
\newblock In {\em 2022 21st IEEE Intersociety Conference on Thermal and Thermomechanical Phenomena in Electronic Systems (iTherm)}, pages 1--6. IEEE, 2022.

\bibitem{sridhar20103dice}
Arvind Sridhar, Alessandro Vincenzi, Martino Ruggiero, Thomas Brunschwiler, and David Atienza.
\newblock 3d-ice: Fast compact transient thermal modeling for 3d ics with inter-tier liquid cooling.
\newblock In {\em 2010 IEEE/ACM International Conference on Computer-Aided Design (ICCAD)}, pages 463--470. IEEE, 2010.

\bibitem{pathania2018hotsniper}
Anuj Pathania and J{\"o}rg Henkel.
\newblock Hotsniper: Sniper-based toolchain for many-core thermal simulations in open systems.
\newblock {\em IEEE Embedded Systems Letters}, 11(2):54--57, 2018.

\bibitem{hankin2021hotgauge}
Alexander Hankin, David Werner, Maziar Amiraski, Julien Sebot, Kaushik Vaidyanathan, and Mark Hempstead.
\newblock Hotgauge: A methodology for characterizing advanced hotspots in modern and next generation processors.
\newblock In {\em 2021 IEEE International Symposium on Workload Characterization (IISWC)}, pages 163--175. IEEE, 2021.

\bibitem{siddhu2022comet}
Lokesh Siddhu, Rajesh Kedia, Shailja Pandey, Martin Rapp, Anuj Pathania, J{\"o}rg Henkel, and Preeti~Ranjan Panda.
\newblock Comet: An integrated interval thermal simulation toolchain for 2d, 2.5 d, and 3d processor-memory systems.
\newblock {\em ACM Transactions on Architecture and Code Optimization (TACO)}, 19(3):1--25, 2022.

\bibitem{wang2023}
Runxi Wang, Jun-Han Han, Mircea~R. Stan, and Xinfei Guo.
\newblock Hot-lego: Architect microfluidic cooling equipped 3dic with pre-rtl thermal simulation.
\newblock In {\em 14th International Green and Sustainable Computing Conference}, 2023.

\bibitem{wang2025cool3d}
Runxi Wang, Ziheng Wang, Ting Lin, Jacob~M. Raby, Mircea~R. Stan, and Xinfei Guo.
\newblock Cool-3d: An end-to-end thermal-aware framework for early-phase design space exploration of microfluidic-cooled 3dics.
\newblock {\em IEEE Journal on Emerging and Selected Topics in Circuits and Systems}, pages 1--1, 2025.

\bibitem{genbrugge2010interval}
Davy Genbrugge, Stijn Eyerman, and Lieven Eeckhout.
\newblock Interval simulation: Raising the level of abstraction in architectural simulation.
\newblock In {\em HPCA-16 2010 The Sixteenth International Symposium on High-Performance Computer Architecture}, pages 1--12. IEEE, 2010.

\bibitem{chakravarthi2025software}
Veena~S Chakravarthi and Shivananda~R Koteshwar.
\newblock Software-defined socs.
\newblock In {\em SOC-Based Solutions in Emerging Application Domains}, pages 27--38. Springer, 2025.

\bibitem{mei2021zigzag}
Linyan Mei, Pouya Houshmand, Vikram Jain, Sebastian Giraldo, and Marian Verhelst.
\newblock Zigzag: Enlarging joint architecture-mapping design space exploration for dnn accelerators.
\newblock {\em IEEE Transactions on Computers}, 70(8):1160--1174, 2021.

\bibitem{dai2024multichiplet}
Zhuoyu Dai, Feibin Xiang, Xiangqu Fu, Yifan He, Wenyu Sun, Yongpan Liu, Guanhua Yang, Feng Zhang, Jinshan Yue, and Ling Li.
\newblock A multichiplet computing-in-memory architecture exploration framework based on various cim devices.
\newblock {\em IEEE Transactions on Computer-Aided Design of Integrated Circuits and Systems}, 43(12):4613--4625, 2024.

\bibitem{yang2022aero}
Simei Yang, Debjyoti Bhattacharjee, Vinay~BY Kumar, Saikat Chatterjee, Sayandip De, Peter Debacker, Diederik Verkest, Arindam Mallik, and Francky Catthoor.
\newblock Aero: Design space exploration framework for resource-constrained cnn mapping on tile-based accelerators.
\newblock {\em IEEE Journal on Emerging and Selected Topics in Circuits and Systems}, 12(2):508--521, 2022.

\bibitem{zhao2017comba}
Jieru Zhao, Liang Feng, Sharad Sinha, Wei Zhang, Yun Liang, and Bingsheng He.
\newblock Comba: A comprehensive model-based analysis framework for high level synthesis of real applications.
\newblock In {\em 2017 IEEE/ACM International Conference on Computer-Aided Design (ICCAD)}, pages 430--437. IEEE, 2017.

\bibitem{wu2021ironman}
Nan Wu, Yuan Xie, and Cong Hao.
\newblock Ironman: Gnn-assisted design space exploration in high-level synthesis via reinforcement learning.
\newblock In {\em Proceedings of the 2021 Great Lakes Symposium on VLSI}, pages 39--44, 2021.

\bibitem{o2018hlspredict}
Kenneth O'Neal, Mitch Liu, Hans Tang, Amin Kalantar, Kennen DeRenard, and Philip Brisk.
\newblock Hlspredict: Cross platform performance prediction for fpga high-level synthesis.
\newblock In {\em Proceedings of the International Conference on Computer-Aided Design}, pages 1--8, 2018.

\bibitem{bilavarn2006design}
Sebastien Bilavarn, Guy Gogniat, J-L Philippe, and Lilian Bossuet.
\newblock Design space pruning through early estimations of area/delay tradeoffs for fpga implementations.
\newblock {\em IEEE Transactions on Computer-Aided Design of Integrated Circuits and Systems}, 25(10):1950--1968, 2006.

\bibitem{choi2017hlscope+}
Young-kyu Choi, Peng Zhang, Peng Li, and Jason Cong.
\newblock Hlscope+: Fast and accurate performance estimation for fpga hls.
\newblock In {\em 2017 IEEE/ACM International Conference on Computer-Aided Design (ICCAD)}, pages 691--698. IEEE, 2017.

\bibitem{makrani2019pyramid}
Hosein~Mohammadi Makrani, Farnoud Farahmand, Hossein Sayadi, Sara Bondi, Sai Manoj~Pudukotai Dinakarrao, Houman Homayoun, and Setareh Rafatirad.
\newblock Pyramid: Machine learning framework to estimate the optimal timing and resource usage of a high-level synthesis design.
\newblock In {\em 2019 29th International Conference on Field Programmable Logic and Applications (FPL)}, pages 397--403. IEEE, 2019.

\bibitem{lin2020hl}
Zhe Lin, Jieru Zhao, Sharad Sinha, and Wei Zhang.
\newblock Hl-pow: A learning-based power modeling framework for high-level synthesis.
\newblock In {\em 2020 25th Asia and South Pacific Design Automation Conference (ASP-DAC)}, pages 574--580. IEEE, 2020.

\bibitem{lee2015dynamic}
Dongwook Lee, Lizy~K John, and Andreas Gerstlauer.
\newblock Dynamic power and performance back-annotation for fast and accurate functional hardware simulation.
\newblock In {\em 2015 Design, Automation \& Test in Europe Conference \& Exhibition (DATE)}, pages 1126--1131. IEEE, 2015.

\bibitem{zhao2019machine}
Jieru Zhao, Tingyuan Liang, Sharad Sinha, and Wei Zhang.
\newblock Machine learning based routing congestion prediction in fpga high-level synthesis.
\newblock In {\em 2019 Design, Automation \& Test in Europe Conference \& Exhibition (DATE)}, pages 1130--1135. IEEE, 2019.

\bibitem{10.1145/3195970.3196026}
Cunxi Yu, Houping Xiao, and Giovanni De~Micheli.
\newblock Developing synthesis flows without human knowledge.
\newblock In {\em Proceedings of the 55th Annual Design Automation Conference}, DAC '18, New York, NY, USA, 2018. Association for Computing Machinery.

\bibitem{10.1145/3380446.3430638}
Cunxi Yu and Wang Zhou.
\newblock Decision making in synthesis cross technologies using lstms and transfer learning.
\newblock In {\em Proceedings of the 2020 ACM/IEEE Workshop on Machine Learning for CAD}, MLCAD '20, page 55–60, New York, NY, USA, 2020. Association for Computing Machinery.

\bibitem{9599868}
Daniela~Sánchez Lopera, Lorenzo Servadei, Vishwa~Priyanka Kasi, Sebastian Prebeck, and Wolfgang Ecker.
\newblock Rtl delay prediction using neural networks.
\newblock In {\em 2021 IEEE Nordic Circuits and Systems Conference (NorCAS)}, pages 1--7, 2021.

\bibitem{wang2025bridging}
Mingjun Wang, Yihan Wen, Bin Sun, Jianan Mu, Juan Li, Xiaoyi Wang, Jing~Justin Ye, Bei Yu, and Huawei Li.
\newblock Bridging layout and {RTL}: Knowledge distillation based timing prediction.
\newblock In {\em Forty-second International Conference on Machine Learning}, 2025.

\bibitem{9218643}
Yanqing Zhang, Haoxing Ren, and Brucek Khailany.
\newblock Grannite: Graph neural network inference for transferable power estimation.
\newblock In {\em 2020 57th ACM/IEEE Design Automation Conference (DAC)}, pages 1--6, 2020.

\bibitem{10.1145/3316781.3317884}
Yuan Zhou, Haoxing Ren, Yanqing Zhang, Ben Keller, Brucek Khailany, and Zhiru Zhang.
\newblock Primal: Power inference using machine learning.
\newblock In {\em Proceedings of the 56th Annual Design Automation Conference 2019}, DAC '19, New York, NY, USA, 2019. Association for Computing Machinery.

\bibitem{10.1145/3352460.3358322}
Donggyu Kim, Jerry Zhao, Jonathan Bachrach, and Krste Asanovi\'{c}.
\newblock Simmani: Runtime power modeling for arbitrary rtl with automatic signal selection.
\newblock In {\em Proceedings of the 52nd Annual IEEE/ACM International Symposium on Microarchitecture}, MICRO-52, page 1050–1062, New York, NY, USA, 2019. Association for Computing Machinery.

\bibitem{ansys_powerartist_2022}
{Ansys, Inc.}
\newblock Ansys powerartist.
\newblock Product Brief, 2022.
\newblock Accessed: September 2, 2025.

\bibitem{8920342}
Robert Kirby, Saad Godil, Rajarshi Roy, and Bryan Catanzaro.
\newblock Congestionnet: Routing congestion prediction using deep graph neural networks.
\newblock In {\em 2019 IFIP/IEEE 27th International Conference on Very Large Scale Integration (VLSI-SoC)}, pages 217--222, 2019.

\bibitem{hemadri2025veriloclineofcodelevelprediction}
Raghu~Vamshi Hemadri, Jitendra Bhandari, Andre Nakkab, Johann Knechtel, Badri~P Gopalan, Ramesh Narayanaswamy, Ramesh Karri, and Siddharth Garg.
\newblock Veriloc: Line-of-code level prediction of hardware design quality from verilog code, 2025.

\bibitem{810630}
Yanbin Jiang and S.S. Sapatnekar.
\newblock An integrated algorithm for combined placement and libraryless technology mapping.
\newblock In {\em 1999 IEEE/ACM International Conference on Computer-Aided Design. Digest of Technical Papers (Cat. No.99CH37051)}, pages 102--105, 1999.

\bibitem{742830}
W.~Gosti, A.~Narayan, R.K. Brayton, and A.L. Sangiovanni-Vincentelli.
\newblock Wireplanning in logic synthesis.
\newblock In {\em 1998 IEEE/ACM International Conference on Computer-Aided Design. Digest of Technical Papers (IEEE Cat. No.98CB36287)}, pages 26--33, 1998.

\bibitem{968622}
W.~Gosti, S.R. Khatri, and A.L. Sangiovanni-Vincentelli.
\newblock Addressing the timing closure problem by integrating logic optimization and placement.
\newblock In {\em IEEE/ACM International Conference on Computer Aided Design. ICCAD 2001. IEEE/ACM Digest of Technical Papers (Cat. No.01CH37281)}, pages 224--231, 2001.

\bibitem{zhu2023delay}
Linyu Zhu and Xinfei Guo.
\newblock Delay-driven physically-aware logic synthesis with informed search.
\newblock In {\em 2023 IEEE 41st International Conference on Computer Design (ICCD)}, pages 327--335. IEEE, 2023.

\bibitem{1382637}
S.~Chatterjee and R.~Brayton.
\newblock A new incremental placement algorithm and its application to congestion-aware divisor extraction.
\newblock In {\em IEEE/ACM International Conference on Computer Aided Design, 2004. ICCAD-2004.}, pages 541--548, 2004.

\bibitem{garg2016study}
Shivani Garg and Neeraj~Kr Shukla.
\newblock A study of floorplanning challenges and analysis of macro placement approaches in physical aware synthesis.
\newblock {\em International Journal of Hybrid Information Technology}, 9(1):279--290, 2016.

\bibitem{10.1145/611817.611836}
Joey~Y. Lin, Ashok Jagannathan, and Jason Cong.
\newblock Placement-driven technology mapping for lut-based fpgas.
\newblock In {\em Proceedings of the 2003 ACM/SIGDA Eleventh International Symposium on Field Programmable Gate Arrays}, FPGA '03, page 121–126, New York, NY, USA, 2003. Association for Computing Machinery.

\bibitem{Joshi2020iSpatial}
Neha Joshi.
\newblock {iSpatial Flow in Genus: A Modern Approach for Physical Synthesis}.
\newblock Cadence Community Blog, July 2020.
\newblock Accessed: 2025-09-01.

\bibitem{Cadence2013Encounter}
{Cadence Design Systems, Inc.}
\newblock {Encounter RTL Compiler Advanced Physical Option}.
\newblock Technical Report 1793, Cadence Design Systems, Inc., December 2013.
\newblock Accessed: 2025-09-02.

\bibitem{979797}
Jen-Pin Weng and A.C. Parker.
\newblock 3d scheduling: high-level synthesis with floorplanning.
\newblock In {\em 28th ACM/IEEE Design Automation Conference}, pages 668--673, 1991.

\bibitem{10.1145/3649329.3658268}
Daniel Xing and Ankur Srivastava.
\newblock A high level approach to co-designing 3d ics.
\newblock In {\em Proceedings of the 61st ACM/IEEE Design Automation Conference}, DAC '24, New York, NY, USA, 2024. Association for Computing Machinery.

\bibitem{10.1145/2554688.2554775}
Hongbin Zheng, Swathi~T. Gurumani, Kyle Rupnow, and Deming Chen.
\newblock Fast and effective placement and routing directed high-level synthesis for fpgas.
\newblock In {\em Proceedings of the 2014 ACM/SIGDA International Symposium on Field-Programmable Gate Arrays}, FPGA '14, page 1–10, New York, NY, USA, 2014. Association for Computing Machinery.

\bibitem{9218718}
Licheng Guo, Jason Lau, Yuze Chi, Jie Wang, Cody~Hao Yu, Zhe Chen, Zhiru Zhang, and Jason Cong.
\newblock Analysis and optimization of the implicit broadcasts in fpga hls to improve maximum frequency.
\newblock In {\em 2020 57th ACM/IEEE Design Automation Conference (DAC)}, pages 1--6, 2020.

\bibitem{prost2013fast}
Adrien Prost-Boucle, Olivier Muller, and Fr{\'e}d{\'e}ric Rousseau.
\newblock A fast and autonomous hls methodology for hardware accelerator generation under resource constraints.
\newblock In {\em 2013 Euromicro Conference on Digital System Design}, pages 201--208. IEEE, 2013.

\bibitem{zhong2014design}
Guanwen Zhong, Vanchinathan Venkataramani, Yun Liang, Tulika Mitra, and Smail Niar.
\newblock Design space exploration of multiple loops on fpgas using high level synthesis.
\newblock In {\em 2014 IEEE 32nd international conference on computer design (ICCD)}, pages 456--463. IEEE, 2014.

\bibitem{liu2012compositional}
Hung-Yi Liu, Michele Petracca, and Luca~P Carloni.
\newblock Compositional system-level design exploration with planning of high-level synthesis.
\newblock In {\em 2012 Design, Automation \& Test in Europe Conference \& Exhibition (DATE)}, pages 641--646. IEEE, 2012.

\bibitem{wang2017flexcl}
Shuo Wang, Yun Liang, and Wei Zhang.
\newblock Flexcl: An analytical performance model for opencl workloads on flexible fpgas.
\newblock In {\em Proceedings of the 54th Annual Design Automation Conference 2017}, pages 1--6, 2017.

\bibitem{wang2016performance}
Zeke Wang, Bingsheng He, Wei Zhang, and Shunning Jiang.
\newblock A performance analysis framework for optimizing opencl applications on fpgas.
\newblock In {\em 2016 IEEE International Symposium on High Performance Computer Architecture (HPCA)}, pages 114--125. IEEE, 2016.

\bibitem{williams2009roofline}
Samuel Williams, Andrew Waterman, and David Patterson.
\newblock Roofline: an insightful visual performance model for multicore architectures.
\newblock {\em Communications of the ACM}, 52(4):65--76, 2009.

\bibitem{da2013performance}
Bruno Da~Silva, An~Braeken, Erik~H D'Hollander, and Abdellah Touhafi.
\newblock Performance modeling for fpgas: Extending the roofline model with high-level synthesis tools.
\newblock {\em International Journal of Reconfigurable Computing}, 2013(1):428078, 2013.

\bibitem{10.1145/3489517.3530408}
Nan Wu, Hang Yang, Yuan Xie, Pan Li, and Cong Hao.
\newblock High-level synthesis performance prediction using gnns: benchmarking, modeling, and advancing.
\newblock In {\em Proceedings of the 59th ACM/IEEE Design Automation Conference}, DAC '22, page 49–54, New York, NY, USA, 2022. Association for Computing Machinery.

\bibitem{zhong2017design}
Guanwen Zhong, Alok Prakash, Siqi Wang, Yun Liang, Tulika Mitra, and Smail Niar.
\newblock Design space exploration of fpga-based accelerators with multi-level parallelism.
\newblock In {\em Design, Automation \& Test in Europe Conference \& Exhibition (DATE), 2017}, pages 1141--1146. IEEE, 2017.

\bibitem{liu2016efficient}
Dong Liu and Benjamin~Carrion Schafer.
\newblock Efficient and reliable high-level synthesis design space explorer for fpgas.
\newblock In {\em 2016 26th International Conference on Field Programmable Logic and Applications (FPL)}, pages 1--8. IEEE, 2016.

\bibitem{liu2013learning}
Hung-Yi Liu and Luca~P Carloni.
\newblock On learning-based methods for design-space exploration with high-level synthesis.
\newblock In {\em Proceedings of the 50th annual design automation conference}, pages 1--7, 2013.

\bibitem{dai2018fast}
Steve Dai, Yuan Zhou, Hang Zhang, Ecenur Ustun, Evangeline~FY Young, and Zhiru Zhang.
\newblock Fast and accurate estimation of quality of results in high-level synthesis with machine learning.
\newblock In {\em 2018 IEEE 26th Annual International Symposium on Field-Programmable Custom Computing Machines (FCCM)}, pages 129--132. IEEE, 2018.

\bibitem{makrani2019xppe}
Hosein~Mohammadi Makrani, Hossein Sayadi, Tinoosh Mohsenin, Setareh Rafatirad, Avesta Sasan, and Houman Homayoun.
\newblock Xppe: cross-platform performance estimation of hardware accelerators using machine learning.
\newblock In {\em Proceedings of the 24th Asia and South Pacific Design Automation Conference}, pages 727--732, 2019.

\bibitem{pham2015exploiting}
Nam~Khanh Pham, Amit~Kumar Singh, Akash Kumar, and Mi~Mi~Aung Khin.
\newblock Exploiting loop-array dependencies to accelerate the design space exploration with high level synthesis.
\newblock In {\em 2015 Design, Automation \& Test in Europe Conference \& Exhibition (DATE)}, pages 157--162. IEEE, 2015.

\bibitem{zhong2016lin}
Guanwen Zhong, Alok Prakash, Yun Liang, Tulika Mitra, and Smail Niar.
\newblock Lin-analyzer: A high-level performance analysis tool for fpga-based accelerators.
\newblock In {\em Proceedings of the 53rd Annual Design Automation Conference}, pages 1--6, 2016.

\bibitem{ustun2020accurate}
Ecenur Ustun, Chenhui Deng, Debjit Pal, Zhijing Li, and Zhiru Zhang.
\newblock Accurate operation delay prediction for fpga hls using graph neural networks.
\newblock In {\em Proceedings of the 39th international conference on computer-aided design}, pages 1--9, 2020.

\bibitem{10.1145/2160916.2160952}
Jason Cong, Bin Liu, Guojie Luo, and Raghu Prabhakar.
\newblock Towards layout-friendly high-level synthesis.
\newblock In {\em Proceedings of the 2012 ACM International Symposium on International Symposium on Physical Design}, ISPD '12, page 165–172, New York, NY, USA, 2012. Association for Computing Machinery.

\bibitem{8399766}
Masato Tatsuoka and Mineo Kaneko.
\newblock Wire congestion aware high level synthesis flow with source code compiler.
\newblock In {\em 2018 International Conference on IC Design \& Technology (ICICDT)}, pages 101--104, 2018.

\bibitem{tariq2021high}
Osama~Bin Tariq, Junnan Shan, Georgios Floros, Christos~P Sotiriou, Mario~R Casu, Mihai~Teodor Lazarescu, and Luciano Lavagno.
\newblock High-level annotation of routing congestion for xilinx vivado hls designs.
\newblock {\em IEEE Access}, 9:54286--54297, 2021.

\bibitem{10323951}
Wenji Fang, Yao Lu, Shang Liu, Qijun Zhang, Ceyu Xu, Lisa~Wu Wills, Hongce Zhang, and Zhiyao Xie.
\newblock Masterrtl: A pre-synthesis ppa estimation framework for any rtl design.
\newblock In {\em 2023 IEEE/ACM International Conference on Computer Aided Design (ICCAD)}, pages 1--9, 2023.

\bibitem{10577671}
Wenji Fang, Yao Lu, Shang Liu, Qijun Zhang, Ceyu Xu, Lisa Wu~Wills, Hongce Zhang, and Zhiyao Xie.
\newblock Transferable presynthesis ppa estimation for rtl designs with data augmentation techniques.
\newblock {\em IEEE Transactions on Computer-Aided Design of Integrated Circuits and Systems}, 44(1):200--213, 2025.

\bibitem{10.1145/3470496.3527444}
Ceyu Xu, Chris Kjellqvist, and Lisa~Wu Wills.
\newblock Sns's not a synthesizer: a deep-learning-based synthesis predictor.
\newblock In {\em Proceedings of the 49th Annual International Symposium on Computer Architecture}, ISCA '22, page 847–859, New York, NY, USA, 2022. Association for Computing Machinery.

\bibitem{10.1145/3316781.3323919}
Jihye Kwon, Matthew~M. Ziegler, and Luca~P. Carloni.
\newblock A learning-based recommender system for autotuning design flows of industrial high-performance processors.
\newblock In {\em Proceedings of the 56th Annual Design Automation Conference 2019}, DAC '19, New York, NY, USA, 2019. Association for Computing Machinery.

\bibitem{9045201}
Zhiyao Xie, Guan-Qi Fang, Yu-Hung Huang, Haoxing Ren, Yanqing Zhang, Brucek Khailany, Shao-Yun Fang, Jiang Hu, Yiran Chen, and Erick~Carvajal Barboza.
\newblock Fist: A feature-importance sampling and tree-based method for automatic design flow parameter tuning.
\newblock In {\em 2020 25th Asia and South Pacific Design Automation Conference (ASP-DAC)}, pages 19--25, 2020.

\bibitem{9643533}
W.~Rhett Davis, Paul Franzon, Luis Francisco, Billy Huggins, and Rajeev Jain.
\newblock Fast and accurate ppa modeling with transfer learning.
\newblock In {\em 2021 IEEE/ACM International Conference On Computer Aided Design (ICCAD)}, pages 1--8, 2021.

\bibitem{10299840}
Yikang Ouyang, Sicheng Li, Dongsheng Zuo, Hanwei Fan, and Yuzhe Ma.
\newblock Asap: Accurate synthesis analysis and prediction with multi-task learning.
\newblock In {\em 2023 ACM/IEEE 5th Workshop on Machine Learning for CAD (MLCAD)}, pages 1--6, 2023.

\bibitem{synopsys_rtlarchitect_2021}
{Synopsys, Inc.}
\newblock Shift left with rtl architect for faster time-to-results.
\newblock Technical report, Synopsys, Inc., June 2021.
\newblock Accessed: September 2, 2025.

\bibitem{fang2025circuitfusionmultimodalcircuitrepresentation}
Wenji Fang, Shang Liu, Jing Wang, and Zhiyao Xie.
\newblock Circuitfusion: Multimodal circuit representation learning for agile chip design, 2025.

\bibitem{moravej2025graphsapprenticeteachingllm}
Reza Moravej, Saurabh Bodhe, Zhanguang Zhang, Didier Chetelat, Dimitrios Tsaras, Yingxue Zhang, Hui-Ling Zhen, Jianye Hao, and Mingxuan Yuan.
\newblock The graph's apprentice: Teaching an llm low level knowledge for circuit quality estimation, 2025.

\bibitem{10.1145/3676536.3676775}
Xufeng Yao, Yiwen Wang, Xing Li, Yingzhao Lian, Ran Chen, Lei Chen, Mingxuan Yuan, Hong Xu, and Bei Yu.
\newblock Rtlrewriter: Methodologies for large models aided rtl code optimization.
\newblock In {\em Proceedings of the 43rd IEEE/ACM International Conference on Computer-Aided Design}, ICCAD '24, New York, NY, USA, 2025. Association for Computing Machinery.

\bibitem{9371553}
Walter~Lau Neto, Matheus Trevisan~Moreira, Luca Amaru, Cunxi Yu, and Pierre-Emmanuel Gaillardon.
\newblock Read your circuit: Leveraging word embedding to guide logic optimization.
\newblock In {\em 2021 26th Asia and South Pacific Design Automation Conference (ASP-DAC)}, pages 530--535, 2021.

\bibitem{10299859}
Daniela Sánchez~Lopera, Ishwor Subedi, and Wolfgang Ecker.
\newblock Using graph neural networks for timing estimations of rtl intermediate representations.
\newblock In {\em 2023 ACM/IEEE 5th Workshop on Machine Learning for CAD (MLCAD)}, pages 1--6, 2023.

\bibitem{10.1145/3508352.3561095}
Daniela~S\'{a}nchez Lopera and Wolfgang Ecker.
\newblock Applying gnns to timing estimation at rtl.
\newblock In {\em Proceedings of the 41st IEEE/ACM International Conference on Computer-Aided Design}, ICCAD '22, New York, NY, USA, 2022. Association for Computing Machinery.

\bibitem{10473925}
Haisheng Zheng, Zhuolun He, Fangzhou Liu, Zehua Pei, and Bei Yu.
\newblock Lstp: A logic synthesis timing predictor.
\newblock In {\em 2024 29th Asia and South Pacific Design Automation Conference (ASP-DAC)}, pages 728--733, 2024.

\bibitem{10.1145/3508352.3549375}
Prianka Sengupta, Aakash Tyagi, Yiran Chen, and Jiang Hu.
\newblock How good is your verilog rtl code? a quick answer from machine learning.
\newblock In {\em Proceedings of the 41st IEEE/ACM International Conference on Computer-Aided Design}, ICCAD '22, New York, NY, USA, 2022. Association for Computing Machinery.

\bibitem{10299879}
Prianka Sengupta, Aakash Tyagi, Yiran Chen, and Jiang Hu.
\newblock Early identification of timing critical rtl components using ml based path delay prediction.
\newblock In {\em 2023 ACM/IEEE 5th Workshop on Machine Learning for CAD (MLCAD)}, pages 1--6, 2023.

\bibitem{10.1145/3649329.3655671}
Wenji Fang, Shang Liu, Hongce Zhang, and Zhiyao Xie.
\newblock Annotating slack directly on your verilog: Fine-grained rtl timing evaluation for early optimization.
\newblock In {\em Proceedings of the 61st ACM/IEEE Design Automation Conference}, DAC '24, New York, NY, USA, 2024. Association for Computing Machinery.

\bibitem{10181823}
Rakesh~M B, Pabitra Das, Anant Terkar, and Amit Acharyya.
\newblock Graspe: Accurate post-synthesis power estimation from rtl using graph representation learning.
\newblock In {\em 2023 IEEE International Symposium on Circuits and Systems (ISCAS)}, pages 1--5, 2023.

\bibitem{10.1145/3466752.3480064}
Zhiyao Xie, Xiaoqing Xu, Matt Walker, Joshua Knebel, Kumaraguru Palaniswamy, Nicolas Hebert, Jiang Hu, Huanrui Yang, Yiran Chen, and Shidhartha Das.
\newblock Apollo: An automated power modeling framework for runtime power introspection in high-volume commercial microprocessors.
\newblock In {\em MICRO-54: 54th Annual IEEE/ACM International Symposium on Microarchitecture}, MICRO '21, page 1–14, New York, NY, USA, 2021. Association for Computing Machinery.

\bibitem{7059105}
Jianlei Yang, Liwei Ma, Kang Zhao, Yici Cai, and Tin-Fook Ngai.
\newblock Early stage real-time soc power estimation using rtl instrumentation.
\newblock In {\em The 20th Asia and South Pacific Design Automation Conference}, pages 779--784, 2015.

\bibitem{ennocad_enfortius_2023}
{ennocad}.
\newblock {EnFortius\textregistered{} RPA empowers flagship AI chip design companies to achieve early power analysis}.
\newblock Press Release, October 2023.
\newblock Accessed: September 2, 2025.

\bibitem{synopsys_rtlpower_2023}
{Synopsys, Inc.}
\newblock Achieving consistent rtl power analysis accuracy.
\newblock Technical report, Synopsys, Inc., March 2023.
\newblock Accessed: September 2, 2025.

\bibitem{10.5555/3600270.3601747}
Shuwen Yang, Zhihao Yang, Dong Li, Yingxue Zhang, Zhanguang Zhang, Guojie Song, and Jianye Hao.
\newblock Versatile multi-stage graph neural network for circuit representation.
\newblock In {\em Proceedings of the 36th International Conference on Neural Information Processing Systems}, NIPS '22, Red Hook, NY, USA, 2022. Curran Associates Inc.

\bibitem{9643446}
Amur Ghose, Vincent Zhang, Yingxue Zhang, Dong Li, Wulong Liu, and Mark Coates.
\newblock Generalizable cross-graph embedding for gnn-based congestion prediction.
\newblock In {\em 2021 IEEE/ACM International Conference On Computer Aided Design (ICCAD)}, pages 1--9, 2021.

\bibitem{9707500}
Zhiyao Xie, Rongjian Liang, Xiaoqing Xu, Jiang Hu, Chen-Chia Chang, Jingyu Pan, and Yiran Chen.
\newblock Preplacement net length and timing estimation by customized graph neural network.
\newblock {\em IEEE Transactions on Computer-Aided Design of Integrated Circuits and Systems}, 41(11):4667--4680, 2022.

\bibitem{10.1145/277044.277072}
Amir~H. Salek, Jinan Lou, and Massoud Pedram.
\newblock A dsm design flow: putting floorplanning, technology-mapping, and gate-placement together.
\newblock In {\em Proceedings of the 35th Annual Design Automation Conference}, DAC '98, page 128–134, New York, NY, USA, 1998. Association for Computing Machinery.

\bibitem{5715609}
Yifang Liu, Rupesh~S. Shelar, and Jiang Hu.
\newblock Simultaneous technology mapping and placement for delay minimization.
\newblock {\em IEEE Transactions on Computer-Aided Design of Integrated Circuits and Systems}, 30(3):416--426, 2011.

\bibitem{406709}
Chau-Shen Chen, Yu-Wen Tsay, TingTing Hwang, A.C.H. Wu, and Youn-Long Lin.
\newblock Combining technology mapping and placement for delay-minimization in fpga designs.
\newblock {\em IEEE Transactions on Computer-Aided Design of Integrated Circuits and Systems}, 14(9):1076--1084, 1995.

\bibitem{979696}
M.~Pedram and N.~Bhat.
\newblock Layout driven technology mapping.
\newblock In {\em 28th ACM/IEEE Design Automation Conference}, pages 99--105, 1991.

\bibitem{10.5555/1509456.1509489}
Yifang Liu, Rupesh~S. Shelar, and Jiang Hu.
\newblock Delay-optimal simultaneous technology mapping and placement with applications to timing optimization.
\newblock In {\em Proceedings of the 2008 IEEE/ACM International Conference on Computer-Aided Design}, ICCAD '08, page 101–106. IEEE Press, 2008.

\bibitem{185212}
M.~Pedram and N.~Bhat.
\newblock Layout driven logic restructuring/decomposition.
\newblock In {\em 1991 IEEE International Conference on Computer-Aided Design Digest of Technical Papers}, pages 134--137, 1991.

\bibitem{4167765}
Charles~J. Alpert, Shrirang~K. Karandikar, Zhuo Li, Gi-Joon Nam, Stephen~T. Quay, Haoxing Ren, C.~N. Sze, Paul~G. Villarrubia, and Mehmet~C. Yildiz.
\newblock Techniques for fast physical synthesis.
\newblock {\em Proceedings of the IEEE}, 95(3):573--599, 2007.

\bibitem{727128}
Aiguo Lu, H.~Eisenmann, G.~Stenz, and F.M. Johannes.
\newblock Combining technology mapping with post-placement resynthesis for performance optimization.
\newblock In {\em Proceedings International Conference on Computer Design. VLSI in Computers and Processors (Cat. No.98CB36273)}, pages 616--621, 1998.

\bibitem{Reis2018}
Andr{\'e}~In{\'a}cio Reis and Jody M.~A. Matos.
\newblock {\em Physical Awareness Starting at Technology-Independent Logic Synthesis}, pages 69--101.
\newblock Springer International Publishing, Cham, 2018.

\bibitem{cai2025revisit}
Yichen Cai, Linyu Zhu, and Xinfei Guo.
\newblock Revisit mbff: Efficient early-stage multi-bit flip-flops clustering with physical and timing awareness.
\newblock In {\em Proceedings of the 30th Asia and South Pacific Design Automation Conference}, pages 1230--1236, 2025.

\bibitem{968621}
T.~Kutzschebauch and L.~Stok.
\newblock Congestion aware layout driven logic synthesis.
\newblock In {\em IEEE/ACM International Conference on Computer Aided Design. ICCAD 2001. IEEE/ACM Digest of Technical Papers (Cat. No.01CH37281)}, pages 216--223, 2001.

\bibitem{998371}
T.~Kutzschebauch and U.~Stok.
\newblock Layout driven decomposition with congestion consideration.
\newblock In {\em Proceedings 2002 Design, Automation and Test in Europe Conference and Exhibition}, pages 672--676, 2002.

\bibitem{998370}
D.~Pandini, L.T. Pileggi, and A.J. Strojwas.
\newblock Congestion-aware logic synthesis.
\newblock In {\em Proceedings 2002 Design, Automation and Test in Europe Conference and Exhibition}, pages 664--671, 2002.

\bibitem{410815}
Tong GaO, Kuang-Chien Chen, J.~Cong, Yuzheng Ding, and C.~Liu.
\newblock Placement and placement driven technology mapping for fpga synthesis.
\newblock In {\em Sixth Annual IEEE International ASIC Conference and Exhibit}, pages 91--94, 1993.

\bibitem{1190987}
D.~Pandini, L.T. Pileggi, and A.J. Strojwas.
\newblock Global and local congestion optimization in technology mapping.
\newblock {\em IEEE Transactions on Computer-Aided Design of Integrated Circuits and Systems}, 22(4):498--505, 2003.

\bibitem{10.1145/3649329.3656243}
Yuan Pu, Fangzhou Liu, Yu~Zhang, Zhuolun He, Yibo Lin, Kai-Yuan Chao, and Bei Yu.
\newblock Lesyn: Placement-aware logic resynthesis for non-integer multiple-cell-height designs.
\newblock In {\em Proceedings of the 61st ACM/IEEE Design Automation Conference}, DAC '24, New York, NY, USA, 2024. Association for Computing Machinery.

\bibitem{7459483}
Matthew~M. Ziegler, Hung-Yi Liu, George Gristede, Bruce Owens, Ricardo Nigaglioni, and Luca~P. Carloni.
\newblock A synthesis-parameter tuning system for autonomous design-space exploration.
\newblock In {\em 2016 Design, Automation \& Test in Europe Conference \& Exhibition (DATE)}, pages 1148--1151, 2016.

\bibitem{geralla2018optimization}
Leo~E Geralla, MJ~Guzman, and Jefferson~A Hora.
\newblock Optimization of physically-aware synthesis for digital implementation flow.
\newblock {\em International Journal of Engineering \& Technology}, 7(2.11):31--34, 2018.

\bibitem{Clarke2011Eliminating}
Mike Clarke, Diego Hammerschlag, Matt Rardon, and Ankush Sood.
\newblock {Eliminating Routing Congestion Issues with Logic Synthesis}.
\newblock Technical Report 21025, Cadence Design Systems, Inc., December 2011.
\newblock Accessed: 2025-09-02.

\bibitem{1167596}
Junhyung Um, Jae hoon Kim, and Taewhan Kim.
\newblock Layout-driven resource sharing in high-level synthesis.
\newblock In {\em IEEE/ACM International Conference on Computer Aided Design, 2002. ICCAD 2002.}, pages 614--618, 2002.

\bibitem{7516905}
Koichi Fujiwara, Kazushi Kawamura, Masao Yanagisawa, and Nozomu Togawa.
\newblock Clock skew estimate modeling for fpga high-level synthesis and its application.
\newblock In {\em 2015 IEEE 11th International Conference on ASIC (ASICON)}, pages 1--4, 2015.

\bibitem{5733713}
Yunfeng~Wang Yunfeng~Wang, Jinian~Bian Jinian~Bian, Qiang~Wu Qiang~Wu, and Heng~Hu Heng~Hu.
\newblock Reallocation and rescheduling after floor-planning for timing optimization.
\newblock In {\em 2003 5th International Conference on ASIC Proceedings}, volume~1, pages 212--215 Vol.1, 2003.

\bibitem{10.1145/2744769.2744893}
Masato Tatsuoka, Ryosuke Watanabe, Tatsushi Otsuka, Takashi Hasegawa, Qiang Zhu, Ryosuke Okamura, Xingri Li, and Tsuyoshi Takabatake.
\newblock Physically aware high level synthesis design flow.
\newblock In {\em Proceedings of the 52nd Annual Design Automation Conference}, DAC '15, New York, NY, USA, 2015. Association for Computing Machinery.

\bibitem{4756913}
Junhua Wu, Chunmei Ma, and Baogui Huang.
\newblock Congestion aware high level synthesis combined with floorplanning.
\newblock In {\em 2008 IEEE Pacific-Asia Workshop on Computational Intelligence and Industrial Application}, volume~2, pages 935--938, 2008.

\bibitem{10.1145/337292.337769}
William~E. Dougherty and Donald~E. Thomas.
\newblock Unifying behavioral synthesis and physical design.
\newblock In {\em Proceedings of the 37th Annual Design Automation Conference}, DAC '00, page 756–761, New York, NY, USA, 2000. Association for Computing Machinery.

\bibitem{10.1145/2744769.2744801}
Ritchie Zhao, Mingxing Tan, Steve Dai, and Zhiru Zhang.
\newblock Area-efficient pipelining for fpga-targeted high-level synthesis.
\newblock In {\em Proceedings of the 52nd Annual Design Automation Conference}, DAC '15, New York, NY, USA, 2015. Association for Computing Machinery.

\bibitem{10.1145/2684746.2689063}
Mingxing Tan, Steve Dai, Udit Gupta, and Zhiru Zhang.
\newblock Mapping-aware constrained scheduling for lut-based fpgas.
\newblock In {\em Proceedings of the 2015 ACM/SIGDA International Symposium on Field-Programmable Gate Arrays}, FPGA '15, page 190–199, New York, NY, USA, 2015. Association for Computing Machinery.

\bibitem{742906}
S.~Tarafdar, M.~Leeser, and Zixin Yin.
\newblock Integrating floorplanning in data-transfer based high-level synthesis.
\newblock In {\em 1998 IEEE/ACM International Conference on Computer-Aided Design. Digest of Technical Papers (IEEE Cat. No.98CB36287)}, pages 412--417, 1998.

\bibitem{10.1109/ISVLSI.2006.8}
Vijay Sundaresan and Ranga Vemuri.
\newblock A novel approach to performance-oriented datapath allocation and floorplanning.
\newblock In {\em Proceedings of the IEEE Computer Society Annual Symposium on Emerging VLSI Technologies and Architectures}, ISVLSI '06, page 323, USA, 2006. IEEE Computer Society.

\bibitem{3697728}
Zirui Li, Kanglin Tian, Jianwang Zhai, Zixuan Li, Shixiong Kai, Siyuan Xu, Bei Yu, and Kang Zhao.
\newblock {\em FTAFP: A Feedthrough-Aware Floorplanner for Hierarchical Design of Large-Scale SoCs}, page 886–892.
\newblock Association for Computing Machinery, New York, NY, USA, 2025.

\bibitem{xu2025stepbeyondfeedthrough}
Zhexuan Xu, Jie Wang, Siyuan Xu, Zijie Geng, Mingxuan Yuan, and Feng Wu.
\newblock One step beyond: Feedthrough \& placement-aware rectilinear floorplanner, 2025.

\bibitem{hong2022feedthrough}
Y.~Hong, C.~Huang, Y.~Gao, and C.~Li.
\newblock Channel based soc feedthrough insertion methodology.
\newblock In {\em International Conference on Communications, Circuits and Systems (ICCCAS)}, pages 125--130, 2022.

\bibitem{sakariya2025decongesting}
Rajanikant Sakariya, Subhadeep Aich, Vivek Joshi, and Roger Griesmer.
\newblock Die area reduction by decongesting top channels using novel feedthrough insertion methodology in hierarchical soc designs.
\newblock In {\em Proceedings of the 2025 International Symposium on Quality Electronic Design (ISQED)}, pages 1--6, 2025.

\bibitem{srinath2021investigation}
B~Srinath, Rajesh Verma, Abdulwasa~Bakr Barnawi, Ramkumar Raja, Mohammed~Abdul Muqeet, Neeraj~Kumar Shukla, A~Ananthi Christy, C~Bharatiraja, and Josiah~Lange Munda.
\newblock An investigation of clock skew using a wirelength-aware floorplanning process in the pre-placement stages of msv layouts.
\newblock {\em Electronics}, 10(22):2795, 2021.

\bibitem{1466067}
Chih-Hung Lee, Chin-Hung Su, Shih-Hsu Huang, Chih-Yuan Lin, and Tsai-Ming Hsieh.
\newblock Floorplanning with clock tree estimation.
\newblock In {\em 2005 IEEE International Symposium on Circuits and Systems (ISCAS)}, pages 6244--6247 Vol. 6, 2005.

\bibitem{6498346}
Ang~Boon Chong.
\newblock Asic clock tree estimation in design planning.
\newblock In {\em 2013 4th International Conference on Intelligent Systems, Modelling and Simulation}, pages 619--626, 2013.

\bibitem{yim1999floorplan}
Joon-Seo Yim, Seong-Ok Bae, and Chong-Min Kyung.
\newblock A floorplan-based planning methodology for power and clock distribution in asics.
\newblock In {\em Proceedings of the 36th Annual Design Automation Conference (DAC)}, pages 504--509, 1999.

\bibitem{donno2004power}
Monica Donno, Enrico Macii, and Luca Mazzoni.
\newblock Power-aware clock tree planning.
\newblock In {\em Proceedings of the 2004 international symposium on Physical design}, pages 138--147, 2004.

\bibitem{butt2007system}
Saif~Ali Butt, Stefan Schmermbeck, Jurij Rosenthal, Alexander Pratsch, and Eike Schmidt.
\newblock System level clock tree synthesis for power optimization.
\newblock In {\em Proceedings of the Design, Automation and Test in Europe Conference (DATE)}, pages 1--6, 2007.

\bibitem{seo2015spine}
Hyungjung Seo, Juyeon Kim, Minseok Kang, and Taewhan Kim.
\newblock Synthesis for power-aware clock spines.
\newblock In {\em Proceedings of the IEEE/ACM International Conference on Computer-Aided Design (ICCAD)}, pages 576--583, 2015.

\bibitem{kim2017spine}
Youngchan Kim and Taewhan Kim.
\newblock Algorithm for synthesis and exploration of clock spines.
\newblock In {\em Proceedings of the Asia and South Pacific Design Automation Conference (ASP-DAC)}, pages 219--224, 2017.

\bibitem{chu1998matrix}
Chris C.~N. Chu and D.~F. Wong.
\newblock A matrix synthesis approach to thermal placement.
\newblock {\em IEEE Transactions on Computer-Aided Design of Integrated Circuits and Systems}, 17(11):1166--1174, 1998.

\bibitem{cong2004thermal}
Jason Cong, Jie Wei, and Yan Zhang.
\newblock A thermal-driven floorplanning algorithm for 3d ics.
\newblock In {\em Proceedings of the IEEE/ACM International Conference on Computer-Aided Design (ICCAD)}, pages 306--313, 2004.

\bibitem{jang2014thermal}
Cheoljon Jang and Jongwha Chong.
\newblock Thermal-aware floorplanning with min-cut die partition for 3d ics.
\newblock {\em ETRI Journal}, 36(4):517--527, 2014.

\bibitem{molter2023thermal}
Michael Molter, Rahul Kumar, Sonja Koller, Osama~Waqar Bhatti, Nikita Ambasana, Elyse Rosenbaum, and Madhavan Swaminathan.
\newblock Thermal-aware soc macro placement and multi-chip module design optimization with bayesian optimization.
\newblock In {\em 2023 IEEE 73rd Electronic Components and Technology Conference (ECTC)}, pages 935--942. IEEE, 2023.

\bibitem{guan2023thermal}
Wenbo Guan, Xiaoyan Tang, Hongliang Lu, Yuming Zhang, and Yimen Zhang.
\newblock Thermal-aware fixed-outline 3-d ic floorplanning: An end-to-end learning-based approach.
\newblock {\em IEEE Transactions on Very Large Scale Integration (VLSI) Systems}, 2023.

\bibitem{wang2024atplace2}
Qipan Wang, Xueqing Li, Tianyu Jia, Yibo Lin, Runsheng Wang, and Ru~Huang.
\newblock Atplace2. 5d: Analytical thermal-aware chiplet placement framework for large-scale 2.5 d-ic.
\newblock In {\em Proceedings of the 43rd IEEE/ACM International Conference on Computer-Aided Design}, pages 1--9, 2024.

\bibitem{lee2006voltage}
Wan-Ping Lee, Hung-Yi Liu, and Yao-Wen Chang.
\newblock Voltage island aware floorplanning for power and timing optimization.
\newblock In {\em Proceedings of the IEEE/ACM International Conference on Computer-Aided Design (ICCAD)}, pages 389--394, 2006.

\bibitem{mohamood2007noise}
Fayez Mohamood, Michael~B. Healy, Sung~Kyu Lim, and Hsien-Hsin~S. Lee.
\newblock Noise-direct: A technique for power supply noise aware floorplanning using microarchitecture profiling.
\newblock In {\em Proceedings of the Asia and South Pacific Design Automation Conference (ASP-DAC)}, pages 786--791, 2007.

\bibitem{liu2001power}
I-Min Liu, Hung-Ming Chen, Tan-Li Chou, Adnan Aziz, and D.~F. Wong.
\newblock Integrated power supply planning and floorplanning.
\newblock In {\em Proceedings of the Asia and South Pacific Design Automation Conference (ASP-DAC)}, pages 151--156, 2001.

\bibitem{chen2005ir}
Jian Chen, Chonghong Zhao, Dian Zhou, and Xiaofang Zhou.
\newblock Floorplanning with ir-drop consideration.
\newblock In {\em Proceedings of the WSEAS/IASME International Conference on Computer Engineering and Applications}, pages 203--208, 2005.

\bibitem{li2009guided}
Li~Li, Yuchun Ma, Ning Xu, Yu~Wang, and Xianlong Hong.
\newblock Floorplan and power/ground network co-design using guided incremental floorplanning.
\newblock In {\em IEEE Asia Pacific Conference on ASIC}, pages 177--180, 2009.

\bibitem{zhao2002noise}
Shiyou Zhao, Kaushik Roy, and Cheng-Kok Koh.
\newblock Power supply noise aware floorplanning and decoupling capacitance placement.
\newblock In {\em Proceedings of the Asia and South Pacific Design Automation Conference (ASP-DAC)}, pages 123--128, 2002.

\bibitem{falkenstern2010pg}
Paul Falkenstern, Yuan Xie, Yao-Wen Chang, and Yu~Wang.
\newblock Three-dimensional ic floorplan and power/ground network co-synthesis.
\newblock In {\em Proceedings of the Asia and South Pacific Design Automation Conference (ASP-DAC)}, pages 169--174, 2010.

\bibitem{chu2013multi}
Zhufei Chu, Yinshui Xia, Lunyao Wang, and Jian Wang.
\newblock Voltage drop aware power pad assignment and floorplanning for multi-voltage soc designs.
\newblock In {\em International Conference on Computer-Aided Design and Computer Graphics (CAD/Graphics)}, pages 8--13, 2013.

\bibitem{basha2022pg}
S.~Mahaboob Basha et~al.
\newblock P/g pin position-aware voltage island floorplanning for ir drop security.
\newblock {\em Analog Integrated Circuits and Signal Processing}, 111(3):441--450, 2022.

\bibitem{zhang2012reclaiming}
Y.~Zhang, A.~Chakraborty, S.~Chowdhury, and D.~Z. Pan.
\newblock Reclaiming over-the-ip-block routing resources with buffering-aware rectilinear steiner minimum tree construction.
\newblock In {\em IEEE/ACM International Conference on Computer-Aided Design (ICCAD)}, pages 137--143, 2012.

\bibitem{sankaranarayanan2005case}
Karthik Sankaranarayanan, Sivakumar Velusamy, Mircea Stan, Kevin Skadron, et~al.
\newblock A case for thermal-aware floorplanning at the microarchitectural level.
\newblock {\em Journal of Instruction-Level Parallelism}, 7(1):8--16, 2005.

\bibitem{winther2015thermal}
Andreas~Thor Winther, Wei Liu, Alberto Nannarelli, and Sarma Vrudhula.
\newblock Thermal aware floorplanning incorporating temperature dependent wire delay estimation.
\newblock {\em Microprocessors and Microsystems}, 39(8):807--815, 2015.

\bibitem{cadence2019celsius}
{Cadence}.
\newblock Cadence launches celsius thermal solver.
\newblock \url{https://www.cadence.com/en_US/home/tools/system-analysis/thermal-solutions/celsius-thermal-solver.html}, 2019.

\bibitem{nature_placement}
Azalia Mirhoseini, Anna Goldie, Mustafa Yazgan, Joe~Wenjie Jiang, Ebrahim Songhori, Shen Wang, Young-Joon Lee, Eric Johnson, Omkar Pathak, Azade Nova, et~al.
\newblock A graph placement methodology for fast chip design.
\newblock {\em Nature}, 594(7862):207--212, 2021.

\bibitem{10.1145/3505170.3506731}
Andrew~B. Kahng, Ravi Varadarajan, and Zhiang Wang.
\newblock Rtl-mp: Toward practical, human-quality chip planning and macro placement.
\newblock In {\em Proceedings of the 2022 International Symposium on Physical Design}, ISPD '22, page 3–11, New York, NY, USA, 2022. Association for Computing Machinery.

\bibitem{10372220}
Andrew~B. Kahng, Ravi Varadarajan, and Zhiang Wang.
\newblock Hier-rtlmp: A hierarchical automatic macro placer for large-scale complex ip blocks.
\newblock {\em IEEE Transactions on Computer-Aided Design of Integrated Circuits and Systems}, 43(5):1552--1565, 2024.

\bibitem{zhao2024standard}
Xiaotian Zhao, Tianju Wang, Run Jiao, and Xinfei Guo.
\newblock Standard cells do matter: Uncovering hidden connections for high-quality macro placement.
\newblock In {\em 2024 Design, Automation \& Test in Europe Conference \& Exhibition (DATE)}, pages 1--6. IEEE, 2024.

\bibitem{zhao2025incredflip}
Xiaotian Zhao, Jiayin Chen, Zixuan Li, Yichen Cai, and Xinfei Guo.
\newblock Incredflip: Incremental dataflow-driven macro flipping for efficient macro placement refinement.
\newblock In {\em 2025 International Symposium of Electronics Design Automation (ISEDA)}, pages 311--316. IEEE, 2025.

\bibitem{zhao2025mp}
Xiaotian Zhao, Zixuan Li, Yichen Cai, Tianju Wang, Yushan Pan, and Xinfei Guo.
\newblock Das-mp: Enabling high-quality macro placement with enhanced dataflow awareness.
\newblock {\em arXiv preprint arXiv:2505.16445}, 2025.

\bibitem{pu2024incremacro}
Yuan Pu, Tinghuan Chen, Zhuolun He, Chen Bai, Haisheng Zheng, Yibo Lin, and Bei Yu.
\newblock Incremacro: Incremental macro placement refinement.
\newblock In {\em Proceedings of the 2024 International Symposium on Physical Design}, pages 169--176, 2024.

\bibitem{liao2022dreamplace}
Peiyu Liao, Siting Liu, Zhitang Chen, Wenlong Lv, Yibo Lin, and Bei Yu.
\newblock Dreamplace 4.0: Timing-driven global placement with momentum-based net weighting.
\newblock In {\em 2022 Design, Automation \& Test in Europe Conference \& Exhibition (DATE)}, pages 939--944. IEEE, 2022.

\bibitem{lin2024timing}
Jai-Ming Lin, You-Yu Chang, and Wei-Lun Huang.
\newblock Timing-driven analytical placement according to expected cell distribution range.
\newblock In {\em Proceedings of the 2024 International Symposium on Physical Design}, pages 177--184, 2024.

\bibitem{shi2025timing}
Yunqi Shi, Siyuan Xu, Shixiong Kai, Xi~Lin, Ke~Xue, Mingxuan Yuan, and Chao Qian.
\newblock Timing-driven global placement by efficient critical path extraction.
\newblock In {\em 2025 Design, Automation \& Test in Europe Conference (DATE)}, pages 1--7. IEEE, 2025.

\bibitem{fu2024hybrid}
Bangqi Fu, Lixin Liu, Martin~DF Wong, and Evangeline~FY Young.
\newblock Hybrid modeling and weighting for timing-driven placement with efficient calibration.
\newblock In {\em Proceedings of the 43rd IEEE/ACM International Conference on Computer-Aided Design}, pages 1--9, 2024.

\bibitem{hamada1993prime}
Takeo Hamada, Chung-Kuan Cheng, and Paul~M Chau.
\newblock Prime: A timing-driven placement tool using a piecewise linear resistive network approach.
\newblock In {\em Proceedings of the 30th international Design Automation Conference}, pages 531--536, 1993.

\bibitem{swartz1995timing}
William Swartz and Carl Sechen.
\newblock Timing driven placement for large standard cell circuits.
\newblock In {\em Proceedings of the 32nd Annual ACM/IEEE Design Automation Conference}, pages 211--215, 1995.

\bibitem{guo2022differentiable}
Zizheng Guo and Yibo Lin.
\newblock Differentiable-timing-driven global placement.
\newblock In {\em Proceedings of the 59th ACM/IEEE Design Automation Conference}, pages 1315--1320, 2022.

\bibitem{he2011ripple}
Xu~He, Tao Huang, Linfu Xiao, Haitong Tian, Guxin Cui, and Evangeline~FY Young.
\newblock Ripple: An effective routability-driven placer by iterative cell movement.
\newblock In {\em 2011 IEEE/ACM International Conference on Computer-Aided Design (ICCAD)}, pages 74--79. IEEE, 2011.

\bibitem{hsu2011routability}
Meng-Kai Hsu, Sheng Chou, Tzu-Hen Lin, and Yao-Wen Chang.
\newblock Routability-driven analytical placement for mixed-size circuit designs.
\newblock In {\em 2011 IEEE/ACM International Conference on Computer-Aided Design (ICCAD)}, pages 80--84. IEEE, 2011.

\bibitem{kim2011simplr}
Myung-Chul Kim, Jin Hu, Dong-Jin Lee, and Igor~L Markov.
\newblock A simplr method for routability-driven placement.
\newblock In {\em 2011 IEEE/ACM International Conference on Computer-Aided Design (ICCAD)}, pages 67--73. IEEE, 2011.

\bibitem{cheng2018replace}
Chung-Kuan Cheng, Andrew~B Kahng, Ilgweon Kang, and Lutong Wang.
\newblock Replace: Advancing solution quality and routability validation in global placement.
\newblock {\em IEEE Transactions on Computer-Aided Design of Integrated Circuits and Systems}, 38(9):1717--1730, 2018.

\bibitem{hsu2014ntuplace4h}
Meng-Kai Hsu, Yi-Fang Chen, Chau-Chin Huang, Sheng Chou, Tzu-Hen Lin, Tung-Chieh Chen, and Yao-Wen Chang.
\newblock Ntuplace4h: A novel routability-driven placement algorithm for hierarchical mixed-size circuit designs.
\newblock {\em IEEE Transactions on Computer-Aided Design of Integrated Circuits and Systems}, 33(12):1914--1927, 2014.

\bibitem{xie2018routenet}
Zhiyao Xie, Yu-Hung Huang, Guan-Qi Fang, Haoxing Ren, Shao-Yun Fang, Yiran Chen, and Jiang Hu.
\newblock Routenet: Routability prediction for mixed-size designs using convolutional neural network.
\newblock In {\em 2018 IEEE/ACM International Conference on Computer-Aided Design (ICCAD)}, pages 1--8. IEEE, 2018.

\bibitem{baek2022pin}
Kyeonghyeon Baek, Hyunbum Park, Suwan Kim, Kyumyung Choi, and Taewhan Kim.
\newblock Pin accessibility and routing congestion aware drc hotspot prediction using graph neural network and u-net.
\newblock In {\em Proceedings of the 41st IEEE/ACM International Conference on Computer-Aided Design}, pages 1--9, 2022.

\bibitem{islam2025pgr}
Riadul Islam and Dhandeep Challagundla.
\newblock Pre-global routing drc violation prediction using unsupervised learning.
\newblock In {\em 2025 23rd IEEE Interregional NEWCAS Conference (NEWCAS)}, pages 450--454. IEEE, 2025.

\bibitem{huang2017ntuplace4dr}
Chau-Chin Huang, Hsin-Ying Lee, Bo-Qiao Lin, Sheng-Wei Yang, Chin-Hao Chang, Szu-To Chen, Yao-Wen Chang, Tung-Chieh Chen, and Ismail Bustany.
\newblock Ntuplace4dr: A detailed-routing-driven placer for mixed-size circuit designs with technology and region constraints.
\newblock {\em IEEE Transactions on Computer-Aided Design of Integrated Circuits and Systems}, 37(3):669--681, 2017.

\bibitem{liu2022xplace}
Lixin Liu, Bangqi Fu, Martin~DF Wong, and Evangeline~FY Young.
\newblock Xplace: An extremely fast and extensible global placement framework.
\newblock In {\em Proceedings of the 59th ACM/IEEE Design Automation Conference}, pages 1309--1314, 2022.

\bibitem{lee2011obstacle}
Dong-Jin Lee and Igor~L Markov.
\newblock Obstacle-aware clock-tree shaping during placement.
\newblock In {\em Proceedings of the 2011 international symposium on Physical design}, pages 123--130, 2011.

\bibitem{lu2015eplace}
Jingwei Lu, Pengwen Chen, Chin-Chih Chang, Lu~Sha, Dennis Jen-Hsin Huang, Chin-Chi Teng, and Chung-Kuan Cheng.
\newblock eplace: Electrostatics-based placement using fast fourier transform and nesterov's method.
\newblock {\em ACM Transactions on Design Automation of Electronic Systems (TODAES)}, 20(2):1--34, 2015.

\bibitem{ding2023clock}
Jinghao Ding, Linhao Lu, Zhaoqi Fu, Jie Ma, Mengshi Gong, Yuanrui Qi, and Wenxin Yu.
\newblock Clock aware low power placement.
\newblock In {\em 2023 IEEE/ACM International Conference on Computer Aided Design (ICCAD)}, pages 01--08. IEEE, 2023.

\bibitem{kim2011simpl}
Myung-Chul Kim, Dong-Jin Lee, and Igor~L Markov.
\newblock Simpl: An effective placement algorithm.
\newblock {\em IEEE Transactions on Computer-Aided Design of Integrated Circuits and Systems}, 31(1):50--60, 2011.

\bibitem{siemens_drc_placement}
Siemens EDA.
\newblock Shifting left in p\&r with in-design signoff fill for faster and more accurate tapeouts.
\newblock Technical report, Siemens, 2021.

\bibitem{iccad2015contest}
Myung-Chul Kim, Jin Hu, Jiajia Li, and Natarajan Viswanathan.
\newblock Iccad-2015 cad contest in incremental timing-driven placement and benchmark suite.
\newblock In {\em 2015 IEEE/ACM International Conference on Computer-Aided Design (ICCAD)}, pages 921--926. IEEE, 2015.

\bibitem{ma2007micro}
Yuchun Ma, Zhuoyuan Li, Jason Cong, Xianlong Hong, Glenn Reinman, Sheqin Dong, and Qiang Zhou.
\newblock Micro-architecture pipelining optimization with throughput-aware floorplanning.
\newblock In {\em 2007 Asia and South Pacific Design Automation Conference}, pages 920--925. IEEE, 2007.

\bibitem{10.1145/1065579.1065731}
Vidyasagar Nookala, Ying Chen, David~J. Lilja, and Sachin~S. Sapatnekar.
\newblock Microarchitecture-aware floorplanning using a statistical design of experiments approach.
\newblock In {\em Proceedings of the 42nd Annual Design Automation Conference}, DAC '05, page 579–584, New York, NY, USA, 2005. Association for Computing Machinery.

\bibitem{9360844}
Jai-Ming Lin, You-Lun Deng, Ya-Chu Yang, Jia-Jian Chen, and Po-Chen Lu.
\newblock Dataflow-aware macro placement based on simulated evolution algorithm for mixed-size designs.
\newblock {\em IEEE Transactions on Very Large Scale Integration (VLSI) Systems}, 29(5):973--984, 2021.

\bibitem{9309318}
Alex Vidal-Obiols, Jordi Cortadella, Jordi Petit, Marc Galceran-Oms, and Ferran Martorell.
\newblock Multilevel dataflow-driven macro placement guided by rtl structure and analytical methods.
\newblock {\em IEEE Transactions on Computer-Aided Design of Integrated Circuits and Systems}, 40(12):2542--2555, 2021.

\bibitem{vidal2019rtl}
Alex Vidal-Obiols, Jordi Cortadella, Jordi Petit, Marc Galceran-Oms, and Ferran Martorell.
\newblock Rtl-aware dataflow-driven macro placement.
\newblock In {\em 2019 Design, Automation \& Test in Europe Conference \& Exhibition (DATE)}, pages 186--191. IEEE, 2019.

\bibitem{lin2019dreamplace}
Yibo Lin, Shounak Dhar, Wuxi Li, Haoxing Ren, Brucek Khailany, and David~Z Pan.
\newblock Dreamplace: Deep learning toolkit-enabled gpu acceleration for modern vlsi placement.
\newblock In {\em Proceedings of the 56th Annual Design Automation Conference 2019}, pages 1--6, 2019.

\bibitem{kong2002novel}
Tim Kong.
\newblock A novel net weighting algorithm for timing-driven placement.
\newblock In {\em Proceedings of the 2002 IEEE/ACM international conference on Computer-aided design}, pages 172--176, 2002.

\bibitem{chang2002net}
H~Chang, Eugene Shragowitz, Jian Liu, Habib Youssef, Bing Lu, and Suphachai Sutanthavibul.
\newblock Net criticality revisited: An effective method to improve timing in physical design.
\newblock In {\em Proceedings of the 2002 international symposium on Physical design}, pages 155--160, 2002.

\bibitem{huang2015opentimer}
Tsung-Wei Huang and Martin~DF Wong.
\newblock Opentimer: A high-performance timing analysis tool.
\newblock In {\em 2015 IEEE/ACM International Conference on Computer-Aided Design (ICCAD)}, pages 895--902. IEEE, 2015.

\bibitem{chow2016placement}
Wing-Kai Chow and Evangeline~FY Young.
\newblock Placement: from wirelength to detailed routability.
\newblock {\em IPSJ Transactions on System and LSI Design Methodology}, 9:2--12, 2016.

\bibitem{wang2000congestion}
Maogang Wang, Xiaojian Yang, and Majid Sarrafzadeh.
\newblock Congestion minimization during placement.
\newblock {\em IEEE Transactions on Computer-Aided Design of Integrated Circuits and Systems}, 19(10):1140--1148, 2000.

\bibitem{viswanathan2011ispd}
Natarajan Viswanathan, Charles~J Alpert, Cliff Sze, Zhuo Li, Gi-Joon Nam, and Jarrod~A Roy.
\newblock The ispd-2011 routability-driven placement contest and benchmark suite.
\newblock In {\em Proceedings of the 2011 international symposium on Physical design}, pages 141--146, 2011.

\bibitem{viswanathan2012iccad}
Natarajan Viswanathan, Charles Alpert, Cliff Sze, Zhuo Li, and Yaoguang Wei.
\newblock Iccad-2012 cad contest in design hierarchy aware routability-driven placement and benchmark suite.
\newblock In {\em Proceedings of the International Conference on Computer-Aided Design}, pages 345--348, 2012.

\bibitem{viswanathan2012dac}
Natarajan Viswanathan, Charles Alpert, Cliff Sze, Zhuo Li, and Yaoguang Wei.
\newblock The dac 2012 routability-driven placement contest and benchmark suite.
\newblock In {\em Proceedings of the 49th Annual Design Automation Conference}, pages 774--782, 2012.

\bibitem{yutsis2014ispd}
Vladimir Yutsis, Ismail~S Bustany, David Chinnery, Joseph~R Shinnerl, and Wen-Hao Liu.
\newblock Ispd 2014 benchmarks with sub-45nm technology rules for detailed-routing-driven placement.
\newblock In {\em Proceedings of the 2014 on International symposium on physical design}, pages 161--168, 2014.

\bibitem{bustany2015ispd}
Ismail~S Bustany, David Chinnery, Joseph~R Shinnerl, and Vladimir Yutsis.
\newblock Ispd 2015 benchmarks with fence regions and routing blockages for detailed-routing-driven placement.
\newblock In {\em Proceedings of the 2015 Symposium on International Symposium on Physical Design}, pages 157--164, 2015.

\bibitem{liu2013optimization}
Wen-Hao Liu, Cheng-Kok Koh, and Yih-Lang Li.
\newblock Optimization of placement solutions for routability.
\newblock In {\em Proceedings of the 50th Annual Design Automation Conference}, pages 1--9, 2013.

\bibitem{li2014analytical}
Shuai Li and Cheng-Kok Koh.
\newblock Analytical placement of mixed-size circuits for better detailed-routability.
\newblock In {\em 2014 19th Asia and South Pacific Design Automation Conference (ASP-DAC)}, pages 41--46. IEEE, 2014.

\bibitem{liu2013nctu}
Wen-Hao Liu, Wei-Chun Kao, Yih-Lang Li, and Kai-Yuan Chao.
\newblock Nctu-gr 2.0: Multithreaded collision-aware global routing with bounded-length maze routing.
\newblock {\em IEEE Transactions on computer-aided design of integrated circuits and systems}, 32(5):709--722, 2013.

\bibitem{alpert2010makes}
Charles~J Alpert, Zhuo Li, Michael~D Moffitt, Gi-Joon Nam, Jarrod~A Roy, and Gustavo Tellez.
\newblock What makes a design difficult to route.
\newblock In {\em Proceedings of the 19th international symposium on Physical design}, pages 7--12, 2010.

\bibitem{liu2013case}
Wen-Hao Liu, Cheng-Kok Koh, and Yih-Lang Li.
\newblock Case study for placement solutions in ispd11 and dac12 routability-driven placement contests.
\newblock In {\em Proceedings of the 2013 ACM International symposium on Physical Design}, pages 114--119, 2013.

\bibitem{liu2021global}
Siting Liu, Qi~Sun, Peiyu Liao, Yibo Lin, and Bei Yu.
\newblock Global placement with deep learning-enabled explicit routability optimization.
\newblock In {\em 2021 Design, Automation \& Test in Europe Conference \& Exhibition (DATE)}, pages 1821--1824. IEEE, 2021.

\bibitem{lin2025early}
Jingui Lin, Shiyan Liang, Wenxiong Lin, Peng Gao, Yan Xing, Tingting Wu, Xiaoming Xiong, and Shuting Cai.
\newblock Early stage drc hotspot prediction for mixed-size designs through an efficient graph-based deep learning.
\newblock {\em ACM Transactions on Design Automation of Electronic Systems}, 30(4):1--21, 2025.

\bibitem{huang2015detailed}
Chau-Chin Huang, Chien-Hsiung Chiou, Kai-Han Tseng, and Yao-Wen Chang.
\newblock Detailed-routing-driven analytical standard-cell placement.
\newblock In {\em The 20th Asia and South Pacific Design Automation Conference}, pages 378--383. IEEE, 2015.

\bibitem{dac25congestion}
Wenchao Li, Hongxi Wu, Xingquan Li, and Wenxing Zhu.
\newblock Differentiable net-moving and local congestion mitigationfor routability-driven global placement.
\newblock In {\em Proceedings of the 62th Annual Design Automation Conference 2025}, 2025.

\bibitem{rabaey2002digital}
Jan~M Rabaey, Anantha Chandrakasan, and Borivoje Nikolic.
\newblock {\em Digital integrated circuits}, volume~2.
\newblock Prentice hall Englewood Cliffs, 2002.

\bibitem{alpert2008handbook}
Charles~J Alpert, Dinesh~P Mehta, and Sachin~S Sapatnekar.
\newblock {\em Handbook of algorithms for physical design automation}.
\newblock CRC press, 2008.

\bibitem{choi2024bti}
Mujun Choi, Deokkeun Oh, and Juho Kim.
\newblock Bti tolerant clock tree synthesis using lp-based supply voltage alignment.
\newblock {\em JOURNAL OF SEMICONDUCTOR TECHNOLOGY AND SCIENCE}, 24(5):424--439, 2024.

\bibitem{oh2020symmetrical}
Deokkeun Oh, Mujun Choi, and Juho Kim.
\newblock Symmetrical buffered clock tree synthesis considering nbti.
\newblock In {\em 2020 IEEE International Symposium on Circuits and Systems (ISCAS)}, pages 1--5. IEEE, 2020.

\bibitem{chen2013novel}
Jifeng Chen and Mohammad Tehranipoor.
\newblock A novel flow for reducing clock skew considering nbti effect and process variations.
\newblock In {\em International Symposium on Quality Electronic Design (ISQED)}, pages 327--334. IEEE, 2013.

\bibitem{chakraborty2010skew}
Ashutosh Chakraborty and David~Z Pan.
\newblock Skew management of nbti impacted gated clock trees.
\newblock In {\em Proceedings of the 19th international symposium on Physical design}, pages 127--133, 2010.

\bibitem{huang2013low}
Shih-Hsu Huang, Wen-Pin Tu, Chia-Ming Chang, and Song-Bin Pan.
\newblock Low-power anti-aging zero skew clock gating.
\newblock {\em ACM Transactions on Design Automation of Electronic Systems (TODAES)}, 18(2):1--37, 2013.

\bibitem{huang2010critical}
Shih-Hsu Huang, Chia-Ming Chang, Wen-Pin Tu, and Song-Bin Pan.
\newblock Critical-pmos-aware clock tree design methodology for anti-aging zero skew clock gating.
\newblock In {\em 2010 15th Asia and South Pacific Design Automation Conference (ASP-DAC)}, pages 480--485. IEEE, 2010.

\bibitem{lin2014buffered}
Chung-Wei Lin, Tzu-Hsuan Hsu, Xin-Wei Shih, and Yao-Wen Chang.
\newblock Buffered clock tree synthesis considering self-heating effects.
\newblock In {\em Proceedings of the 2014 international symposium on Low power electronics and design}, pages 111--116, 2014.

\bibitem{vishnu2019clock}
Priya~V Vishnu, AR~Priyarenjini, and Naveen Kotha.
\newblock Clock tree synthesis techniques for optimal power and timing convergence in soc partitions.
\newblock In {\em 2019 4th International Conference on Recent Trends on Electronics, Information, Communication \& Technology (RTEICT)}, pages 276--280. IEEE, 2019.

\bibitem{patel2013novel}
Nikhil Patel.
\newblock A novel clock distribution technology-multisource clock tree system (mcts).
\newblock {\em International Journal of Advanced Research in Electrical, Electronics and Instrumentation Engineering}, 2(6), 2013.

\bibitem{mohammadi2011hybrid}
Zohre Mohammadi-Arfa and Ali Jahanian.
\newblock A hybrid rf/metal clock routing algorithm to improve clock delay and routing congestion.
\newblock In {\em 2011 IEEE Computer Society Annual Symposium on VLSI}, pages 138--143. IEEE, 2011.

\bibitem{farhangi2010pattern}
Ali~M Farhangi, Asim~J Al-Khalili, and Dhamin Al-Khalili.
\newblock Pattern-driven clock tree routing with via minimization.
\newblock In {\em 2010 IEEE Computer Society Annual Symposium on VLSI}, pages 216--221. IEEE, 2010.

\bibitem{lu2011fast}
Jingwei Lu, Wing-Kai Chow, and Chiu-Wing Sham.
\newblock Fast power-and slew-aware gated clock tree synthesis.
\newblock {\em IEEE Transactions on very large scale integration (VLSI) Systems}, 20(11):2094--2103, 2011.

\bibitem{oh2019thermal}
Deok~Keun Oh, Mu~Jun Choi, and Ju~Ho Kim.
\newblock Thermal-aware 3d symmetrical buffered clock tree synthesis.
\newblock {\em ACM Transactions on Design Automation of Electronic Systems (TODAES)}, 24(3):1--22, 2019.

\bibitem{rajaram2011robust}
Anand Rajaram and David~Z Pan.
\newblock Robust chip-level clock tree synthesis.
\newblock {\em IEEE transactions on computer-aided design of integrated circuits and systems}, 30(6):877--890, 2011.

\bibitem{lu2003process}
Bing Lu, Jiang Hu, Gary Ellis, and Haihua Su.
\newblock Process variation aware clock tree routing.
\newblock In {\em Proceedings of the 2003 international symposium on Physical design}, pages 174--181, 2003.

\bibitem{sivaswamy2008statistical}
Satish Sivaswamy and Kia Bazargan.
\newblock Statistical analysis and process variation-aware routing and skew assignment for fpgas.
\newblock {\em ACM Transactions on Reconfigurable Technology and Systems (TRETS)}, 1(1):1--35, 2008.

\bibitem{chan2014ocv}
Tuck-Boon Chan, Kwangsoo Han, Andrew~B Kahng, Jae-Gon Lee, and Siddhartha Nath.
\newblock Ocv-aware top-level clock tree optimization.
\newblock In {\em Proceedings of the 24th edition of the great lakes symposium on VLSI}, pages 33--38, 2014.

\bibitem{10.1145/1960397.1960407}
Tarun Mittal and Cheng-Kok Koh.
\newblock Cross link insertion for improving tolerance to variations in clock network synthesis.
\newblock In {\em Proceedings of the 2011 International Symposium on Physical Design}, ISPD '11, page 29–36, New York, NY, USA, 2011. Association for Computing Machinery.

\bibitem{yang2011robust}
Jae-Seok Yang, Jiwoo Pak, Xin Zhao, Sung~Kyu Lim, and David~Z Pan.
\newblock Robust clock tree synthesis with timing yield optimization for 3d-ics.
\newblock In {\em 16th Asia and South Pacific Design Automation Conference (ASP-DAC 2011)}, pages 621--626. IEEE, 2011.

\bibitem{singh2023reliability}
Karan Singh and Shruti Kalra.
\newblock Reliability forecasting and accelerated lifetime testing in advanced cmos technologies.
\newblock {\em Microelectronics Reliability}, 151:115261, 2023.

\bibitem{amrouch2016reliability}
Hussam Amrouch, Behnam Khaleghi, Andreas Gerstlauer, and J{\"o}rg Henkel.
\newblock Reliability-aware design to suppress aging.
\newblock In {\em Proceedings of the 53rd Annual Design Automation Conference}, pages 1--6, 2016.

\bibitem{guo2017towards}
Shaofeng Guo, Runsheng Wang, Zhuoqing Yu, Peng Hao, Pengpeng Ren, Yangyuan Wang, Siyu Liao, Chunyi Huang, Tianlei Guo, Alvin Chen, et~al.
\newblock Towards reliability-aware circuit design in nanoscale finfet technology:—new-generation aging model and circuit reliability simulator.
\newblock In {\em 2017 IEEE/ACM International Conference on Computer-Aided Design (ICCAD)}, pages 780--785. IEEE, 2017.

\bibitem{ho1989electromigration}
Paul~S Ho and Thomas Kwok.
\newblock Electromigration in metals.
\newblock {\em Reports on Progress in Physics}, 52(3):301, 1989.

\bibitem{lienig2018fundamentals}
Jens Lienig and Matthias Thiele.
\newblock Fundamentals of electromigration.
\newblock In {\em Fundamentals of electromigration-aware integrated circuit design}, pages 13--60. Springer, 2018.

\bibitem{kim2018systematic}
Hyunjin Kim, Minjung Jin, Hyunchul Sagong, Jinju Kim, Ukjin Jung, Minhyuck Choi, Junekyun Park, Sangchul Shin, and Sangwoo Pae.
\newblock A systematic study of gate dielectric tddb in finfet technology.
\newblock In {\em 2018 IEEE International Reliability Physics Symposium (IRPS)}, pages 4A--4. IEEE, 2018.

\bibitem{ramadan2024impact}
Firas Ramadan, Majd Ganaeim, Maayan Ella, and Freddy Gabbay.
\newblock The impact of asymmetric transistor aging on clock tree design considerations.
\newblock {\em IEEE Access}, 2024.

\bibitem{chakraborty2009analysis}
Ashutosh Chakraborty, Gokul Ganesan, Anand Rajaram, and David~Z Pan.
\newblock Analysis and optimization of nbti induced clock skew in gated clock trees.
\newblock In {\em 2009 Design, Automation \& Test in Europe Conference \& Exhibition}, pages 296--299. IEEE, 2009.

\bibitem{lu2015electromigration}
Tiantao Lu and Ankur Srivastava.
\newblock Electromigration-aware clock tree synthesis for tsv-based 3d-ics.
\newblock In {\em Proceedings of the 25th edition on Great Lakes Symposium on VLSI}, pages 27--32, 2015.

\bibitem{westra2009congestion}
Hylke Jurjen~Lijsbert Westra.
\newblock Congestion analysis and management.
\newblock 2009.

\bibitem{9215244}
Sajja~Krishna Kishore, Tulasi~Radhika Patnala, Arun~S Tigadi, and Aatif Jamshed.
\newblock An on-chip analysis of the vlsi designs under process variations.
\newblock In {\em 2020 International Conference on Smart Electronics and Communication (ICOSEC)}, pages 1273--1277, 2020.

\bibitem{velenis2003reduced}
Dimitrios Velenis, Marios~C Papaefthymiou, and Eby~G Friedman.
\newblock Reduced delay uncertainty in high performance clock distribution networks.
\newblock In {\em 2003 Design, Automation and Test in Europe Conference and Exhibition}, pages 68--73. IEEE, 2003.

\bibitem{tseng2014power}
Tsu-Wei Tseng, Chang-Tzu Lin, Chia-Hsin Lee, Yung-Fa Chou, and Ding-Ming Kwai.
\newblock A power delivery network (pdn) engineering change order (eco) approach for repairing ir-drop failures after the routing stage.
\newblock In {\em Technical Papers of 2014 International Symposium on VLSI Design, Automation and Test}, pages 1--4. IEEE, 2014.

\bibitem{cheng2021core}
Chung-Kuan Cheng, Andrew~B Kahng, Ilgweon Kang, Minsoo Kim, Daeyeal Lee, Bill Lin, Dongwon Park, and Mingyu Woo.
\newblock Core-eco: Concurrent refinement of detailed place-and-route for an efficient eco automation.
\newblock In {\em 2021 IEEE 39th International Conference on Computer Design (ICCD)}, pages 366--373. IEEE, 2021.

\bibitem{cong2001multilevel}
Jason Cong, Jie Fang, and Yan Zhang.
\newblock Multilevel approach to full-chip gridless routing.
\newblock In {\em IEEE/ACM International Conference on Computer Aided Design. ICCAD 2001. IEEE/ACM Digest of Technical Papers (Cat. No. 01CH37281)}, pages 396--403. IEEE, 2001.

\bibitem{wei2012eco}
Xing Wei, Wai-Chung Tang, Yi~Diao, and Yu-Liang Wu.
\newblock Eco timing optimization with negotiation-based re-routing and logic re-structuring using spare cells.
\newblock In {\em 17th Asia and South Pacific Design Automation Conference}, pages 511--516. IEEE, 2012.

\bibitem{zhu2024timing}
Yuhan Zhu, Genggeng Liu, Ren Lu, Xing Huang, Min Gan, and Wenzhong Guo.
\newblock Timing-driven obstacle-avoiding x-architecture steiner minimum tree algorithm with slack constraints.
\newblock {\em IEEE Transactions on Systems, Man, and Cybernetics: Systems}, 54(5):2927--2940, 2024.

\bibitem{bhardwaj2002quantifying}
Manish Bhardwaj, Rex Min, and Anantha~P Chandrakasan.
\newblock Quantifying and enhancing power awareness of vlsi systems.
\newblock {\em IEEE Transactions on Very Large Scale Integration (VLSI) Systems}, 9(6):757--772, 2002.

\bibitem{dwibedi2024hybrid}
Rajat~Kumar Dwibedi, V~Senthil Kumar, K~Rajmohan, A~Durga Bhavani, S~Murugesan, and Mohit Tiwari.
\newblock Hybrid optimization algorithms for dynamic and static power reduction in low-power vlsi circuits.
\newblock In {\em 2024 International Conference on Advances in Computing, Communication and Materials (ICACCM)}, pages 1--6. IEEE, 2024.

\bibitem{youssef2005pomr}
Ahmed Youssef, Mohab Anis, and Mohamed Elmasry.
\newblock Pomr: A power-aware interconnect optimization methodology.
\newblock {\em IEEE transactions on very large scale integration (VLSI) systems}, 13(3):297--307, 2005.

\bibitem{cong2005thermal}
Jason Cong and Yan Zhang.
\newblock Thermal-driven multilevel routing for 3-d ics.
\newblock In {\em Proceedings of the 2005 Asia and South Pacific Design Automation Conference}, pages 121--126, 2005.

\bibitem{pak2012electromigration}
Jiwoo Pak, Sung~Kyu Lim, and David~Z Pan.
\newblock Electromigration-aware routing for 3d ics with stress-aware em modeling.
\newblock In {\em Proceedings of the International Conference on Computer-Aided Design}, pages 325--332, 2012.

\bibitem{jia2018electromigration}
Xiaotao Jia, Jing Wang, Yici Cai, and Qiang Zhou.
\newblock Electromigration design rule aware global and detailed routing algorithm.
\newblock In {\em Proceedings of the 2018 Great Lakes Symposium on VLSI}, pages 267--272, 2018.

\bibitem{zhou2019aware}
Han Zhou, Zeyu Sun, Sheriff Sadiqbatcha, Naehyuck Chang, and Sheldon X-D Tan.
\newblock Em-aware and lifetime-constrained optimization for multisegment power grid networks.
\newblock {\em IEEE Transactions on Very Large Scale Integration (VLSI) Systems}, 27(4):940--953, 2019.

\bibitem{axelou2024electromigration}
Olympia Axelou, Kostas Kolomvatsos, George Floros, Nestor Evmorfopoulos, Georg Georgakos, and George Stamoulis.
\newblock An electromigration-aware wire sizing methodology via particle swarm optimization.
\newblock In {\em Proceedings of the Great Lakes Symposium on VLSI 2024}, pages 403--408, 2024.

\bibitem{bigalke2018increasing}
Steve Bigalke, Jens Lienig, Thorben Casper, and Sebastian Sch{\"o}ps.
\newblock Increasing em robustness of placement and routing solutions based on layout-driven discretization.
\newblock In {\em 2018 14th Conference on Ph. D. Research in Microelectronics and Electronics (PRIME)}, pages 89--92. IEEE, 2018.

\bibitem{zhu2024one}
Linyu Zhu, Yichen Cai, and Xinfei Guo.
\newblock One-for-all: An unified learning-based framework for efficient cross-corner timing signoff.
\newblock In {\em Proceedings of the 43rd IEEE/ACM International Conference on Computer-Aided Design}, pages 1--9, 2024.

\bibitem{guo2024harnessing}
Xinfei Guo, Linyu Zhu, and Yichen Cai.
\newblock Harnessing graph learning for efficient timing signoff.
\newblock In {\em AI-Enabled Electronic Circuit and System Design: From Ideation to Utilization}, pages 155--187. Springer, 2024.

\bibitem{zhu2023rc}
Linyu Zhu, Yue Gu, and Xinfei Guo.
\newblock Rc-gnn: Fast and accurate signoff wire delay estimation with customized graph neural networks.
\newblock In {\em 2023 IEEE 5th International Conference on Artificial Intelligence Circuits and Systems (AICAS)}, pages 1--5. IEEE, 2023.

\bibitem{lu2023eco}
Yi-Chen Lu, Siddhartha Nath, Sai Pentapati, and Sung~Kyu Lim.
\newblock Eco-gnn: Signoff power prediction using graph neural networks with subgraph approximation.
\newblock {\em ACM Transactions on Design Automation of Electronic Systems}, 28(4):1--22, 2023.

\bibitem{wang2022graph}
Kai Wang and Peng Cao.
\newblock A graph neural network method for fast eco leakage power optimization.
\newblock In {\em 2022 27th Asia and South Pacific Design Automation Conference (ASP-DAC)}, pages 196--201. IEEE, 2022.

\bibitem{fang2018machine}
Yen-Chun Fang, Heng-Yi Lin, Min-Yan Sui, Chien-Mo Li, and Eric Jia-Wei Fang.
\newblock Machine-learning-based dynamic ir drop prediction for eco.
\newblock In {\em 2018 IEEE/ACM International Conference on Computer-Aided Design (ICCAD)}, pages 1--7. IEEE, 2018.

\bibitem{lin2025graph}
Yen-Lin Lin, Yung-Chih Chen, and Ming-Chao Lee.
\newblock Graph neural network-based glitch rate prediction at the signoff stage.
\newblock {\em ACM Transactions on Design Automation of Electronic Systems}, 2025.

\bibitem{oye2024deep}
Emma Oye, James Dillian, and Mark Wright.
\newblock A deep dive into formal verification techniques for ensuring functional correctness in custom ai accelerator architectures at the register-transfer level (rtl).
\newblock 2024.

\bibitem{anton2024vericheri}
Anna Lena~Duque Ant{\'o}n, Johannes M{\"u}ller, Philipp Schmitz, Tobias Jauch, Alex Wezel, Lucas Deutschmann, Mohammad~Rahmani Fadiheh, Dominik Stoffel, and Wolfgang Kunz.
\newblock Vericheri: Exhaustive formal security verification of cheri at the rtl.
\newblock {\em arXiv preprint arXiv:2407.18679}, 2024.

\bibitem{herklotz2021formal}
Yann Herklotz, James~D Pollard, Nadesh Ramanathan, and John Wickerson.
\newblock Formal verification of high-level synthesis.
\newblock {\em Proceedings of the ACM on Programming Languages}, 5(OOPSLA):1--30, 2021.

\bibitem{lee2013equivalence}
Jong-Hoon Lee, Junbeom Yoo, Jong~Gyun Choi, and Jang-Soo Lee.
\newblock Equivalence checking between pre-synthesis and post-synthesis programs by using vis.
\newblock In {\em Proceedings of the KNS 2013 spring meeting}, pages 1020--1021, 2013.

\bibitem{lu2025hierarchical}
Huaixi Lu, Yue Xing, Aarti Gupta, and Sharad Malik.
\newblock Hierarchical formal verification of hardware.
\newblock {\em IEEE Transactions on Computer-Aided Design of Integrated Circuits and Systems}, 2025.

\bibitem{marquez2013formal}
Carlos Ivan~Castro Marquez, Marius Strum, and Wang~Jiang Chau.
\newblock Formal equivalence checking between high-level and rtl hardware designs.
\newblock In {\em 2013 14th Latin American Test Workshop-LATW}, pages 1--6. IEEE, 2013.

\bibitem{raia2024case}
Gaetano Raia, Gianluca Rigano, David Vincenzoni, and Maurizio Martina.
\newblock A case study on formal equivalence verification between a c/c++ model and its rtl design.
\newblock In {\em International Symposium on Formal Methods}, pages 373--389. Springer, 2024.

\bibitem{tan2025rtl}
Qinhan Tan, Yuheng Yang, Thomas Bourgeat, Sharad Malik, and Mengjia Yan.
\newblock Rtl verification for secure speculation using contract shadow logic.
\newblock In {\em Proceedings of the 30th ACM International Conference on Architectural Support for Programming Languages and Operating Systems, Volume 1}, pages 970--986, 2025.

\bibitem{grimm2018survey}
Tom{\'a}s Grimm, Djones Lettnin, and Michael H{\"u}bner.
\newblock A survey on formal verification techniques for safety-critical systems-on-chip.
\newblock {\em Electronics}, 7(6):81, 2018.

\bibitem{bai2025assertionforge}
Yunsheng Bai, Ghaith~Bany Hamad, Syed Suhaib, and Haoxing Ren.
\newblock Assertionforge: Enhancing formal verification assertion generation with structured representation of specifications and rtl.
\newblock {\em arXiv preprint arXiv:2503.19174}, 2025.

\bibitem{maddala2024laag}
Karthik Maddala, Bhabesh Mali, and Chandan Karfa.
\newblock Laag-rv: Llm assisted assertion generation for rtl design verification.
\newblock In {\em 2024 IEEE 8th International Test Conference India (ITC India)}, pages 1--6. IEEE, 2024.

\bibitem{huang2024towards}
Hanxian Huang, Zhenghan Lin, Zixuan Wang, Xin Chen, Ke~Ding, and Jishen Zhao.
\newblock Towards llm-powered verilog rtl assistant: Self-verification and self-correction.
\newblock {\em arXiv preprint arXiv:2406.00115}, 2024.

\bibitem{liu2024openllm}
Shang Liu, Yao Lu, Wenji Fang, Mengming Li, and Zhiyao Xie.
\newblock Openllm-rtl: Open dataset and benchmark for llm-aided design rtl generation.
\newblock In {\em Proceedings of the 43rd IEEE/ACM International Conference on Computer-Aided Design}, pages 1--9, 2024.

\bibitem{liu2024rtlcoder}
Shang Liu, Wenji Fang, Yao Lu, Jing Wang, Qijun Zhang, Hongce Zhang, and Zhiyao Xie.
\newblock Rtlcoder: Fully open-source and efficient llm-assisted rtl code generation technique.
\newblock {\em IEEE Transactions on Computer-Aided Design of Integrated Circuits and Systems}, 2024.

\bibitem{saha2025sv}
Dipayan Saha, Shams Tarek, Hasan~Al Shaikh, Khan~Thamid Hasan, Pavan~Sai Nalluri, Md~Ajoad Hasan, Nashmin Alam, Jingbo Zhou, Sujan~Kumar Saha, Mark Tehranipoor, et~al.
\newblock Sv-llm: An agentic approach for soc security verification using large language models.
\newblock {\em arXiv preprint arXiv:2506.20415}, 2025.

\bibitem{kang2025fveval}
Minwoo Kang, Mingjie Liu, Ghaith~Bany Hamad, Syed~M Suhaib, and Haoxing Ren.
\newblock Fveval: Understanding language model capabilities in formal verification of digital hardware.
\newblock In {\em 2025 Design, Automation \& Test in Europe Conference (DATE)}, pages 1--6. IEEE, 2025.

\bibitem{SynopsysExplorerLVS}
{Synopsys, Inc.}
\newblock {Faster and Smarter LVS for the SoC Era: Introducing Explorer LVS}.
\newblock Technical report, Synopsys, Inc., 2019.
\newblock White Paper.

\bibitem{SiemensDesignStageVerification}
{Siemens Digital Industries Software}.
\newblock {A Game Changer for IP Designers: Design-Stage Verification}, 2021.
\newblock Technical Paper.

\bibitem{CadenceInnovus}
{Cadence Design Systems, Inc.}
\newblock {Innovus Implementation System}, 2024.
\newblock Product page.

\bibitem{sanghavi2010formal}
Alok Sanghavi.
\newblock What is formal verification?
\newblock {\em EE Times\_Asia}, 2010.

\bibitem{brinkmann2017formal}
Raik Brinkmann and Dave Kelf.
\newblock Formal verification—the industrial perspective.
\newblock In {\em Formal System Verification: State-of the-Art and Future Trends}, pages 155--182. Springer, 2017.

\bibitem{hsiao2021synthesizing}
Yao Hsiao, Dominic~P Mulligan, Nikos Nikoleris, Gustavo Petri, and Caroline Trippel.
\newblock Synthesizing formal models of hardware from rtl for efficient verification of memory model implementations.
\newblock In {\em MICRO-54: 54th annual IEEE/ACM international symposium on microarchitecture}, pages 679--694, 2021.

\bibitem{devarajegowda2019formal}
Keerthikumara Devarajegowda, Lorenzo Servadei, Zhao Han, Michael Werner, and Wolfgang Ecker.
\newblock Formal verification methodology in an industrial setup.
\newblock In {\em 2019 22nd Euromicro Conference on Digital System Design (DSD)}, pages 610--614. IEEE, 2019.

\bibitem{hasan2015formal}
Osman Hasan and Sofiene Tahar.
\newblock Formal verification methods.
\newblock In {\em Encyclopedia of Information Science and Technology, Third Edition}, pages 7162--7170. IGI Global Scientific Publishing, 2015.

\bibitem{jha2025large}
Chandan~Kumar Jha, Muhammad Hassan, Khushboo Qayyum, Sallar Ahmadi-Pour, Kangwei Xu, Ruidi Qiu, Jason Blocklove, Luca Collini, Andre Nakkab, Ulf Schlichtmann, et~al.
\newblock Large language models (llms) for verification, testing, and design.
\newblock In {\em 2025 IEEE European Test Symposium (ETS)}, pages 1--10. IEEE, 2025.

\bibitem{siemens_shift_left_calibre}
Siemens EDA.
\newblock Calibre shift left solutions optimize ic design flow productivity.
\newblock Technical report, Siemens, 2024.
\newblock Accessed: [Current Date, e.g., 26 August 2025].

\end{thebibliography}

\end{document}